\documentclass[a4paper,11pt]{article}
\pdfoutput=1
\usepackage{jheppub}
\usepackage{amsmath,amssymb}
\usepackage{graphicx}
\usepackage{booktabs}
\usepackage{CJKutf8}
\usepackage{orcidlink}

\title{Horizon resonances, near-extremal thermodynamics and singular non-BPS black holes in Calabi--Yau compactifications}
\author[a,b]{Alan Zhang\orcidlink{0009-0000-1187-0755},}
\author[c]{Ye Zhou (\begin{CJK*}{UTF8}{gbsn}周烨\end{CJK*})\orcidlink{0009-0004-7050-7736}}
\affiliation[a]{Department of Physics and Astronomy, McMaster University,\\Hamilton, Ontario L8S 4M1, Canada}
\affiliation[b]{Perimeter Institute for Theoretical Physics,\\Waterloo, Ontario N2L 2Y5, Canada}
\affiliation[c]{Institute of Systems and Physical Biology, Shenzhen Bay Laboratory,\\Shenzhen 518055, China}
\emailAdd{zhana142@mcmaster.ca}
\emailAdd{\\ye.zhou.horizon@gmail.com (zhouye@szbl.ac.cn)}

\abstract{The standard BPS and non-BPS black holes of supergravity have integer near-horizon scaling exponents, and their horizons have been expected to be smooth. We show that integer exponents are not enough. A resonance between modes of exponents one and two forces a logarithm at the extremal horizon, and this obstruction is visible from infinity: it fixes a universal, sign-definite $T^5\log T$ term in the near-extremal free energy, the thermal trace of resonant renormalization-group running in the near-horizon throat. In $\mathcal N=2$ supergravity with a cubic prepotential, a selection rule of special geometry removes the resonance on the aligned branch of non-BPS attractors, which explains why the non-BPS black holes of the symmetric models are smooth. Off this branch the integer spectrum itself is lost: we prove that every other non-degenerate stable non-BPS attractor with nonzero central charge has an exponent between zero and one. In every two-modulus Calabi--Yau compactification with a non-degenerate cubic form, such attractors fill open cones of charges, with exponents as small as desired, and their extremal horizons are singular for generic asymptotic moduli. At vanishing central charge the resonance can be the leading obstruction, as a cubic prepotential with integer coefficients shows.}

\keywords{Black Holes, Supergravity Models, Spacetime Singularities, Black Holes in String Theory}

\begin{document}
\maketitle
\flushbottom
\section{Introduction}

An extremal black hole has a near-horizon throat in which the moduli are attracted to values fixed by the charges~\cite{FerraraKalloshStrominger1995,Strominger1996,FerraraKallosh1996,Goldstein2005,Sen2005}. Whether the fields extend smoothly through the horizon is a separate question. A scalar perturbation of the attractor decays as $z^{c}$, where $z=r/R$, $R$ is the area radius, $r$ is an affine radial coordinate and $c(c+1)$ is an eigenvalue of the rescaled Hessian of the black-hole potential. When $c$ is not an integer, a derivative of the scalar diverges at the horizon, and for $c<1$ so do parallelly propagated tidal forces. Horowitz, Kolanowski and Santos showed that this happens generically in anti-de Sitter space, and that it leaves an anomalous power of the temperature in the near-extremal specific heat~\cite{HKS2022AdS}. Related obstructions appear in effective-field-theory corrections to Kerr and Kerr--Newman black holes~\cite{HKRS2023Kerr,HKRS2024KN}, with axions~\cite{HKRS2026Axions}, and in five-dimensional rotating solutions away from supersymmetric couplings~\cite{HorowitzSantos2024}. For four-dimensional Einstein--Maxwell theory without a cosmological constant, Horowitz, Kolanowski and Santos also noted that all exponents are positive integers, so the extremal horizon is smooth~\cite{HKS2022AdS} (see also Ref.~\cite{ChenDeRhamTolley2025}). Static multi-center horizons are analytic in four dimensions~\cite{Hartle:1972ya} but not smooth in five or more~\cite{Welch1995,CandlishReall2007,Kimura:2014uaa}. A companion Letter shows that static tides of extremal Reissner--Nordstr\"om still resonate at nonlinear order: generic tides without a common symmetry axis make the horizon non-smooth, through a term $r^4\log r$ in the Maxwell invariant at third order in the tidal amplitude, while tides with a common axis leave it analytic because the static equations are integrable~\cite{ZhangZhouLetter}.

With scalar fields the exponents are fixed by the attractor and are generically not integers. A dyonic dilaton black hole with coupling $e^{-2a\phi}F^2$ has $c(c+1)=2a^2$~\cite{GibbonsMaeda1988,Garfinkle:1990qj,Poletti1995}, so its exponent is an integer only for special couplings such as the supergravity values $a=1$ and $a=\sqrt3$, and the horizon is analytic only for these ``triangular'' values~\cite{Poletti1995,Galtsov2015}. Integer spectra are typical of the standard supergravity black holes, and the question is whether they suffice.

We show that integer exponents are not enough. With several scalars, a slow mode ($c=1$) squared has exactly the radial weight of a fast mode ($c=2$). A quadratic coupling then drives the fast mode at its own falloff and produces $z^2\log z$ at the extremal horizon, where the metric is $C^{3,\alpha}$ and the scalars are $C^{1,\alpha}$, but neither is smooth. The obstruction is a covariant tensor $\Gamma$, built from the third covariant derivative of the potential at the attractor and evaluated on the nonlinear slow amplitude $A(\xi)$ that the full exterior selects from the boundary data $\xi$. When these two exponents exhaust the spectrum, the horizon is smooth if and only if $\Gamma(A,A)=0$.

This local obstruction has a precise imprint at infinity. At fixed charges and physical temperature, and at every order in the boundary sources, the leading classical non-analytic term of the free energy is
\begin{equation}
F\supset-\frac{4Q}{9}\big\|\Gamma\big(A(\xi),A(\xi)\big)\big\|^2\,\vartheta^5\log\vartheta,\qquad\vartheta=2\pi QT,
\end{equation}
where $Q$ is the radius of the extremal horizon. The coefficient is minus a square, so no tuning of the slow amplitudes can make the signal change sign. Read in the near-horizon $\mathrm{AdS}_2$ throat, the slow and fast modes are irrelevant operators of dimensions two and three. Since $3=2\cdot2-1$, their operator product drives a logarithmic running of the dimension-three coupling, and $\Gamma$ is, up to sign, the corresponding operator-product coefficient. The thermal logarithm is the finite remainder of a pole cancellation between the infrared response and an exterior contact term, so its coefficient depends only on infrared data: $\Gamma$ and the amplitude $A(\xi)$ that the exterior selects.

\begin{figure}[t]
\centering
\includegraphics[width=\textwidth]{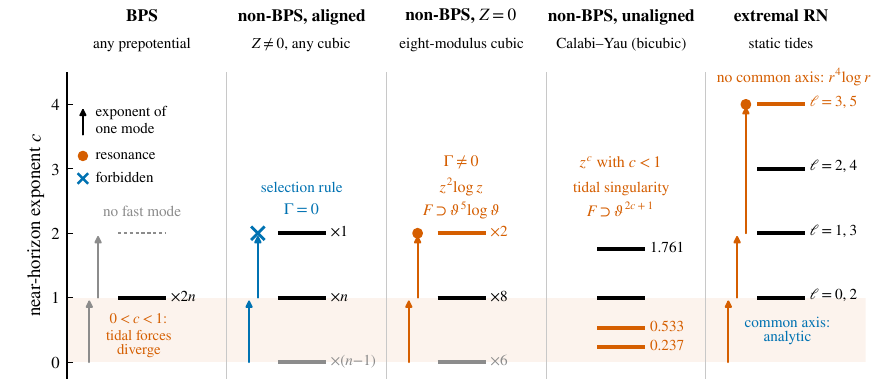}
\caption{Near-horizon exponents $c$ of extremal black holes, drawn as levels ($\Delta=c+1$ in the $\mathrm{AdS}_2$ throat; $\times k$: multiplicity; grey: flat directions). Stacked arrows add the exponents of the modes in a product; when the sum lands on a level, the product can resonate with that mode, and a nonzero coupling then forces a logarithm. The resonance $1+1\to2$ is absent for BPS attractors (no mode with $c=2$), forbidden on the aligned non-BPS branch by the selection rule of Sec.~\ref{Sec:Selection}, and realized at the $Z=0$ attractors of an eight-modulus cubic prepotential. Unaligned attractors, here of the bicubic Calabi--Yau threefold, have exponents in $0<c<1$ (shaded), where tidal forces diverge at the horizon. Reissner--Nordstr\"om tides have $c=\ell\pm1$; the product $1+1+2\to4$ resonates for generic tides without a common axis, and its logarithm cancels for tides with a common axis~\cite{ZhangZhouLetter}.}
\label{fig:spectra}
\end{figure}

Three further results, summarized in Fig.~\ref{fig:spectra}, determine where this mechanism matters. First, in $\mathcal N=2$ supergravity, where integer spectra arise naturally on non-BPS branches~\cite{LarsenZeng2018,CastroVerheijden2021}, the resonance obeys a selection rule: $\Gamma=0$ on the aligned non-BPS branch of an arbitrary cubic prepotential, a branch that contains the D0--D4 charges with D4 class inside the K\"ahler cone, and hence, by duality, on every non-BPS $Z\neq0$ attractor of the symmetric models. The same holds at every attractor with spectrum $\{1,2\}$ of the circle-reduced five-dimensional Einstein--Maxwell--Chern--Simons theory, for every Chern--Simons coupling. The ordinary cubic couplings do not vanish; the target-space connection cancels them.

Second, the selection rule protects only what the integer spectrum already provides, and off the aligned branch the spectrum itself is not protected. We prove that every stable unaligned non-BPS attractor with $Z\neq0$ of a cubic prepotential has an exponent below one, strictly between zero and one when its Hessian is non-degenerate. In every Calabi--Yau compactification with $h^{1,1}=2$ and a non-degenerate cubic form, such minima fill open cones of charges in every sector of the K\"ahler cone, with exponents as small as desired, and in random cubics with more moduli they occupy open sets of charges; the corresponding extremal horizons are singular for generic asymptotic moduli. Even on the aligned branch of a non-symmetric cubic, where $\Gamma=0$, the flat directions are generically lifted only at quartic order; where this lift stabilizes the attractor, the classical extremal horizon is again singular for generic moduli, as the metric backreaction found in Ref.~\cite{TripathyTrivedi2005} already implies.

Third, at vanishing central charge the resonance is realized. A stable non-BPS attractor with $Z=0$ and exponents one and two must have flat directions, and we construct a cubic prepotential with integer coefficients that has such attractors with $\Gamma\neq0$; for generic asymptotic moduli the resonance is then the leading obstruction, and the free energy carries its thermal term.

Near-extremal thermodynamics distinguishes these mechanisms: a resonance gives $\vartheta^5\log\vartheta$, a non-integer exponent $c$ gives $\vartheta^{2c+1}$, and a quartically lifted flat direction gives $1/\log^2T$ in the entropy; complete finite-temperature solutions confirm the first two coefficients at fixed nonzero sources.

Section~\ref{Sec:Resonance} sets up the resonance, the smoothness criterion and its infrared reading. Section~\ref{Sec:Thermal} derives the thermal signature and tests it at finite source. Section~\ref{Sec:Selection} proves the selection rule, delimits it and constructs attractors with $Z=0$ at which the resonance leads, and Section~\ref{Sec:CY} applies these results to Calabi--Yau compactifications. Section~\ref{Sec:Discussion} states the scope and outlook. The main text can be read on its own. The appendices give the complete proofs and checks: Appendices~\ref{App:Branches}--\ref{App:Exchange} for Secs.~\ref{Sec:Resonance} and~\ref{Sec:Thermal}, Appendix~\ref{App:Special} for the special geometry of Sec.~\ref{Sec:Selection}, Appendix~\ref{App:CY} for the Calabi--Yau data, exterior solutions and corrections, Appendix~\ref{App:Related} for two related models, Appendix~\ref{App:Detuning} for the detuned response and the infrared dictionary, and Appendices~\ref{App:Numerics}--\ref{App:Reproducibility} for the finite-source numerics and the availability of code and data.
\section{Integer exponents and horizon resonances}\label{Sec:Resonance}

We work with four-dimensional Einstein gravity coupled to Maxwell fields and neutral scalars, in units $G_4=1$,
\begin{equation}
\begin{aligned}
I_4=\frac1{16\pi}\int d^4x\sqrt{-g}\,\Big[&\mathcal R-2G_{ij}(\phi)\,\partial_\mu\phi^i\partial^\mu\phi^j-f_{AB}(\phi)F^A_{\mu\nu}F^{B\mu\nu}\Big],
\end{aligned}
\end{equation}
with positive analytic $G$ and $f$ and no scalar potential. For static spherical black holes at fixed charges, the Maxwell fields can be eliminated in favor of the effective potential $V(\phi)=q_A(f^{-1})^{AB}q_B$; magnetic charges and theta terms enter through the same positive analytic potential after a Routhian reduction. In $\mathcal N=2$ supergravity, $V=|Z|^2+g^{a\bar b}D_aZ\,\overline{D_bZ}$ is the familiar black-hole potential~\cite{FerraraGibbonsKallosh1997}.

\paragraph{Attractor exponents.} An extremal horizon sits at a critical point $p$ of $V$, with area $4\pi Q^2$, where $Q^2=V(p)$~\cite{FerraraKalloshStrominger1995,FerraraGibbonsKallosh1997,Goldstein2005,Sen2005}. Linear scalar perturbations decay toward the horizon as $z^{c}$, where $z=r/R$ with $R$ the area radius and $r$ an affine radial coordinate, and
\begin{equation}
c(c+1)\in\operatorname{spec}\mathsf H,\qquad \mathsf H=\frac{G^{-1}\nabla^2V}{2Q^2}\Big|_p .
\end{equation}
Every BPS attractor of $\mathcal N=2$ supergravity has $c=1$ in all vector-multiplet directions~\cite{FerraraGibbonsKallosh1997}; hypermultiplet scalars decouple as exact moduli. Non-integer $c$ makes the scalar non-smooth at the horizon already at linear order~\cite{HKS2022AdS}. We are interested in the opposite case, where the exponents are integers. The simplest non-trivial spectrum, which appears on the non-BPS branches of supergravity~\cite{LarsenZeng2018,CastroVerheijden2021}, contains slow modes $E_1=\ker(\mathsf H-2)$ with $c=1$ and fast modes $E_2=\ker(\mathsf H-6)$ with $c=2$. In this section and the next we assume that these exhaust the scalar spectrum at an isolated attractor, $T_p=E_1\oplus E_2$, with $G$ and $V$ analytic near $p$; flat directions, which also occur in supergravity, are treated in Secs.~\ref{Sec:Selection} and~\ref{Sec:CY}.

\paragraph{The resonance.} Near the horizon a slow mode squared has radial weight $z^2$, the weight of a fast mode. The extremal flow therefore has one possible resonant term. In analytic normal coordinates, with $s=-\log z$,
\begin{equation}
\frac{dX}{ds}=-X,\qquad \frac{dY}{ds}=-2Y-\Gamma(X,X),
\label{eq:normalform}
\end{equation}
where the resonance map $\Gamma:\mathrm{Sym}^2E_1\to E_2$ is
\begin{equation}
\big\langle w,\Gamma(u,v)\big\rangle_G=\frac{(\nabla^3V)_p(w,u,v)}{20\,Q^2}.
\label{eq:Gammadef}
\end{equation}
Only the covariant third derivative enters: ordinary third derivatives of $V$ can be nonzero while $\Gamma$ vanishes, and in supergravity this is the typical situation (Sec.~\ref{Sec:Selection}). Integrating~\eqref{eq:normalform} gives
\begin{equation}
\begin{aligned}
X&=z\,A(\xi),\\
Y&=z^2\big[B(\xi)+\Gamma\big(A(\xi),A(\xi)\big)\log z\big],
\end{aligned}
\label{eq:profiles}
\end{equation}
where the amplitudes $A$ and $B$ are fixed by the boundary values $\xi=\phi_\infty$ of the scalars through the complete extremal exterior. The slow amplitude $A$ is unique in this normalization. Under these assumptions the full configuration (metric, scalars and field strengths) extends smoothly through the horizon if and only if
\begin{equation}
\Gamma\big(A(\xi),A(\xi)\big)=0,
\label{eq:smooth}
\end{equation}
in which case the extension is analytic (Appendix~\ref{App:Branches}).

\paragraph{Regularity class.} A nonzero resonance makes the horizon non-smooth but only mildly so. Along an ingoing radial null geodesic with affine parameter $\lambda=r$ and tangent $k$, the scalar $\phi$, a vector on the target space, has the component $(r/Q)^2\Gamma(A,A)\log(r/Q)$ along $E_2$ in normal-form coordinates. The scalars, and in general the field strengths, are therefore $C^{1,\alpha}$ for every $\alpha<1$; a single purely electric field strength is $C^{3,\alpha}$, like the metric (Appendix~\ref{App:Branches}). The focusing equation $R''=-R|\phi'|_G^2$ makes the area radius, and hence the metric, $C^{3,\alpha}$ but not $C^4$. All components of the Riemann tensor and of its first covariant derivative in a parallelly propagated frame stay finite. The first divergent quantities are
\begin{equation}
\begin{aligned}
k^ak^b\nabla_a\nabla_b\phi&\simeq\frac{2\,\Gamma(A,A)}{Q^2}\log|\lambda|,\\
\nabla_k\nabla_kR_{kk}&\simeq\frac{16\,\|\Gamma(A,A)\|^2}{Q^4}\log^2|\lambda|,
\end{aligned}
\label{eq:firstdivergence}
\end{equation}
where the first line is $D^2\phi/d\lambda^2$ with the target connection. These divergences are invariant: in an extension with $C^2$ scalars and a $C^1$ metric, $D^2\phi/d\lambda^2$ would be finite along the radial geodesics; in a $C^4$ extension these geodesics are $C^5$ and $d^2R_{kk}/d\lambda^2$ would be finite. No extension of higher regularity therefore exists. The leading curvature divergence is controlled by $\|\Gamma(A,A)\|^2$, the invariant that also fixes the thermal coefficient of Sec.~\ref{Sec:Thermal}.

\paragraph{Infrared reading.} The throat of an extremal black hole is $\mathrm{AdS}_2\times S^2$, and a mode of exponent $c$ corresponds to an operator of dimension $\Delta=c+1$. We use this language as a reading of the classical throat computation (Appendix~\ref{App:IR}). The Jackiw--Teitelboim mode, which moves the area radius, also has dimension two, but its cubic couplings to the matter fields are diagonal in the mass basis and contain no term quadratic in the Jackiw--Teitelboim field~\cite{CastroVerheijden2021}, so it cannot feed the resonance. Slow and fast modes are irrelevant operators $\mathcal O_2$ and $\mathcal O_3$; their couplings at radial scale $z$ are $A$ and $B+\Gamma(A,A)\log z$ in Eq.~\eqref{eq:profiles}. Because $\Delta_3=2\Delta_2-1$, the operator product $\mathcal O_2\times\mathcal O_2\to\mathcal O_3$ produces a logarithmic divergence in conformal perturbation theory, and the dimension-three coupling runs. The resonance map is this operator-product coefficient up to sign: with sources coupled as $\exp(+\int d\tau\,g\,\mathcal O)$ and the throat two-point functions of Appendix~\ref{App:IR}, $\mathcal O_{2a}(t)\,\mathcal O_{2b}(0)\supset\mathcal C_{ab}{}^i\,|t|^{-1}\mathcal O_{3i}(0)$ with $\mathcal C_{ab}{}^i=-\Gamma^i{}_{ab}$, and
\begin{equation}
\mu\,\partial_\mu g_3=\Gamma(g_2,g_2),\qquad g_2=A(\xi),
\end{equation}
which integrates to the profile~\eqref{eq:profiles} with $\mu\propto z$. The horizon logarithm is therefore the classical renormalization-group mixing familiar from holographic renormalization in higher dimensions~\cite{BanadosSchwimmerTheisen2006,vanRees2011b,BMS2015}, here at the end of the throat rather than at a conformal boundary. The full exterior enters only through $A(\xi)$, which plays the role of the infrared couplings reached from the ultraviolet data $\xi$.
\section{The thermodynamic signature}\label{Sec:Thermal}

At positive temperature the throat ends at a finite radial scale, and the running of the dimension-three coupling is cut off there. We hold the charges, the asymptotic time normalization and the physical temperature fixed, and use the source coordinates given by the exponential map at the attractor. Write $\vartheta=2\pi QT$ and let $F_\alpha(\vartheta)$ be the Taylor coefficient of the free energy $F=M-TS$ at a fixed multi-index $\alpha$ of the boundary sources.

\paragraph{Main result.} For a complete scalar spectrum $\{1,2\}$ and every fixed $\alpha$,
\begin{equation}
\begin{aligned}
F_\alpha(\vartheta)={}&P_{\alpha,5}(\vartheta)+C_\alpha\,\vartheta^5\log\vartheta+O_\alpha\!\big(\vartheta^6[1+|\log\vartheta|^{p_\alpha}]\big),\\
C_\alpha={}&-\frac{4Q}{9}\,[\xi^\alpha]\,\big\|\Gamma\big(A(\xi),A(\xi)\big)\big\|^2,
\end{aligned}
\label{eq:main}
\end{equation}
where $P_{\alpha,5}$ is a polynomial of degree five and $p_\alpha$ is a finite power. There are no earlier thermal logarithms and no squared logarithm at order five. The leading classical non-analytic term of the free energy is therefore fixed by the same tensor that obstructs horizon smoothness, evaluated on the infrared amplitude rather than on the boundary data. Correspondingly $S\supset\frac{40\pi Q^2}{9}\|\Gamma(A,A)\|^2\vartheta^4\log\vartheta$ and $M\supset\frac{16Q}{9}\|\Gamma(A,A)\|^2\vartheta^5\log\vartheta$, with the same amplitude $A(\xi)$ at every order in the sources. Because the coefficient is minus a sum of squares, orthogonal fast channels cannot cancel, and at a given source the sign of the signal is fixed; individual mixed Taylor coefficients can have either sign. The coefficient can also be read at the horizon: $C_\alpha=-\frac{Q^5}{36}[\xi^\alpha]K_{\rm ext}=-Q[\xi^\alpha]\|\mathsf D_h\|^2$. Here $K_{\rm ext}$ is the coefficient of $\log^2|\lambda|$ in $\nabla_k\nabla_kR_{kk}$ at the extremal horizon, Eq.~\eqref{eq:firstdivergence}, and $\mathsf D_h=\frac23\Gamma(A,A)$ is the coefficient of $\vartheta^2\log\vartheta$ in the fast component of the scalar at the warm horizon (Appendix~\ref{App:IRresponse}).

\paragraph{The quartic exchange.} The mechanism is visible at fourth order in a straight source $\epsilon(a+b)$, $a\in E_1$, $b\in E_2$. In the compact coordinate $u\in[k,1]$ of the Reissner--Nordstr\"om background (horizon at $u=k$, infinity at $u=1$), the static scalar operator is exactly the $\mathrm{AdS}_2$ black-hole operator $L_c=\partial_u((u^2-k^2)\partial_u)-c(c+1)$, negative definite on profiles regular at the horizon and vanishing at infinity. Its negative definiteness fixes the sign of the exchange below. Here $k$ is the non-extremality parameter, $\vartheta=k+O(k^2)$, and $M_0=Q/\sqrt{1-k^2}$ is the background mass. The slow source produces the fast forcing $5\,\Gamma(a,a)\,u^2$, whose regular inverse contains $\log[(u+k)/(1+k)]$. Scalar work at fixed charges gives the non-analytic part
\begin{equation}
F_4\big|_{\rm non\text{-}an.}=\frac{M_0}{2}\sum_{c=1,2}\int_k^1J_c\,L_c^{-1}J_c\,du\Big|_{\rm non\text{-}an.},
\end{equation}
where $J_c=\tfrac12\Pi_c\nabla v_3(\phi_1)$ is the quadratic forcing in the channel of exponent $c$: $\Pi_c$ projects onto that channel, $v_3$ is the cubic term of $v=V/Q^2$ in target normal coordinates and $\phi_1$ is the linear solution (Appendix~\ref{App:Exchange}). In the fast channel this forcing is the term $5\,\Gamma(a,a)\,u^2$ found above, with response $\frac52\Gamma(a,a)\mathcal Y$, where $\mathcal Y$ is the regular solution of $L_2\mathcal Y=2u^2$ that vanishes at $u=1$. The single endpoint residue $[\int_k^1u^2\mathcal Y\,du]_{\log k}=-\tfrac{16}{225}k^5$ yields $-\frac{4Q}{9}\|\Gamma(a,a)\|^2\vartheta^5\log\vartheta$. All other quadratic forcings are either non-resonant or of degree three or four, and contribute only at higher orders in $k$ (Appendix~\ref{App:Exchange}). At higher source orders the boundary tangent $a$ is replaced by the nonlinear amplitude $A(\xi)$; the identification follows from a matched expansion of the thermal hierarchy onto the extremal stable flow and a symplectic-flux argument (Appendix~\ref{App:Matching}).

\paragraph{Infrared interpretation of the coefficient.} Detuning the fast exponent to $c=2+\delta$ makes the mechanism transparent (Appendix~\ref{App:Detuning}). The infrared response of the forced fast mode then contributes a term $\propto k^{2c+1}$ with a coefficient that has a simple pole at $\delta=0$, because the static $\mathrm{AdS}_2$ response kernel vanishes linearly at integer dimension while the square of the resonantly induced amplitude has a double pole. The exterior contributes an analytic $k^5$ term with the opposite pole. Their sum is finite and produces $\vartheta^5\log\vartheta$, with a coefficient set by the residue, i.e.\ by the throat theory once the infrared amplitude $A$ is given. The result has a renormalization-group form (Appendix~\ref{App:RGform}): at fixed non-extremality $\mathsf b=k/\sqrt{1-k^2}=\vartheta[1+O(\vartheta)]$, the fifth-order scalar work is $Q\,\mathsf b^5[q_4(A)-\frac49\langle\Gamma(A,A),B+\Gamma(A,A)\log\mathsf b\rangle]$, with $q_4$ a scheme-dependent contact quartic. The thermal scale evaluates the running dimension-three coupling, and no $\log^2$ appears because the static response vanishes at integer dimension. Small corrections can shift the exponents to $1+\delta_1$ and $2+\delta_2$, as higher-derivative terms do in the examples of Refs.~\cite{HKRS2023Kerr,ChenDeRhamTolley2025}. The horizon is then non-smooth already at linear order, the resonance is detuned by $\delta_2-2\delta_1$, and the shifted modes add their own non-analytic linear responses. To leading order in the detunings and for $|\delta_i\log\vartheta|\ll1$, the resonant logarithm keeps the coefficient of Eq.~\eqref{eq:main} in the window of Eq.~\eqref{eq:window} (Appendix~\ref{App:Detuning}). This window is wide for corrections suppressed by $\ell_{\rm hd}^2/Q^2$, where $\ell_{\rm hd}$ is the length scale of the higher-derivative terms.

\paragraph{A delayed signal.} The nonlinear amplitude matters. In the three-scalar model $W/Q=1+\tfrac12(x_1^2+x_2^2)+y^2+\gamma y(x_1^2-x_2^2)+\lambda x_1^3$, the straight source $(\epsilon,\epsilon,0)$ has $\Gamma(a,a)=0$, so the quartic signal vanishes, yet $\Gamma(A,A)=-6\gamma\lambda\epsilon^3+O(\epsilon^4)$: the extremal horizon is not smooth. A direct sixth-order thermal calculation confirms $[[\epsilon^6](F_\gamma-F_0)]_{\vartheta^5\log\vartheta}=-16Q\gamma^2\lambda^2$, the square of the cubic obstruction.

\begin{figure}[t]
\centering
\includegraphics[width=\textwidth]{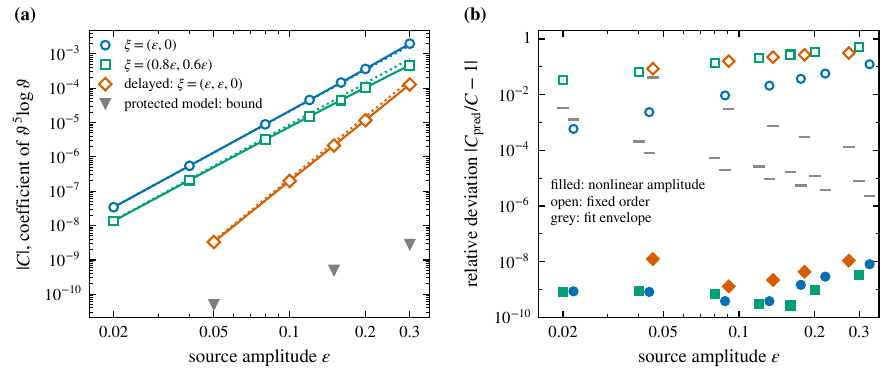}
\caption{Thermal logarithm at fixed nonzero boundary source. (a) Magnitude of the coefficient $C$ of $\vartheta^5\log\vartheta$ in the free energy, extracted from complete nonlinear solutions at fixed charges and physical temperature (open symbols), compared with the prediction $-\frac{4Q}{9}\|\Gamma(A(\xi),A(\xi))\|^2$ evaluated with the extremal infrared amplitude $A(\xi)$ (solid lines). Circles and squares: the exponential model, $v=\frac12[e^{2\sqrt3y}\cosh(2\sqrt2x)+e^{-2\sqrt3y}]$ with flat target metric, for the sources $\xi=(\epsilon,0)$ and $(0.8\epsilon,0.6\epsilon)$ in the coordinates $(x,y)$; diamonds: the three-scalar model of Sec.~\ref{Sec:Thermal} with $\gamma=\frac35$, $\lambda=\frac15$ along $(\epsilon,\epsilon,0)$, where the quartic term vanishes and the signal starts at order $\epsilon^6$; triangles: upper bounds in a supergravity model ($t^3$ with D0--D6 charges) in which the resonance is forbidden. Dotted lines: the leading term of the prediction expanded in the boundary source. (b) Relative deviation $|C_{\rm pred}/C-1|$ of the prediction from the extracted coefficient, evaluated with the nonlinear amplitude $A(\xi)$ (filled symbols) and at fixed order, i.e.\ as its leading term in the boundary source (open symbols); symbols are slightly displaced horizontally for visibility. Grey ticks mark the conservative fitting envelopes of Appendix~\ref{App:Numerics}, relative to $C$.}
\label{fig:finite}
\end{figure}

\paragraph{Beyond the source expansion.} Equation~\eqref{eq:main} holds coefficientwise. To test it at fixed nonzero source we solved the complete nonlinear static equations at fixed charges and physical temperature, for 22 boundary sources and 48 temperatures each in $10^{-3}\le\vartheta\le10^{-1}$, with spectral collocation at 80-digit precision (Appendix~\ref{App:Numerics}). The coefficient of $\vartheta^5\log\vartheta$ is extracted from the free energy and compared with $-\frac{4Q}{9}\|\Gamma(A(\xi),A(\xi))\|^2$, where the extremal amplitude is computed independently from the Hamilton--Jacobi flow. For the 19 sources with a nonzero prediction the central coefficients differ from the prediction by at most $1.3\times10^{-8}$ in relative terms (Fig.~\ref{fig:finite}), separate fits of the entropy agree, and so does the fast horizon field through $-Q\|\mathsf D_h\|^2$. The numerical error is far below these discrepancies, fits that allow logarithms at orders two to four find none, and other fitting windows and bases move the coefficient by less than $0.4\%$ of the signal, except for the smallest delayed signal ($4\%$). Along the direction where $\Gamma(a,a)=0$ the signal appears at sixth order, as predicted. In a supergravity model where the resonance is forbidden, the extracted coefficient is zero to within $3\times10^{-9}$. The fixed-order expression in the boundary source, by contrast, misses the coefficient by more than $50\%$ at $\epsilon=0.3$ along the generic exponential direction: the thermal logarithm measures the infrared amplitude, not the boundary data.

\paragraph{Regime of validity.} The signal is classical. In our ensemble the angular momentum is not fixed, and the Schwarzian and $SO(3)$ rotational zero modes each add $-\tfrac32T\log T$ to the free energy, $-3T\log T$ in total, for $T_q\lesssim T\ll1/Q$~\cite{IliesiuTuriaci2020,IliesiuMurthyTuriaci2022}, where $T_q=M_{SL(2)}$ is $1/Q^3$ in Planck units for Reissner--Nordstr\"om; at finite source this scale is set by the linear entropy coefficient of the exterior. Terms linear in $T$ only shift the extremal entropy, so only $T\log T$ competes with the classical non-analytic term. The classical term dominates when
\begin{equation}
\vartheta\gtrsim\Big(\frac{27}{8\pi}\Big)^{1/4}\big(Q\|\Gamma(A,A)\|\big)^{-1/2},
\end{equation}
where $(27/8\pi)^{1/4}\simeq1.02$. Together with $\vartheta\ll1$, this condition leaves a wide window when $Q\|\Gamma(A,A)\|\gg1$, as for macroscopic black holes away from protected directions. Other quantum non-analytic terms, from the one-loop determinants of the light fields and from $1/Q^2$ corrections to the Schwarzian coupling, are not estimated here.

\section{Selection rule and spectral dichotomy in \texorpdfstring{$\mathcal N=2$}{N=2} supergravity}\label{Sec:Selection}

Integer spectra are common in supergravity. Every BPS attractor has $c=1$ in all vector-multiplet directions, so it has no fast mode and no resonance. The non-BPS branches with $Z\neq0$~\cite{Kallosh:2005ax,Kallosh:2006bt} are different: the scalar spectrum contains flat directions~\cite{TripathyTrivedi2005,Ferrara:2007pc,FerraraMarrani2007}, slow modes and one fast mode, for instance $c=(0^2,1^3,2)$ for the STU model~\cite{CastroVerheijden2021} and $c=(0^{42},1^{27},2)$ for $\mathcal N=8$~\cite{Ferrara:2007pc,LarsenZeng2018}. The cubic couplings of these modes do not vanish in general. Given the mechanism of Sec.~\ref{Sec:Resonance}, one might therefore expect resonant logarithms at the horizons of non-BPS supergravity black holes. On the aligned branch defined below, and hence for every non-BPS black hole with $Z\neq0$ of the symmetric models, these logarithms are absent.

\paragraph{Theorem (selection rule).} Let $p$ be a regular non-BPS $Z\neq0$ attractor of four-dimensional $\mathcal N=2$ supergravity with a cubic prepotential $F=-\frac16d_{abc}X^aX^bX^c/X^0$ and positive-definite metric, on the aligned branch, defined by $\mathcal D_aZ=\zeta L_a$ with $\zeta\neq0$ (equivalently, the matter charges have no component tangent to the surface of constant volume; Appendix~\ref{App:Special}). Here $\mathcal D_a$ is the K\"ahler-covariant derivative, $L_a=\partial_{y^a}D/(2D)$ and $D=\frac16d_{abc}y^ay^by^c$ is the volume of the moduli $t^a=x^a+iy^a$. Then
\begin{equation}
\Gamma\big|_{E_2\otimes\mathrm{Sym}^2E_1}=0 .
\end{equation}
For the symmetric models (the very special geometries with covariantly constant $C_{abc}=e^Kd_{abc}$~\cite{Cremmer:1984hc,deWit:1991nm}, including $t^3$, $st^2$ and STU), every regular non-BPS $Z\neq0$ attractor is related by a duality transformation to an aligned one~\cite{BellucciEtAl2006}. Alignment refers to the chosen cubic frame and is not preserved by these dualities, but $\Gamma$ is a tensor and is transported by them, so $\Gamma=0$ at every such attractor. For a generic non-symmetric cubic, whose continuous dualities are the axion shifts, dilations and automorphisms of $d_{abc}$, alignment is duality invariant. The homogeneous non-symmetric models, which exist for $n\ge3$, have further isometries~\cite{deWit:1991nm,deWit:1992wf} that we do not consider. The flat directions of the symmetric models are generated by the stabilizer of the charges and are exact moduli along which the spectrum is constant~\cite{FerraraMarrani2007}, so Appendix~\ref{App:flatthermal} applies with $\Gamma_p=0$ at every attractor: the horizons are analytic for all asymptotic moduli, and the free energy has no $\vartheta^5\log\vartheta$ term. For STU the flat directions can also be removed by a Noether reduction (Appendix~\ref{App:Special}). For $\mathcal N=8$, the $USp(8)$ stabilizer of the attractor splits the 70 scalars into $42+27+1$ (flat, slow, fast); since $\mathrm{Sym}^2\mathbf{27}$ contains a single singlet, $\Gamma$ is proportional to the invariant metric on $E_1$, and its vanishing in the STU truncation gives $\Gamma=0$.

\paragraph{Mechanism.} For D0--D4 charges with D4 class inside the K\"ahler cone the attractor has vanishing axions, $E_1$ consists of the saxion (volume) directions and $E_2$ is the axion direction along the magnetic charge vector. Reversing all axions is an isometry of the target that preserves $V$, so a coupling with one axion and two saxions vanishes. The aligned branch of a general cubic has two phases, separated by the D0--D6 charges. Up to axion shifts, the phase that contains the D0--D4 charges is the family of non-supersymmetric D0--D4--D6 attractors with moduli proportional to the magnetic charges studied in Ref.~\cite{NTT2007}, for which the Hessian has $n+1$ positive and $n-1$ vanishing eigenvalues and the zero modes have no cubic self-couplings~\cite{TripathyTrivedi2005,NTT2007}. The other phase, which contains the D2--D6 charges, is not of this form (Appendix~\ref{App:CY}). Equation~\eqref{eq:alignedjet} below, derived in Appendix~\ref{App:Special}, extends this eigenvalue count and the absence of cubic zero-mode self-couplings to the whole branch. The resonance requires the complete covariant cubic tensor, including its components along the slow and fast directions, and this tensor follows from cubic homogeneity alone. Writing a displacement as $\delta t=(\delta x,\delta y)$ and rotating it by an angle set by the phase of $\zeta$ into components $\hat a$ and $\hat b$, one finds at an aligned attractor
\begin{equation}
\begin{aligned}
\nabla^2V(\delta t,\delta t)&=4Q^2\big[(L\cdot\hat a)^2+g(\hat b,\hat b)\big],\\
\nabla^3V(\delta t,\delta t,\delta t)&=-\frac{Q^2}{D}\,d_{abc}\hat b^a\hat b^b\hat b^c .
\end{aligned}
\label{eq:alignedjet}
\end{equation}
The slow space is $\{\hat a=0\}$, while the fast and flat directions have $\hat b=0$; hence the whole covariant cubic tensor vanishes as soon as one argument is fast or flat, and in particular $\Gamma=0$. No symmetry of the cubic form is used. The ordinary third derivative of $V$ need not vanish: for the $t^3$ model with D0--D6 charges its resonant projection is $2/(5\sqrt3)$, and the target-space connection cancels it exactly. In the language of Sec.~\ref{Sec:Resonance}, the operator-product coefficient that would drive resonant running vanishes by a selection rule of special geometry.

\paragraph{Off the aligned branch.} The algebraic identity behind the selection rule needs alignment. An unaligned critical point of a three-modulus cubic can have slow and fast exponents with $\Gamma=(9-\sqrt{33})/15\neq0$, although it also has exponents below one; a non-cubic term $(t-i)^5$ added to the $t^3$ prepotential keeps the spectrum $\{1,2\}$ while giving $\Gamma\neq0$ (Appendix~\ref{App:Special}). Such non-cubic terms are absent at large volume. What holds off the aligned branch is a statement about the spectrum itself.

\paragraph{Theorem (unaligned branches).} Let $p$ be a regular non-BPS $Z\neq0$ critical point of a cubic prepotential, with $D>0$ and positive-definite metric, that is not aligned. Then the smallest eigenvalue of $\mathsf H$ is smaller than $2$. If $p$ is a local minimum with non-degenerate Hessian, it therefore has an exponent $0<c<1$.

\paragraph{Proof.} In an orthonormal frame whose first vector is the radial direction $y$, write $\mathcal D_aZ=(u,v)$, with $v$ the components transverse to $y$; alignment means $v=0$. The radial component of the attractor equations gives $|2u^2-v\cdot v|=2\sqrt3\,|Zu|$, and the smallest Rayleigh quotient $\varrho$ of $\mathsf H$ on the two real radial directions is
\begin{equation}
\varrho=\frac{2|Z|^2+\frac{14}{3}|u|^2+\frac23\|v\|^2-\frac{8}{\sqrt3}|Zu|}{Q^2},
\end{equation}
so that
\begin{equation}
\varrho-2=\frac{4}{3Q^2}\Big(2|u|^2-\|v\|^2-\big|2u^2-v\cdot v\big|\Big)\le0 .
\end{equation}
If the inequality is strict, the theorem follows. Equality forces $u$ and $v$ to be real up to a common phase, with $0<\|v\|^2<2|u|^2$ because $Z\neq0$. On the complex plane spanned by the radial direction and $v$, the restriction of $\mathsf H$ is then the Hessian of a two-modulus family that we solve in closed form. This Hessian has the exact eigenvalue $2$ and a $3\times3$ block $S$ with $\det(2I-S)>0$, so $S$ has an eigenvalue below $2$, and the min--max principle again gives $h_{\min}<2$ (Appendix~\ref{App:Special}). The bound uses no symmetry and no compactification data.

Since aligned attractors carry $n-1$ flat directions, for $n\ge2$ every regular non-BPS $Z\neq0$ local minimum of a cubic prepotential with non-degenerate Hessian is unaligned and has an exponent strictly between zero and one. Together with the selection rule, this result gives a dichotomy for the non-BPS $Z\neq0$ branches of cubic prepotentials. On the aligned branch the spectrum is the integer one and the resonance is forbidden, although for a non-symmetric cubic the flat directions are generically lifted only at quartic order, which makes the horizon non-smooth in another way (Sec.~\ref{Sec:CY}). At a non-degenerate unaligned local minimum an exponent between zero and one is present, generic asymptotic moduli excite it (Sec.~\ref{Sec:CY}), and the horizon is singular. On these branches, smooth extremal horizons with generic asymptotic moduli therefore require alignment or a degenerate attractor, and for generic boundary data the resonance is never the only obstruction, except possibly at degenerate unaligned minima. Tuned data remain smooth: when the asymptotic moduli take their attractor values, the scalars are constant and the horizon is that of Reissner--Nordstr\"om. For STU, Castro and Verheijden found that the extremal couplings of the non-BPS branch, with $\Delta_i=\Delta_j+\Delta_k$, vanish~\cite{CastroVerheijden2021}. The resonant coupling, with $\Delta_3=2\Delta_2-1$, is not of this type, and the selection rule extends their observation to it. The smoothness of the symmetric-model horizons is consistent with the known explicit non-BPS solutions~\cite{GimonLarsenSimon2008,BossardMichelPioline2010}.

\paragraph{Vanishing central charge.} These statements assume $Z\neq0$. At non-BPS attractors with $Z=0$~\cite{Bellucci:2007zi} the radial bound still gives $h_{\min}\le2$, and a stable attractor whose positive exponents are exactly $1$ and $2$ must have flat directions (Appendix~\ref{special:z0}). Such attractors exist, and at them the resonance is the leading obstruction. An eight-modulus cubic prepotential with integer coefficients $d_{abc}$ and integer charges has a manifold of stable non-BPS attractors with $Z=0$ on which the spectrum is constant, $c=(0^6,1^8,2^2)$, and $\Gamma\neq0$. Its flat modes couple neither to two slow nor to two fast modes. With the endpoint $p$ on this manifold held fixed, the analysis of Sec.~\ref{Sec:Resonance} goes through with $\Gamma_p$, and so does the thermal theorem of Sec.~\ref{Sec:Thermal}, with $\Gamma$ evaluated at the endpoint selected by the asymptotic moduli (Appendix~\ref{App:flatthermal}). For an open dense set of nearby asymptotic moduli the first non-analytic term at the extremal horizon is therefore $z^2\log z$, and the free energy contains the term~\eqref{eq:main}. Integrating the full nonlinear flow reproduces the coefficient of the horizon logarithm to $2.4\times10^{-14}$. The masses in this example are tuned by the coefficients of the cubic, and we do not know whether such a cubic occurs as the intersection form of a Calabi--Yau threefold.

\paragraph{A non-supersymmetric family.} Away from supergravity the integer spectrum itself becomes rare. Consider five-dimensional Einstein--Maxwell--Chern--Simons theory with Chern--Simons coupling $\lambda$ ($\lambda=1$ is minimal supergravity), reduced on a circle. At fixed charges the effective potential is
\begin{equation}
\begin{aligned}
2V_\lambda={}&y^3(p^0)^2+3y(p^1-xp^0)^2+\frac{[q_0+xq_1+\lambda(3x^2p^1-x^3p^0)]^2}{y^3}\\
&+\frac{[q_1+\lambda(6xp^1-3x^2p^0)]^2}{3y},
\end{aligned}
\end{equation}
which reduces to the $t^3$ potential at $\lambda=1$. For every real $\lambda$, every regular attractor whose spectrum is exactly $\{1,2\}$ has $\Gamma=0$; this follows from an exact elimination after using the shift and scaling isometries to place the attractor at $(x,y)=(0,1)$ (Appendix~\ref{special:emcs}). If $p^1-x_*p^0\neq0$, with $x_*$ the axion at the attractor, such a spectrum occurs only for $\lambda\in\{\pm1,\pm\frac32\}$, and at $\lambda=\pm\frac32$ the zero is enforced by axion parity. For $\lambda\neq1$ the spectrum is generically not integer: on the non-BPS D0--D4 branch with $\lambda\ge0$ the axion exponent is $c=2\lambda$, and for the charges $(q_0,q_1;p^0,p^1)=(1,0.7;0.3,1)$ one finds $c=(1,2)+(0.129,1.855)(\lambda-1)+O((\lambda-1)^2)$. This is a static four-dimensional counterpart of the sensitivity to the Chern--Simons coupling found for rotating five-dimensional horizons~\cite{HorowitzSantos2024}: away from supergravity it is the integrality of the spectrum, not the selection rule, that fails first.
\begin{figure}[t]
\centering
\includegraphics[width=\textwidth]{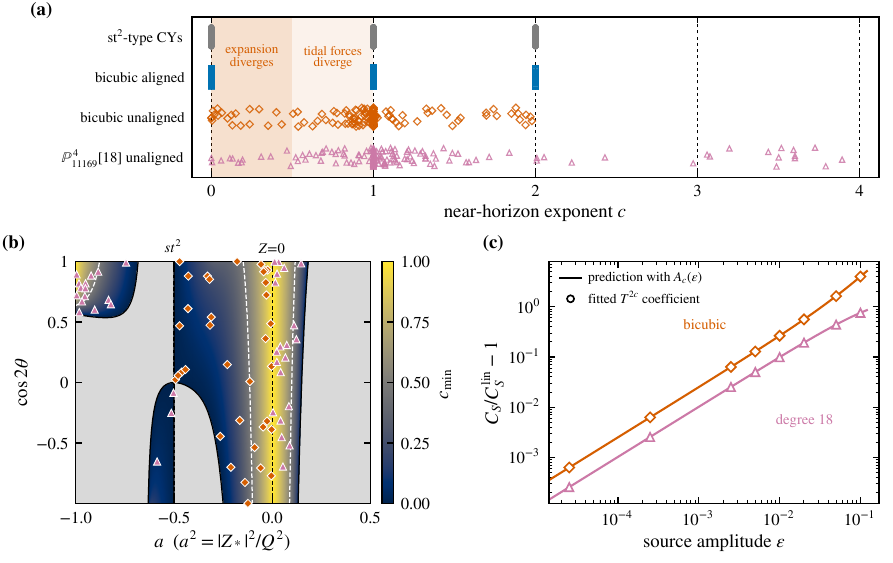}
\caption{Near-horizon exponents and near-extremal thermodynamics of stable non-BPS attractors with $Z\neq0$ in Calabi--Yau compactifications at large volume. (a) All exponents $c$ of each attractor, by threefold and branch. For the two threefolds whose cubic form is equivalent to $st^2$ (degree-8 and degree-12 hypersurfaces) and on the aligned branch of the bicubic, the exponents are $0$, $1$ and $2$. On the unaligned branches of the bicubic and of the degree-18 hypersurface in $\mathbb P^4_{1,1,1,6,9}$ the spectra contain non-integer exponents besides an exact $c=1$. The shaded band marks $0<c<1$, where an excited mode makes tidal forces diverge at the horizon; in its darker part, $0<c\le\frac12$, the expansion of radial null geodesics diverges as well. Points at $c=0$ are flat directions: exact moduli for the $st^2$-type threefolds, and directions lifted only at quartic order on the aligned bicubic branch (filled squares), which when excited also make the expansion diverge. The numbers of points reflect our sampling and are not a measure on the space of charges. (b) The two-modulus unaligned family in closed form (Appendix~\ref{App:Special}): the region of strict minima, $\Delta(a,\cos2\theta)>0$, colored by the smallest exponent $c_{\min}$; grey areas contain saddles; $\theta$ fixes the relative phase of $Z$ and $\mathcal D_aZ$. The white dashed curve is $c_{\min}=\frac12$. Black dashed lines mark the degenerate ($st^2$-type) cubic, $a=-\frac12$, and the limit $Z\to0$. Filled symbols: the sampled unaligned minima of panel (a), bicubic (diamonds) and degree 18 (triangles). (c) Finite-temperature test of Eq.~\eqref{eq:CYlaw} at the unaligned representatives of the two threefolds, along a family of asymptotic moduli at distance $\epsilon$ from the attractor: the relative excess $C_S/C_S^{\rm lin}-1$ of the coefficient $C_S$ of $\vartheta^{2c}$ in $S-S_0$, fitted from complete solutions down to $\vartheta\approx10^{-18}$ (open symbols), over its prediction $C_S^{\rm lin}$ with the linear source amplitude. Lines: the prediction with the nonlinear extremal amplitude $A_c(\epsilon)$, computed from extremal solutions alone (Appendix~\ref{App:NumericsCY}).}
\label{fig:landscape}
\end{figure}

\section{Singular horizons in Calabi--Yau compactifications}\label{Sec:CY}

{\emergencystretch=1.5em The selection rule protects what the integer spectrum provides, so the question is how often the spectrum is integer to begin with (Fig.~\ref{fig:landscape}). We consider the vector-multiplet sector of type IIA string theory on a Calabi--Yau threefold at large volume, $F=-\frac16\kappa_{abc}X^aX^bX^c/X^0$, with the triple intersection numbers $\kappa_{abc}$ of the threefold.\par}

\paragraph{Two-modulus classification.} For $h^{1,1}=2$ the cubic form $D(y)=\frac16\kappa_{abc}y^ay^by^c$ is a real binary cubic. Up to real linear changes of basis, which act on the special geometry together with a symplectic transformation of the charges, the cubic is of one of two non-degenerate types, $uv(u+v)$ or $u(u^2+v^2)$, distinguished by the sign of its discriminant. A double root gives the symmetric $st^2$ model. We study one threefold of each non-degenerate type: the bicubic in $\mathbb P^2\times\mathbb P^2$, with $\kappa_{112}=\kappa_{122}=3$ and Euler number $\chi=-162$~\cite{Candelas:1987kf,Hosono:1994ax,LarforsLukasRuehleSchneider2022,LeeLukas2025}, and the degree-18 hypersurface in $\mathbb P^4_{1,1,1,6,9}$, with $(\kappa_{111},\kappa_{112},\kappa_{122})=(9,3,1)$ and $\chi=-540$~\cite{CandelasFontKatzMorrison1994,Hosono:1993qy}. As controls we use the degree-8 and degree-12 hypersurfaces in $\mathbb P^4_{1,1,2,2,2}$ and $\mathbb P^4_{1,1,2,2,6}$~\cite{CandelasDeLaOssaFontKatzMorrison1994}, whose cubic forms are linear reparametrizations of $st^2$. All attractors below lie inside the K\"ahler cone at large volume, as the exact dilation $t\mapsto Lt$, $(q_0,q_a,p^0,p^a)\mapsto(L^3q_0,L^2q_a,p^0,Lp^a)$, which preserves the exponents, can always arrange.

\paragraph{Two branches.} Non-BPS attractors with $Z\neq0$ in these models~\cite{Saraikin:2007jc} fall into two kinds. On the \emph{aligned} branch, which contains the D0--D4 charges with D4 class inside the K\"ahler cone, the spectrum is exactly $\{0^{n-1},1^n,2\}$ for every cubic, and the selection rule of Sec.~\ref{Sec:Selection} applies. In the $st^2$-type controls, every non-BPS $Z\neq0$ attractor is dual to an aligned one and has this spectrum. For the two threefolds of non-degenerate type there are also \emph{unaligned} branches, on which the attractor can be a strict local minimum (Fig.~\ref{fig:landscape}). By the theorem of Sec.~\ref{Sec:Selection}, every such minimum with a non-degenerate Hessian has an exponent $0<c<1$; all minima in our samples have a non-degenerate Hessian. For two moduli the unaligned family is known in closed form (Appendix~\ref{App:Special}). Up to real changes of frame and charge scale, it is parametrized by a number $a$, with $a^2=|Z|^2/Q^2$, and by the relative phase $\theta$ of $Z$ and $\mathcal D_aZ$. At $\cos2\theta=-1$ the charges are D0--D4: an M5-brane string on a non-holomorphic divisor class~\cite{LSVY2021,MMT2022} with momentum; on the bicubic its orientation decides stability, while on the degree-18 threefold both orientations can give minima (Appendix~\ref{special:fived}). The spectrum always contains $c=1$ exactly, the minima form the region where a single polynomial $\Delta(a,\cos2\theta)$ is positive, and the smallest exponent takes every value in $(0,1)$ (Fig.~\ref{fig:landscape}b). This family contains every unaligned non-BPS $Z\neq0$ critical point of two moduli, and it gives arbitrarily soft minima within each compactification: for every cubic of non-degenerate type and every $\epsilon>0$, each open sector of the K\"ahler cone contains, above any volume, minima with $0<c_{\min}<\epsilon$, and they fill open cones of charges, which contain infinitely many integral rays (Appendix~\ref{special:unaligned}). On the bicubic, for example, the integral D0--D4 charges $(2KN,0,0;0,2N,2-2N)$, $K\ge1$, $N\ge2$, give minima with $c_{\min}\simeq9\sqrt6/(56N)$ at large $N$, bound against their separated D0- and D4-brane constituents (Appendix~\ref{special:fived}). All $73$ strict unaligned minima of our samples lie on the family, and the integer spectrum occurs at all $114$ aligned non-BPS critical points and at none of the $301$ unaligned ones. The bicubic with charges $(q_0,q_1,q_2;p^0,p^1,p^2)=(-5000,600,-400;6,30,-20)$, for instance, has an unaligned attractor with $c=(0.236765,0.532568,1,1.761248)$.

\paragraph{Singular horizons.} A mode with $0<c<1$ that is excited by the asymptotic moduli makes $\phi-\phi_*\propto r^{c}$ near the horizon, so the parallelly propagated tidal component $R''/R=-|\phi'|^2\propto r^{2c-2}$ diverges. The strength of this divergence depends on $c$. For $1/2<c<1$ the area radius is $C^{1,2c-1}$ and the tidal component diverges integrably. For $0<c\le1/2$ its integral diverges as well, so the expansion of the radial null congruence, of magnitude $2R'/R$, diverges at the horizon and the metric is only $C^{0,2c}$ (at $c=1/2$, where $R-Q\propto r\log r$, it is $C^{0,\alpha}$ for every $\alpha<1$). For $c\le1/4$, as for both exteriors constructed below, $R'$ is not even square integrable, so in the ingoing $(v,r)$ chart the metric is not in $H^1_{\rm loc}$ and falls outside the class of Geroch and Traschen, for which the curvature is well defined as a distribution~\cite{Geroch:1986jjl}. Curvature invariants stay finite throughout, and along radial null geodesics, where the Weyl tensor exerts no transverse tidal force, the strength of these singularities is set by $R_{kk}=2|\phi'|^2$: since $\int\lambda R_{kk}\,d\lambda$ converges, they are weak in Tipler's sense, and the doubly integrated tidal distortion of an infalling body stays finite, while for $c\le1/2$ the divergence of $\int R_{kk}\,d\lambda$ makes them strong in Kr\'olak's sense~\cite{Tipler:1977zza,Clarke:1985ck,Krolak:1986pno,Nolan:1999tw}. The backreaction stays finite for the deformed extremal horizons of Ref.~\cite{DiFilippo:2025plx}, but not here: $R''/R$ diverges for all $c<1$, and $R'/R$ as well for $c\le1/2$. The excitation is generic: the map from the stable amplitudes at the horizon to the asymptotic moduli is locally invertible. We constructed complete asymptotically flat extremal solutions with generic moduli and found decay exponents $c=0.236765\pm10^{-6}$ (bicubic) and $0.187045\pm4\times10^{-6}$ (degree 18), with tidal slopes approaching $2c-2$.

The aligned branch of a non-symmetric cubic fails differently. For a generic charge phase its massless directions are not moduli: eliminating the massive fields lifts them at quartic order, as found for D0--D4 charges in Ref.~\cite{NTT2007}. The resulting potential is $V_{\rm eff}=V_*+\kappa f^4+\cdots$ for a unit-normalized zero-mode amplitude $f$, with
\begin{equation}
\hat\kappa\equiv\frac{\kappa}{V_*}=k_0(y)\left(\frac{1-v^2}{1+v^2}\right)^3
\end{equation}
in a parametrization of the aligned charges by $(y,v)$ (Appendix~\ref{App:CY}). For $|v|<1$, the phase of the D0--D4--D6 family of Ref.~\cite{NTT2007} (Sec.~\ref{Sec:Selection}), the phase factor fixes a coefficient that this reference left undetermined; in the other phase, $|v|>1$, the quartic coefficient is new. The sign of the quartic coefficient depends on the charge phase, and in the D0--D6 limit $v\to\pm1$ the flat directions become exact moduli~\cite{NTT2007}. When $\hat\kappa>0$ the attractor is stable but degenerate. The zero mode then approaches the horizon only logarithmically, $f\simeq[4\hat\kappa\,s]^{-1/2}$ with $s=-\log(r/Q)$, as shown in Sec.~2.2 of Ref.~\cite{TripathyTrivedi2005}. The metric backreaction in that reference already implies that the tidal component diverges, $R''/R\simeq-1/(16\hat\kappa\,r^2s^3)$. The area radius then has divergent slope, $R'\simeq Q/(16\hat\kappa\,r\,s^3)$, and $R/Q-1\simeq1/(32\hat\kappa s^2)$, so the metric is continuous but not H\"older continuous. As for $c\le1/2$, the singularity is strong in Kr\'olak's sense and weak in Tipler's.

\paragraph{Thermodynamic signatures.} The two mechanisms leave distinct traces at low temperature (Appendices~\ref{App:IRresponse} and~\ref{App:marginal}). A smallest exponent $0<c<1$, $c\neq\frac12$, gives
\begin{equation}
\begin{gathered}
F\supset-\tfrac12(2c+1)\,\mathcal R(c)\,Q\,|A_c(\xi)|^2\,\vartheta^{2c+1},\\
\mathcal R(c)=\frac{\Gamma_{\mathrm E}(-c-\frac12)\,\Gamma_{\mathrm E}(c+1)}{2^{2c+1}\,\Gamma_{\mathrm E}(-c)\,\Gamma_{\mathrm E}(c+\frac12)},
\end{gathered}
\label{eq:CYlaw}
\end{equation}
at every order in the sources (Appendix~\ref{App:IRresponse}), where $A_c(\xi)$ is the nonlinear amplitude of the mode in the extremal exterior and $\Gamma_{\mathrm E}$ is Euler's gamma function. Complete finite-temperature solutions of the two Calabi--Yau exteriors, down to $\vartheta\approx10^{-18}$, confirm it: the fitted coefficients agree with the prediction to $2.3\times10^{-8}$, while the linear source amplitude misses them by factors of up to about five (Fig.~\ref{fig:landscape}c and Appendix~\ref{App:NumericsCY}). For $c<1/2$ this term dominates the linear term of the entropy, $2\pi Q^2\vartheta$ at leading order in the sources, once $\vartheta^{1-2c}\ll\frac12(2c+1)^2\mathcal R(c)|A_c|^2$, and $S-S_0\propto T^{2c}$. The operator dimension $1<\Delta=c+1<\frac32$ then lies in the window where the near-$\mathrm{AdS}_2$ dynamics is not dominated by the Schwarzian mode~\cite{MSY2016,Milekhin2021}. The horizon tidal component $R_{kk}\sim|A_c|^2\vartheta^{2c-2}/Q^2$ grows at low temperature, and the entropy measures it without reference to the amplitude: for $0<c<\frac12$,
\begin{equation}
\frac{S-S_0}{S_0\,Q^2\vartheta^2R_{kk}|_h}\longrightarrow H(c)=\frac{2(2c+1)\tan\pi c}{\pi c^2(c+1)^2},
\label{eq:Hc}
\end{equation}
while at $c=\frac12$ the ratio grows as $\frac{128}{9\pi^2}\log(1/\vartheta)$ (Appendix~\ref{App:IRresponse}); the same solutions reproduce $H(c)$ to $2\times10^{-6}$. The tidal component reaches $\ell^{-2}$ for a short-distance length $\ell$ at $\vartheta\sim(|A_c|\ell/Q)^{1/(1-c)}$, which for $c<1/2$ and $|A_c|$ of order one lies above the Schwarzian scale $\vartheta\sim Q^{-2}$ when $\ell$ is the Planck length. Below this temperature, infalling probes of size $\ell$ resolve the tidal field, and higher-derivative corrections must be evaluated on the solution~\cite{ChenDeRhamTolley2025}. At $c=\frac12$, $F\supset\frac Q8|A_{1/2}|^2\vartheta^2\log\vartheta$; for $1/2<c<1$ the term is a subleading non-analytic correction. A quartically lifted marginal direction gives $S-S_0\simeq\pi Q^2/[16\hat\kappa\log^2(T_*/T)]$, with $T_*$ set by the exterior (Eq.~\eqref{eq:IRmarginalentropy}), until a small quadratic lift $h_0$ of the flat direction turns it into $T^{2c_0}$ at $\log(T_*/T)\gtrsim1/h_0$. Both behaviors are anomalous compared with the linear approach $S-S_0\propto T$ of Reissner--Nordstr\"om, and both are absent in the symmetric models. The scaling $T^{2c}$ for $c<1/2$ and the $T\log T$ at $c=1/2$ parallel the anti-de Sitter results of Ref.~\cite{HKS2022AdS}.

\paragraph{Corrections.} The classical exponents are invariant under the dilation of moduli and charges (Appendix~\ref{App:CY}) and under a uniform rescaling of the charges, so they are the leading terms of a large-charge expansion. The perturbative $\alpha'$ correction $F\supset i\epsilon_{\alpha'}(X^0)^2$, $\epsilon_{\alpha'}=\chi\zeta(3)/(16\pi^3)$~\cite{CandelasDeLaOssaGreenParkes1991,AlexandrovKetovWakimoto2016}, enters at relative order $|\epsilon_{\alpha'}|/D$. Four-derivative couplings, such as the Weyl-squared term proportional to the second Chern class, enter at relative order $c_2\cdot y/(6Q^2)$ by the parametric estimate of the corrected entropy of supersymmetric black holes~\cite{Maldacena:1997de,LopesCardoso:1998tkj}; the non-BPS coefficients differ~\cite{Sahoo:2006rp}, and we do not include these terms. At the unaligned representatives, $|\epsilon_{\alpha'}|/D\approx(2\text{--}7)\times10^{-3}$ and $c_2\cdot y/(6Q^2)\approx3\times10^{-3}$, compared with deviations of the exponents from integers of order one. Of the 73 sampled unaligned minima, 72 remain strict minima under the full $\alpha'$ correction when dilated to classical volume $D\ge100$, with $c_{\min}$ shifted by at most $0.06$ (the exception, with classical $c_{\min}=3\times10^{-4}$, crosses zero); at $D\ge1000$ all 73 survive. On the aligned branch, by contrast, any quadratic lift competes with the classical quartic one. The $\alpha'$ correction alone gives $c_0(c_0+1)=-\frac{3\epsilon_{\alpha'}}{2D}\frac{1-v^2}{1+v^2}+O((\epsilon_{\alpha'}/D)^2)$, which is nonzero for the D0--D4 and D2--D6 flat directions and vanishes for the D0--D6 ones, as found for $st^2$ in Ref.~\cite{Bellucci:2008tx} and, for D0--D4 and D0--D6 charges, for general cubics in Ref.~\cite{Dominic:2011st}. For $\chi<0$ this correction stabilizes the phase $|v|<1$, which contains the D0--D4 charges, and destabilizes the phase $|v|>1$, which contains the D2--D6 charges. At the aligned bicubic representative, however, the four-derivative estimate exceeds the $\alpha'$ one unless the charges are rescaled by $N\gtrsim8$ (Appendix~\ref{App:CY}), so the fate of the aligned flat directions at given finite charges is open; the classical logarithmic approach holds where the quartic term dominates.

\section{Discussion}\label{Sec:Discussion}

Smoothness of an extremal horizon is decided by more than the linearized spectrum. For generic data, which excite the scalar modes, integer exponents are necessary, and they are common for the standard supergravity black holes. A quadratic resonance between these modes can still force a logarithm. The resulting obstruction is visible from infinity: it fixes a universal, sign-definite term $-\frac{4Q}{9}\|\Gamma(A,A)\|^2\vartheta^5\log\vartheta$ in the free energy. Our finite-source solutions reproduce this coefficient at fixed nonzero boundary data, not only order by order. In supergravity a selection rule switches off the resonance on the aligned branch and, by duality, on every non-BPS attractor with $Z\neq0$ of the symmetric models. This is why the familiar non-BPS black holes of these models have analytic horizons despite nonvanishing cubic couplings. In Calabi--Yau compactifications the selection rule still holds on the aligned branch, but every other stable non-BPS attractor with $Z\neq0$ has an exponent below one, between zero and one when its Hessian is non-degenerate, and for two moduli and a non-degenerate cubic form such attractors fill open cones of charges in every sector of the K\"ahler cone, with exponents as small as desired. On the aligned branch, for a generic charge phase, the flat directions are lifted only at quartic order. On these branches, extremal horizons are singular for generic asymptotic moduli, and near-extremal thermodynamics distinguishes the mechanisms: $\vartheta^5\log\vartheta$ for a resonance in a complete spectrum $\{1,2\}$, $\vartheta^{2c+1}$ for an exponent $c$ below one, and $1/\log^2T$ in the entropy for a quartically lifted direction. With vanishing central charge, finally, the resonance itself appears: an eight-modulus cubic with integer coefficients has stable attractors at which it is the leading obstruction, and the free energy then carries the term $\vartheta^5\log\vartheta$.

\paragraph{Scope.} Our statements concern static, spherically symmetric, asymptotically flat black holes of two-derivative Einstein--Maxwell--scalar theories at fixed charges, treated classically. The thermal theorem for the $(1,2)$ resonance holds at every fixed order in the boundary sources, and its two horizon readings hold at fixed small source; for the free energy itself the fixed-source evidence is numerical. For $0<c\le\frac12$ the leading entropy and free-energy laws hold at fixed small source (Appendix~\ref{App:IRrealexpansion}). For spectra beyond $\{1,2\}$ the quartic exchange identity still holds, but a thermal logarithm can arise without a horizon logarithm, for example $\vartheta^9\log\vartheta$ for a single scalar of exponent two with a cubic coupling. The equivalence between horizon and thermal logarithms is therefore specific to the lowest resonance. This equivalence extends to manifolds of attractors along which the spectrum is constant, such as those of Sec.~\ref{Sec:Selection}, with $\Gamma$ evaluated at the endpoint (Appendix~\ref{App:flatthermal}). A cubic coupling of one flat and two slow modes, which makes the slow exponent vary along such a manifold, produces $\vartheta^5\log\vartheta$ through the $c=0$ channel instead, but for generic data the flat mode then moves the attractor and shifts the slow exponent away from one, so the horizon is non-smooth already at linear order; on the aligned supergravity branch such couplings vanish. The classical resonant signal dominates the one-loop zero-mode correction for $(Q\|\Gamma(A,A)\|)^{-1/2}\lesssim\vartheta\ll1$ (Sec.~\ref{Sec:Thermal}). Corrections that shift the exponents restrict this window further; see Eq.~\eqref{eq:window}. The Calabi--Yau results are the leading terms of a large-charge expansion of the two-derivative vector-multiplet theory (Sec.~\ref{Sec:CY}).

\paragraph{Outlook.} The infrared formulation suggests that the same mechanisms operate in any near-horizon $\mathrm{AdS}_2$ throat, for example in rotating extremal black holes, whose near-horizon geometries contain $\mathrm{AdS}_2$ factors. In the $SO(8)$ gauged STU model of M-theory on $\mathrm{AdS}_4\times S^7$, however, no static attractor with nonzero gauging has a fast mode: all six scalars have $m^2L_2^2<6$, so the resonance cannot occur, and on the electric branch the dilatons have $0<c<1$ (Appendix~\ref{special:gauged}). The obstruction there is a non-integer exponent, as in Ref.~\cite{HKS2022AdS}. The resonant running parallels the classical logarithmic running of the static Love numbers of higher-dimensional Schwarzschild black holes, which occurs when $\ell/(d-3)$ is half-integer in $d$ spacetime dimensions~\cite{KolSmolkin2011,HuiEtAl2020}. Here the running is nonlinear, sits at the horizon end of the throat, and is read off from thermodynamics. For extremal Reissner--Nordstr\"om without scalars it is the integrability of the static axisymmetric equations, not the integer spectrum, that keeps the horizon analytic under coaxial tides; four-derivative corrections leave the exponents unshifted but break the integrability, and then even linear tides produce a logarithm~\cite{ZhangZhouLetter}. Whether integrability plays the same role for rotating extremal horizons is open. Whether attractors like those of Sec.~\ref{Sec:Selection}, with $Z=0$ and a leading resonance, occur in Calabi--Yau compactifications remains open, as do higher-derivative and quantum corrections and the fate of the singular horizons found here once the full string-theory description is included.

\acknowledgments
We thank Hong Guo for helpful comments on an earlier version of this work. A.Z. acknowledges graduate scholarship and research support through McMaster University. Research at Perimeter Institute is supported in part by the Government of Canada through the Department of Innovation, Science and Economic Development and by the Province of Ontario through the Ministry of Colleges, Universities, Research Excellence and Security. Some of the derivations and numerical code in this work were developed with AI tools, mainly Anthropic Claude and OpenAI Codex; separately prompted instances of these tools performed the independent reimplementations and checks documented in the code repository. The tools also assisted with literature screening, internal review, drafting the manuscript and editing the English. The authors checked all derivations, numbers and conclusions and take full responsibility for this work.

\appendix

\section{Extremal and thermal branches}\label{App:Branches}
\subsection{The constrained stable manifold}
At extremality the radial equations are
\begin{equation}
\begin{gathered}
D_r(r^2\phi')=\frac{\operatorname{grad}V}{2R^2},\qquad \frac{R''}{R}=-|\phi'|_G^2,\qquad V=(R-rR')^2+r^2R^2|\phi'|_G^2.
\end{gathered}\label{fnd:radial}
\end{equation}
Call an endpoint controlled when $R\to Q$, $\phi\to p$, $rR'/R\to0$, and $r\phi'\to0$. Smooth finite-area horizons and the logarithmic endpoints considered here satisfy these conditions.

Set $\tau=1/r$, $Y=R/r=e^{-U}$, and use dots for $\tau$ derivatives. The equations become $\ddot U=e^{2U}V$, $D_\tau\dot\phi=\frac12e^{2U}\operatorname{grad}V$, and $\dot U^2+|\dot\phi|_G^2=e^{2U}V$. Introduce $d\sigma/d\tau=Y^{-1}$, $h=\dot Y$, $\pi_i=-YG_{ij}\dot\phi^j$, and $\mathcal H=(h^2+G^{ij}\pi_i\pi_j-V)/2$. The constrained autonomous system is
\begin{align}
\frac{d\phi^i}{d\sigma}&=-G^{ij}\pi_j,\qquad \frac{dh}{d\sigma}=-G^{ij}\pi_i\pi_j,\nonumber\\
\frac{d\pi_i}{d\sigma}&=h\pi_i+\tfrac12(\partial_iG^{jk})\pi_j\pi_k-\tfrac12\partial_iV.\label{fnd:contact}
\end{align}
On the positive root $h=(V-|\pi|^2)^{1/2}$, its scalar blocks at $(p,0,Q)$ have eigenvalues $-Qc_i$ and $Q(c_i+1)$. The stable tangent plane is $\pi_i=Qc_i\phi^i$ in an orthonormal Hessian basis.

The analytic stable-manifold construction \cite{HirschPughShub1977} gives a local graph. In stable and unstable coordinates $(a,b)$, with nonlinearities $f,g=O(2)$ and positive $\Lambda_s,\Lambda_u$, the decaying trajectory solves
\begin{equation}
\begin{aligned}
a(\sigma)&=e^{-\Lambda_s\sigma}a_0+\int_0^\sigma e^{-\Lambda_s(\sigma-t)}f(a,b)\,dt,\\
b(\sigma)&=-\int_\sigma^\infty e^{\Lambda_u(\sigma-t)}g(a,b)\,dt.
\end{aligned}\label{fnd:LP}
\end{equation}
In the norm $\sup_{\sigma\ge0}e^{\eta\sigma}|(a,b)|$, with $0<\eta<\min\operatorname{spec}\Lambda_s$, the nonlinear Lipschitz constant is $O(\delta)$ on a radius-$\delta$ ball. Complexification gives the same contraction uniformly on a smaller parameter ball; uniform limits of holomorphic iterates prove analyticity. Invertibility of the stable projection gives $(h(\phi),\pi(\phi))$.

\subsection{Contact identity and Lagrangian graph}
To prove the gradient property of the stable graph, define $\omega=dh-\pi_i d\phi^i$. Direct differentiation of Eq.~\eqref{fnd:contact} yields $\mathcal L_{\mathcal X}\omega=h\omega-d\mathcal H$. On a constrained stable tangent, $y(\sigma)=\omega(D\Phi_\sigma v)$ therefore obeys $y'=hy$. The propagated tangent decays while $h\to Q>0$; hence $y=0$. Thus $\pi=dh$, and $W=h$ obeys
\begin{equation}
\begin{gathered}
V=W^2+|\nabla W|_G^2,\qquad W(p)=Q,\qquad dW(p)=0,\qquad G^{-1}\nabla^2W(p)=Q\operatorname{diag}(c_i).
\end{gathered}\label{fnd:W}
\end{equation}
This $W$ is the fake superpotential, or Hamilton--Jacobi function, of extremal flows~\cite{Ceresole2007,Andrianopoli:2007gt,LopesCardoso:2007qid,Perz2009,BossardMichelPioline2010}. Uniqueness of the stable manifold proves uniqueness of this local positive-Hessian solution. Every controlled endpoint approaches its fixed point with $\sigma\sim Q^{-1}\log\tau$ and lies on the graph. Consequently $\dot Y=W$ and $\dot\phi=-Y^{-1}\operatorname{grad}W$. For the full radial Lagrangian, the canonical one-form on this branch is
\begin{equation}
2\dot U\,dU+2G_{ij}\dot\phi^j d\phi^i=d(-2e^UW).\label{fnd:lagrangian}
\end{equation}
Thus the full metric--scalar graph is Lagrangian.

The fixed-energy mode count follows without an extra first-order assumption. With $a_R=\log(R/Q)$, $p_r=r\phi'$, and $\tau_r=\log(r/Q)$, the constraint eliminates $q_R=rR'/R$ on the root $q_R=1-\sqrt{e^{-2a_R}V/Q^2-|p_r|^2}$. The remaining system is
\begin{equation}
\begin{gathered}
a_{R,\tau_r}=q_R,\qquad \phi_{,\tau_r}=p_r,\qquad D_{\tau_r}p_r=-p_r+\frac{e^{-2a_R}}{2Q^2}\operatorname{grad}V.
\end{gathered}\label{fnd:fuchsian}
\end{equation}
Its spectrum is $\{1,c_i,-c_i-1\}$. The exponent one is the radius-slope mode; there is no independent unstable metric mode at fixed energy. A Jacobi perturbation with no scalar unstable amplitudes is tangent to the stable graph. At RN the extremal scalar boundary map has the leading form $(A,B)=(\xi_1,\xi_2)+O(\xi^2)$, so the normalized boundary map is locally invertible. If its asymptotic scalar data and time normalization vanish, uniqueness of the normalized first-order flow makes the Jacobi perturbation zero. The same triangular argument applies coefficientwise to every finite Taylor quotient. The assertion uses the radial constraint: dropping it would introduce a spurious metric integration constant.

\subsection{Exterior reconstruction and analytic normal form}
Normalize $Y=1$ at infinity, and set $s=\log Y$, $z=Y^{-1}=r/R$. The flow is $d\phi/ds=-\operatorname{grad}\log(W/Q)$. Positive Hessian makes a sufficiently small source ball forward invariant, and $1+Q\tau\le Y\le1+W(\xi)\tau$. These bounds continue the small-source exterior to all $\tau\ge0$. Reconstruction is
\begin{equation}
\tau(z)=\int_z^1\frac{dt}{t^2W(\phi(t))},\qquad r=\tau^{-1},\qquad R=r/z.\label{fnd:reconstruction}
\end{equation}
It gives $R\to Q$, asymptotic flatness, $M=W(\xi)$, and $\Sigma=-\operatorname{grad}W(\xi)$.

In an orthonormal normal chart, the $yxx$ term of $W/Q$ is $\langle y,\Gamma(x,x)\rangle$. Equation~\eqref{fnd:W} multiplies its coefficient by $2(1+2+1+1)=10$. Remove the non-resonant quadratic vector-field terms by a finite analytic change. The resulting flow is $\dot q=N(q)+\mathcal R(q)$, where $N=(-x,\allowbreak -2y-\Gamma(x,x))$ and $\mathcal R=O(3)$. The normal flow is
\begin{equation}
N_t(x,y)=(e^{-t}x,e^{-2t}[y-t\Gamma(x,x)]).
\end{equation}
Convergence of the conjugacy can be seen explicitly \cite{Arnold1988}. For $0<\varepsilon<1/3$, the full flow satisfies $|\Phi_t(q)|\le Ce^{-(1-\varepsilon)t}|q|$, while the inverse normal-flow derivative along it is bounded by $C(1+t)e^{2t}$. Hence
\begin{equation}
\begin{aligned}
\Theta(q)&=\lim_{t\to\infty}N_{-t}(\Phi_t(q))=q+\int_0^\infty DN_{-t}(\Phi_t(q))\mathcal R(\Phi_t(q))\,dt
\end{aligned}\label{fnd:conjugacy}
\end{equation}
converges uniformly on a small complex ball. It is holomorphic, tangent to the identity, and conjugates the flows. Composing with the quadratic change gives $\chi$, so $A=\Pi_1\chi(\xi)$ and $B=\Pi_2\chi(\xi)$. The slow coordinate is unique: a first nonlinear difference between two choices would need weight one, impossible for a nonlinear monomial. The fast freedom $Y\mapsto Y+C(X,X)$ changes only $B$.

\subsection{Full-field smoothness}
For full-field smoothness, use the ingoing metric $ds^2=-r^2R^{-2}dv^2+2dv\,dr+R^2d\Omega_2^2$. Here $r$ is affine along radial null geodesics. A smooth extension makes $z=r/R$ a smooth defining coordinate; the $z^2\log z$ term in Eq.~\eqref{eq:profiles} must vanish. Conversely, if $\Gamma(A,A)=0$, the normal fields and their analytic inverse are analytic in $z$. Since $W=Q+O(z^2)$, Eq.~\eqref{fnd:reconstruction} gives $\tau=(Qz)^{-1}+c_0+c(z)$ with analytic $c$. Therefore $z(r)$ and $R(r)$ are analytic. The electric fields $(f^{-1})^{AB}(\phi)q_B R^{-2}dv\wedge dr$ are analytic as well; fixed magnetic charges add regular sphere-volume terms, with analytic theta-dependent numerators. Thus the criterion controls the complete configuration.

When $\Gamma(A,A)\neq0$ the regularity of the field strengths is set by the electric numerators $N^A=(f^{-1})^{AB}q_B$. The coefficient of $(r/Q)^2\log(r/Q)$ in $F^A_{vr}=N^A(\phi)/R^2$ is $Q^{-2}(dN^A)_p\Gamma(A,A)$, so a component for which this does not vanish is $C^{1,\alpha}$ but not $C^2$. Stationarity of $V=q_AN^A$ imposes only $q_A(dN^A)_p=0$. For a single purely electric field, $N=V/q$ and $F_{vr}=V/(qR^2)$. Then $V/Q^2=1+2|x|^2+6|y|^2+\cdots$ in the normal coordinates of Sec.~\ref{Sec:Resonance} has no logarithm below $(r/Q)^4$, where $6|y|^2$ contributes a squared logarithm, and focusing gives the coefficient $-\frac13\|\Gamma(A,A)\|^2$ of $(r/Q)^4\log^2(r/Q)$ in $R/Q$. Hence
\begin{equation}
\frac{F_{vr}}{F_{vr}(0)}\supset\frac{20}{3}\,\|\Gamma(A,A)\|^2\Big(\frac rQ\Big)^4\log^2\frac rQ ,
\end{equation}
with no earlier non-smooth term, so this field strength is $C^{3,\alpha}$ for every $\alpha<1$, like the metric. With a magnetic charge and a scalar-dependent theta coupling the electric numerator is no longer a function of $V$, and a single field strength can again be only $C^{1,\alpha}$.

\subsection{The thermal branch and physical temperature}
The compact radial variables used below are
\begin{align}
M_0&=\frac{Q}{\sqrt{1-k^2}},\quad B_0=Q\sqrt{1-k^2},\quad b_{\rm ne}=kM_0,\nonumber\\
u&=1-\frac{B_0}{\rho+M_0},\qquad R=(\rho+M_0)g.\label{eq:RNcoords}
\end{align}
With $v=V/Q^2$, $D=u^2-k^2$, $u\in[k,1]$, and primes denoting $u$ derivatives, the exact equations are
\begin{align}
D_u(D\phi')&=\frac{\operatorname{grad}v}{2g^2},\qquad g''=-g|\phi'|_G^2,\nonumber\\
v&=g^2-2ugg'+D(g'^2+g^2|\phi'|_G^2).\label{eq:compactfull}
\end{align}
Here $D_u$ is the target-covariant derivative. The linear operator and normalized regular modes are
\begin{equation}
\begin{gathered}
L_c=\partial_u(D\partial_u)-c(c+1),\qquad \phi_1=h_1a+h_2b,\qquad h_1=u,\qquad h_2=\frac{3u^2-k^2}{3-k^2}.
\end{gathered}\label{eq:linearprofiles}
\end{equation}
The zero-source temperature is $\vartheta_0(k)=k\sqrt{1-k^2}/(1+k)^2$, where $\vartheta=2\pi QT$. The fixed-charge scalar work is
\begin{equation}
dF=-S\,dT-G_{ij}(\phi_\infty)\Sigma^i d\phi_\infty^j,\qquad \Sigma=-B_0\phi'(1),\label{eq:sourcework}
\end{equation}
with $\phi=\xi+\Sigma/\rho+O(\rho^{-2})$; the sign agrees with $M_{\rm ext}=W$ and $\Sigma=-\operatorname{grad}W$ \cite{GibbonsHawking1977,IyerWald1994,GibbonsKalloshKol1996,Pacilio2018}.

At positive $k$, use Eq.~\eqref{eq:compactfull} with $g(1)=1$, $\phi(1)=\xi$. Regular horizon data satisfy
\begin{equation}
\phi'_h=\frac{\operatorname{grad}v(\phi_h)}{4kg_h^2},\qquad g'_h=\frac{g_h^2-v(\phi_h)}{2kg_h}.\label{fnd:horizondata}
\end{equation}
The scalar Taylor recursion has nonzero denominators $2kn^2$, yielding a convergent analytic horizon series; continuation reaches infinity near RN. The linear scalar shooting map is multiplication by $P_c(1/k)>0$. Equivalently its zero-Dirichlet kernel is trivial by $\int_k^1[D(f')^2+c(c+1)f^2]du=0$. The metric linearization is $\delta g=(u/k)\delta g_h$, also invertible at infinity. The analytic inverse-function theorem gives an actual branch analytic in every source at this fixed $k$. Finally,
\begin{equation}
\vartheta=\frac{\vartheta_0(k)}{g_h^2},\qquad \frac{\vartheta_0'(k)}{\vartheta_0(k)}=\frac{1-2k}{k(1-k^2)}\label{fnd:physicalT}
\end{equation}
allows conversion to fixed physical temperature for $0<k<1/2$; $\vartheta_0(k)$ is maximal at $k=1/2$, which separates the two Reissner--Nordstr\"om branches. These pointwise arguments do not supply a common source radius as $k\to0$.

\section{Proof of the matching relation}\label{App:Matching}
Fix an arbitrary finite quotient of the source Taylor ring. Throughout this appendix the charges are fixed, $G$ and $V$ are analytic, the attractor is isolated, and its complete scalar spectrum is $\{1,2\}$. The dependencies are:
\begin{itemize}
\item L1 (finite-order expansion) and L2 (extremal identification) supply L3 (coefficient connection).
\item L3 and constrained stable-graph uniqueness give L4 (the first logarithmic order).
\item L5 computes the unstable tail; L6 shows that conversion to physical temperature preserves the coefficient.
\item L7 pairs the tail with extremal variations and fixes the integration constant at Reissner--Nordstr\"om (RN), completing Eq.~\eqref{eq:main}.
\end{itemize}
All estimates hold at fixed source order; they do not establish a source neighborhood uniform as $T\to0$.

\subsection{Lemma L1: Finite-order expansion}\label{lem:L1}
\emph{Hypotheses.} Work in the fixed finite source quotient under the assumptions above, with regular horizon data and fixed outer sources.
\emph{Conclusion.} The differentiated thermal hierarchy has a finite separated representation in $t=u/k$ and $L=1/k$. Each free-energy coefficient has an integer-power expansion with finite logarithmic degree and differentiated remainder bounds at every fixed truncation.

\emph{Proof.} Work throughout in an arbitrary finite quotient of the source Taylor ring, so that only finitely many analytic target-space jets occur. Differentiating the actual thermal branch gives $L_c\phi_\alpha=f_\alpha$ and $g_\alpha''=h_\alpha$, with only lower source coefficients on the right. The electric charges have already been Legendre transformed and held fixed; there are no independent Maxwell boundary-source variations. The outer constraint fixes the metric slope:
\begin{equation}
2g_\alpha'(1)=\left[(1-k^2)(g'^2+|\phi'|_{G(\xi)}^2)-(v(\xi)-1)\right]_\alpha.\label{eq:matching-metric-slope}
\end{equation}
Its right side is triangular because the background scalar derivative and radius slope vanish.

To construct the low-temperature expansion, set $t=u/k$ and $L=1/k$. The equations contain no explicit $k$, only the outer endpoint $L$. Let $\mathcal A$ be the algebra generated by polynomials, rational functions with poles at $t=\pm1$, and finite iterated integrals of $dt/(t\mp1)$. Products reduce by shuffle identities; partial fractions and integration by parts show closure under integration against rational functions. The integer Legendre solutions obey
\begin{equation}
\begin{gathered}
Q_c=\tfrac12P_c\log\frac{t+1}{t-1}-W_{c-1},\qquad (t^2-1)(P_cQ_c^\prime-P_c^\prime Q_c)=-1.
\end{gathered}
\end{equation}
The regular particular inverse is
\begin{equation}
\begin{aligned}
Y_f(t)={}&P_c(t)\int_1^tQ_c(s)f(s)\,ds-Q_c(t)\int_1^tP_c(s)f(s)\,ds;
\end{aligned}\label{eq:matching-PQ-inverse}
\end{equation}
subtracting $P_c(t)Y_f(L)/P_c(L)$ imposes zero outer data. The two lower-endpoint terms remain combined; their regular Taylor expansion follows from the nonzero Frobenius factors $2n^2$.

Induction gives a finite separated representation $\sum_r A_r(L)B_r(t)$ with rational endpoint factors and elements of $\mathcal A$. Analytic target couplings use only finite jets at each source order. The metric requires two integrations and Eq.~\eqref{eq:matching-metric-slope}. The representation is used only after grouping terms that share a linearly independent endpoint factor: regularity of the actual horizon solution for every sufficiently large $L$ makes each grouped radial coefficient regular at $t=1$. In particular, the logarithmic singularities of the separate Legendre $Q_c$ pieces cancel before the lower endpoint is evaluated.

At infinity every finite element of $\mathcal A$ has the convergent form
\begin{equation}
H(1/w)=w^{-M}\sum_{p=0}^{P}f_p(w)\log^p w,\label{eq:matching-convergent-log-germ}
\end{equation}
with finite $M,P$ and holomorphic $f_p$ near $w=0$. Indeed $dt/(t\mp1)=-dw/[w(1\mp w)]$. Induction on word length expands the holomorphic numerator in a convergent Taylor series; the constant term raises the logarithm degree, while integrating $w^{n-1}\log^p w$ divides by positive powers of $n$ and preserves convergence on smaller disks. Rational factors add only finite Laurent poles. Taylor bounds give differentiated asymptotic remainders at any fixed truncation. Endpoint inverses $P_c(L)^{-1}$, the analytic temperature factors, and finite-order implicit temperature inversion preserve this class. Scalar work then proves an actual integer-power, finite-log expansion of $F_\alpha$ before any logarithmic variation is introduced.

\subsection{Lemma L2: Identification of the extremal amplitude}\label{lem:L2}
\emph{Hypotheses.} Use the hierarchy of L1, including the outer metric constraint, and the unique normalized extremal exterior of Appendix~\ref{App:Branches}.
\emph{Conclusion.} The thermal coefficients obey Eq.~\eqref{eq:matching-coefficient-bounds}, converge on every $u\ge\delta>0$ to the extremal coefficients, and have leading slow amplitude $A(\xi)$.

\emph{Proof.} We next identify the extremal limit of each thermal coefficient. For monic $h_1=u$, $h_2=u^2-k^2/3$, the exact zero-Dirichlet Green inverse is
\begin{equation}
\begin{aligned}
y(u)&=-h_c(u)\int_u^1\frac{J(s)}{(s^2-k^2)h_c(s)^2}\,ds,\\
J(s)&=\int_k^s h_c(v)f(v)\,dv.
\end{aligned}\label{eq:matching-positive-Green}
\end{equation}
For $s\in[k,2k]$, the numerator contains $s-k$, canceling the horizon zero; for $s\ge2k$, the denominator is comparable to $s^{2c+2}$. If $f=O(u^2\Lambda_u^p)$, $\Lambda_u=1+|\log u|$, this gives $L_1^{-1}f=O(u)$, $(L_1^{-1}f)'=O(1)$, $L_2^{-1}f=O(u^2\Lambda_u^{p+1})$, and $(L_2^{-1}f)'=O(u\Lambda_u^{p+1})$. The slow bound uses $\int_0^1\Lambda_u^pdu<\infty$.

The induction closes because every scalar nonlinearity is $O(u^2\Lambda_u^p)$. This includes the connection term $D\mathcal C(\phi',\phi')$ and the radius terms. Indeed, the scalar fields are $O(u\Lambda_u^p)$ and $g-1=O(u\Lambda_u^p)$ at lower source orders; target connections multiply $D=O(u^2)$ and two bounded scalar derivatives. The potential is stationary at the attractor, so the nonlinear part of its gradient starts at quadratic scalar order. Scalar first derivatives remain bounded by a finite logarithmic power. Focusing and Eq.~\eqref{eq:matching-metric-slope} bound $g_\alpha'$, while the horizon constraint gives
\begin{equation*}
2g_\alpha(k)-2kg_\alpha'(k)=O(k^2\Lambda_k^p).
\end{equation*}
Hence $g_\alpha=O(u)$ with a possibly higher finite logarithmic power. Thus at each fixed source order
\begin{equation}
\begin{aligned}
x_\alpha&=O(u),&x_\alpha^\prime&=O(1),\\
y_\alpha&=O(u^2\Lambda_u^{p_\alpha}),&y_\alpha^\prime&=O(u\Lambda_u^{p_\alpha}),\\
g_\alpha&=O(u),&g_\alpha^\prime&=O(1).
\end{aligned}\label{eq:matching-coefficient-bounds}
\end{equation}
Dominated convergence on every $u\ge\delta>0$ identifies the limit with the controlled extremal coefficient and its unique normalized stable branch. This identification uses both the prescribed outer source and the constrained metric slope; the limiting endpoint alone would not select a unique global exterior. Since $z=u/g$ and $g\to1$ at the extremal horizon, its leading slow coefficient is the independently defined $A(\xi)$.

\subsection{Lemma L3: Coefficient connection}\label{lem:L3}
\emph{Hypotheses.} Use the separated representation of L1 and the extremal identification of L2.
\emph{Conclusion.} Equation~\eqref{eq:matching-connection} connects finitely many inner degrees to each outer degree, with controlled remainders. A slow unstable term $k^ju^{-2}$ comes only from inner degree $j-2$; a fast term $k^ju^{-3}$ comes only from degree $j-3$.

\emph{Proof.} For one separated term, let $D_{\min}$ be the lowest Laurent degree of its endpoint factor and $M_{\max}$ the highest radial degree of its inner factor. The identity
\begin{equation}
k^d(\log k)^p t^m(\log t)^q=k^{d-m}u^m(\log k)^p(\log u-\log k)^q\label{eq:matching-connection}
\end{equation}
shows that outer order $k^j$ receives only $D_{\min}-j\le m\le M_{\max}$, with $d=j+m$. There are finitely many separated terms, so truncating Eq.~\eqref{eq:matching-convergent-log-germ} beyond these bounds proves the coefficient connection with controlled remainders. A slow unstable $k^ju^{-2}$ tail can only originate at inner degree $j-2$, and a fast $k^ju^{-3}$ tail at degree $j-3$.

\subsection{Lemma L4: The first logarithmic order}\label{lem:L4}
\emph{Hypotheses.} Use L3 and constrained stable-graph uniqueness, with zero outer data for the first logarithmic variation.
\emph{Conclusion.} There are no thermal logarithms below order five, and at order five there is at most one power of the logarithm.

\emph{Proof.}
\begin{enumerate}
\item \emph{Degrees and endpoint data.}
Give $u,k$ degree one, $\partial_u$ degree $-1$, and logarithms degree zero.
At $u=kt$ an inner block of degree $n$ is $k^n$ times a function of $t$
polynomial in $\ell=\log k$. Its coefficients belong to the fixed source
quotient. In Riemann normal coordinates let $q=\phi-p$ denote the scalar
displacement, with components $x=\Pi_1q$ and $y=\Pi_2q$. Write
$(x,y,g-1)=\sum_n k^n(X_n,Y_n,G_n)$.
Equation~\eqref{eq:matching-coefficient-bounds}, at fixed $t$, starts $x,g-1$
at degree one and $y$ at degree two. By \eqref{eq:matching-connection},
the primary powers $u^{-2},u^{-3}$ at outer order $j$ require $n=j-2,j-3$,
respectively. Thus the last scalar blocks needed at $j=5$ are $X_3,Y_2$.
In this proof $L_i=\partial_t((t^2-1)\partial_t)
-i(i+1)$ in the inner variable.

\item \emph{Degrees zero, one and two.}
The background is $(0,0,1)$. At degree one the scalar and focusing forcings
vanish. Regularity selects $H_1=t$, $H_2=t^2-1/3$ in the two scalar kernels.
The linear constraint gives $G_1-tG_1'=0$. Thus
$(X_1,Y_1,G_1)=(At,0,D_*t)$, where L2 fixes $A$ and $D_*$
is the coefficient of $u$ in the extremal $g-1$.
Primes on inner blocks denote $t$ derivatives. Put $s_A=\|A\|^2$,
$P=\Gamma(A,A)$, and define the symmetric bilinear map
$\mathsf T(v_1,v_2)=\tfrac14(\nabla^3 v)_p(\,\cdot\,,v_1,v_2)^\sharp$,
where $v=V/Q^2$. The linear force is $\mathsf Hq=2x+6y$.
Equation~\eqref{eq:Gammadef} gives $\Pi_2\mathsf T(A,A)=5P$. Define
$4Z=\Pi_1\mathsf T(A,A)-4D_*A$. Expansion of $g^{-2}=1-2kG_1+O(k^2)$
in \eqref{eq:compactfull} gives
\begin{equation*}
\begin{aligned}
 L_1X_2&=4Zt^2, & X_2&=Z(t^2-1)+a_2 t,\\
 L_2Y_2&=5Pt^2, & Y_2&=P Y_*+\beta H_2,\\
 G_2''&=-s_A, & G_2&=-s_A t^2/2+Et+e_0 .
\end{aligned}\tag{\theequation a}\label{pf:second}
\end{equation*}
Indeed $L_1(t^2-1)=4t^2$. The constant term of the polynomial constraint
at this degree fixes $e_0=(D_*^2+s_A)/2$.
With $\lambda=\log(t+1)$, $Y_*=H_2\lambda+7t^2/6-t-4/9$.
For every polynomial $b$ the product rule gives
\begin{equation*}
 L_i(b\lambda)=(L_i b)\lambda+\mathcal D b,\qquad
 \mathcal D b=b+2(t-1)b'.
 \tag{\theequation b}\label{pf:logproduct}
\end{equation*}
This yields $L_2Y_*=5t^2$. The $t^2$ terms of $Y_*$ are
$t^2(\log t+7/6)$. By L2 and \eqref{eq:profiles} the extremal fast field
has $u^2(\widetilde B+P\log u)$ at this power. Indeed $z/u\to1$, and the
quadratic part of the inverse normal-form change can add a bilinear function
of $(A,A)$ to the polynomial coefficient $\widetilde B$.
Matching gives $\beta=\widetilde B+P(\ell-7/6)$.
The homogeneous coefficients $a_2,E$ may provisionally be polynomials in $\ell$.

\item \emph{Degree three.}
The cubic force on the ray $q=\epsilon A$ has three sources. The potential
gives $U_A=\tfrac1{12}(\nabla^4v)_{p,\mathrm{sym}}(\,\cdot\,,A,A,A)^\sharp$.
With the convention $R^i{}_{jkl}=-R^i{}_{jlk}$, the connection term of
$D_u(D\phi')$ is proportional to $(R^i{}_{jkl}+R^i{}_{kjl})A^jA^kA^l=0$.
The quadratic inverse-metric correction to the linear force is proportional to
$R_{iajb}A^a(\mathsf H A)^j A^b=0$, since $\mathsf H A=2A$.
These are all curvature terms at this degree. A fast vector in either
contraction costs at least one further degree.

Put $K v_1=2\mathsf T(A,v_1)-2D_*\mathsf H v_1$ and $q_2=X_2+Y_2$.
The two scalar equations have right sides $\Pi_i F_3$. Multiplying the
potential jet by $1-2kG_1+k^2(3G_1^2-2G_2)$ gives
\begin{equation*}
\begin{aligned}
 F_3={}&tKq_2+t^3\{U_A-2D_*\mathsf T(A,A)\}+(3D_*^2t^2-2G_2)t\,\mathsf H A\\
     ={}&f_3t^3+f_2t^2+f_1t+(K\widetilde B)tH_2+(KP)\omega,\\
 f_3={}&KZ+U_A-2D_*\mathsf T(A,A)+(3D_*^2+s_A)\mathsf H A,\\
 f_2={}&Ka_2-2E\mathsf H A,\qquad f_1=-KZ-2e_0\mathsf H A,\\
 \omega={}&t\{H_2(\ell+\lambda)-t-1/18\}.
\end{aligned}\tag{\theequation c}\label{pf:third-force}
\end{equation*}
Since $\langle A,Y_2'\rangle=0$ and the target metric has no linear jet,
cubic focusing gives $G_3''=-D_*s_A t-2\langle A,X_2'\rangle$.
This is affine in $t$, so the cubic focusing used in L6 is polynomial.

For a regular particular inverse $I_i$, set $I_i[1]=-1/[i(i+1)]$
for the constant forcing and extend linearly over the vector coefficients
in \eqref{pf:third-force}. With $\varrho=(t\lambda-1)/3$, the slow inverses are
\begin{equation*}
\begin{aligned}
 I_1[t]&=\varrho, & I_1[t^2]&=(t^2-1)/4,\\
 I_1[t^3]&=t^3/10+3\varrho/5,\\
 I_1[tH_2]&=t^3/10+4\varrho/15 .
\end{aligned}\tag{\theequation d}\label{pf:inverses}
\end{equation*}
The entries follow from \eqref{pf:logproduct} and
$L_i t^j=[j(j+1)-i(i+1)]t^j-j(j-1)t^{j-2}$.
For the last forcing put $p_1=t^3/10$ and $\lambda_0=\lambda-\log 2$.
Then $L_1p_1=tH_2-4t/15$, and a regular particular inverse is
\begin{equation*}
\begin{aligned}
 3\Xi={}&-t\operatorname{Li}_2((1-t)/2)+\lambda_0-I_1[t+1]+3(\log 2)I_1[t],\\
 I_1[\omega]={}&\ell I_1[tH_2]+p_1\lambda-I_1[\mathcal D p_1]+4\Xi/15-I_1[t^2]-I_1[t]/18 .
\end{aligned}\tag{\theequation e}\label{pf:log-inverses}
\end{equation*}
Indeed $\partial_t\operatorname{Li}_2((1-t)/2)=-\lambda_0/(t-1)$, so
$L_1[t\operatorname{Li}_2((1-t)/2)]=-(3t+2)\lambda_0-t$ and $L_1\Xi=t\lambda$.
All entries are real analytic at $t=1$. Thus
$X_3=\Pi_1 I_1[F_3]+a_3H_1$, with $a_3$ fixed by matching at the outer endpoint.

\item \emph{The homogeneous constrained problem.}
At fixed outer $u>0$, L2 excludes a logarithm at order zero.
Let $j\ge1$ and suppose that the outer coefficients of orders $1,\dots,j-1$
contain no power of $\ell$. For each $p>0$ take the coefficient of
$k^j\ell^p$ in the equations and constraint. Products of lower orders and
the explicit $k^2$ in $D$ and in \eqref{eq:matching-metric-slope} contribute
no $\ell^p$, and $u$ derivatives preserve $\ell^p$. The resulting system is
the homogeneous extremal Jacobi system. The same is true of the whole
coefficient of $k^1$, since $k$ enters only through $k^2$.
In each case the energy variation is zero because
$b_{\rm ne}^2=Q^2k^2/(1-k^2)$, and the outer scalar and metric data vanish.
The positive-root elimination in \eqref{fnd:fuchsian} gives one radius
exponent $1$ and scalar pairs $(1,-2),(2,-3)$ on $E_1,E_2$.
After the fast $u^{-3}$ amplitude is set to zero, the only possible independent
unstable amplitude is the slow $u^{-2}$ one. With both zero the remaining
scalar modes decay. In the linearization of \eqref{fnd:LP} the unstable
component is then fixed by its integral from infinity. For prescribed stable
initial projection the integral operator has norm $O(\delta)<1$ on the
weighted space of Appendix~\ref{App:Branches}. Differentiating the stable
trajectory gives its unique solution. Thus the stable tangent
characterization following \eqref{fnd:fuchsian} places the variation on
$h=W,\ \pi=dW$, with the radius-slope mode stable. On the graph the variation
satisfies the linearization of $\dot Y=W(\phi)$,
$\dot\phi=-Y^{-1}\operatorname{grad}W$. At $\tau=0$ its initial data are
$\delta Y=0$ by time normalization and $\delta\phi=0$ by fixed sources.
Uniqueness of this first-order initial-value problem kills the radius variation
as well as both scalar variations. In the finite quotient apply this to the
lowest source coefficient that is nonzero. There the system is the homogeneous
Reissner--Nordstr\"om one, with solutions
$x=\alpha u+\beta_0 u^{-2}$, $y=\gamma u^2+\delta u^{-3}$ and
$\delta g=c_1u$. By \eqref{eq:matching-connection}, $\beta_0$ and $\delta$
are the coefficients of $t^{-2}$ and $t^{-3}$ in $X_{j-2}$ and $Y_{j-3}$
at the selected order in $\ell$ after substituting $\log t=\log u-\ell$.
If they vanish, the zero outer data give $\alpha=\gamma=c_1=0$.

\item \emph{Logarithmic degree of the tails.}
For $j<5$ the possible slow blocks have $n\le2$ and are polynomial in $t$.
The possible fast blocks have $n\le1$ and vanish. Step 4 therefore eliminates
every positive power of $\ell$ at those orders. As in step 4, the entire
$k^1$ outer equation is homogeneous. It has zero boundary data and no
unstable tails, so $a_2=E=0$ by $k^2t=ku$.
At $j=5$, $H_2(\ell+\lambda)$ has only
log-free negative powers at infinity. In $X_3$ the entries of
\eqref{pf:inverses} have at most a constant coefficient at $t^{-2}$,
and the provisional $a_2,E$ occur only in the forcing proportional to $t^2$,
whose inverse is polynomial. For $\omega$, reduction of
order gives $(X_3/t)'=J/[(t^2-1)t^2]$ for its particular contribution,
with $J$ the integral of $t$ times its forcing. Integration by parts on
the polynomial times $\lambda$ in $J$ gives at most one logarithm.
A second integration can produce a squared logarithm only from
$t^{-1}\log t$, giving $t\log^2t$ after multiplication by $t$.
At $t^{-2}$ the primitive instead comes from $t^{-4}$ and preserves
logarithmic degree. The part proportional to $\ell$ is
$\ell I_1[tH_2]$, with log-free decaying coefficient.
Hence the coefficient of $t^{-2}$ in $X_3$ is affine in $(\ell,\log t)$
jointly, with no product $\ell\log t$.
Consequently substitution of $\log t=\log u-\ell$ gives degree at most
one in $\ell$ in either unstable amplitude. Step 4 applied separately to
each $p\ge2$ at $j=5$ proves the assertion. The thermal order here is
that of the outer solution and its scalar work. L6 gives the corresponding
temperature conversion, and L7 fixes the free-energy integration constant.

Inner blocks do contain earlier logarithms. $Y_2(1)$ has the term
$\tfrac23P\ell$. Put $C=\Gamma_A^*P$ as in L5. The growing term
$\tfrac89C\,t(\tfrac12\log^2t+\ell\log t)$ of the particular solution
forces a compensating $\ell^2$ term in $a_3$ when $C\ne0$, so that the
outer coefficient of $k^2$ is free of $\ell$. Since $a_3H_1$ has no
decaying power, the tails are unaffected. The lemma concerns the outer
coefficients.
\end{enumerate}

\subsection{Lemma L5: The unstable-tail coefficient}\label{lem:L5}
\emph{Hypotheses.} Use L3--L4 and match the regular inner fields to the extremal slow amplitude. Set $P=\Gamma(A,A)$ and $\Gamma_Av=\Gamma(A,v)$; $\Gamma_A^*$ is its adjoint in $G_p$.
\emph{Conclusion.} The fast logarithmic unstable coefficient vanishes, while Eq.~\eqref{eq:matching-unstable} fixes the slow coefficient to $(16/27)\Gamma_A^*P$.

\emph{Proof.}
For the degree-two fast field in \eqref{pf:second} the regular particular
profile is
\begin{equation}
Y_*=H_2\log(t+1)+\frac76t^2-t-\frac49,\qquad L_2Y_*=5t^2.
\end{equation}
Combining it with $\beta=\widetilde B+P(\ell-7/6)$ gives
\begin{equation}
 \begin{aligned}
 y_2={}&k^2P\big[H_2\{\log k+\log(t+1)\}{}-t-1/18\big]+k^2\widetilde B H_2 .
 \end{aligned}
 \label{eq:matching-fast-core}
\end{equation}
Here $y_2=k^2Y_2$, since $7t^2/6-4/9-7H_2/6=-1/18$.
In $\lambda=\log t+\sum_{m\ge1}(-1)^{m+1}/(mt^m)$ the $m=5,3$ terms
give $[t^{-3}]H_2\lambda=1/5-1/9=4/45$. The terms with $\log t$ or
$\ell$ multiply $H_2$. Thus the fast tail at this order is
$4P k^5u^{-3}/45$ and has zero logarithmic coefficient.

For $v_2\in E_2$, the symmetry of $(\nabla^3v)_p$, which holds because
$dv(p)=0$, and \eqref{eq:Gammadef} give $\Pi_1 K v_2=10\Gamma_A^*v_2$.
Put $C=\Gamma_A^*P$.
The first four forcings in \eqref{pf:third-force}, projected by $\Pi_1$,
are the full polynomial list. The $t,t^3$ terms have log-free coefficients
and inverses with respective $t^{-2}$ coefficients $1/9,1/15$ by
\eqref{pf:inverses}. The $t^2$ term contains all provisional $a_2,E$
and its inverse has no negative power. With the monic $H_3=t^3-3t/5$,
$tH_2=H_3+4H_1/15$. Its resonant component gives $4\varrho/15$, whose
$t^{-2}$ coefficient is $4/135$, also log-free. This deals with the
independent $\widetilde B$. The vanishing curvature contractions in step 3 of
L4 exhaust the derivative couplings at this degree. The remaining forcing is
\begin{equation}
 f(t)=10Ct\big[H_2\{\log k+\log(t+1)\}-t-1/18\big].
 \label{eq:matching-slow-core}
\end{equation}
For $L_1X=f$ put $X=t v_1$. The product rule gives
$[(t^2-1)t^2v_1']'=tf$. Its horizon integration constant is zero by
regularity. Define $\mathcal B(t)=t^5/5-t^3/9$, so $\mathcal B'=t^2H_2$ and
$\mathcal B(1)=4/45$. Division gives
\begin{equation*}
 \begin{aligned}
 \mathcal B(t)&=(t+1)\mathcal D_1(t)-4/45,\\
 45\mathcal D_1(t)&=9t^4-9t^3+4t^2-4t+4 .
 \end{aligned}
 \tag{\theequation a}\label{pf:division}
\end{equation*}
Take $\mathcal D_0(t)=t^5/25-t^4/20+4t^3/135-2t^2/45+4t/45$,
so $\mathcal D_0'=\mathcal D_1$. Integration by parts gives the full numerator
\begin{equation*}
\begin{aligned}
 J(t)&=\int_1^t s f(s)\,ds=10C\,[\,\mathcal J(s)\,]_{s=1}^{s=t},\\
 \mathcal J(s)&=\mathcal B(s)(\ell+\log(s+1))+\tfrac4{45}\log(s+1)-\mathcal D_0(s)-s^4/4-s^3/54,\\
 X(t)&=t\left[\mathrm{const}+\int_1^t\frac{J(s)}{(s^2-1)s^2}\,ds\right].
\end{aligned}\tag{\theequation b}\label{pf:J}
\end{equation*}
The integrand has a removable singularity at one. The constant-power
logarithms of $J$ are $10C[(4/45)\log t-\mathcal B(1)\ell]
=(8/9)C(\log t-\ell)$. The other logarithms are $10C\mathcal B(t)(\log t+\ell)$,
with odd powers $t^5,t^3$. Their product with
$[(t^2-1)t^2]^{-1}=\sum_{n\ge0}t^{-4-2n}$ has odd powers only.
Its $t^{-1}$ term has coefficient $10C(1/5-1/9)(\log t+\ell)$
and hence produces a squared logarithm only on the growing power $t$.
There is no $t^{-4}$ term in this product. Terms with
$\log(1+1/t)$ have ordinary Taylor series and carry no $\ell$.
Since a primitive of $t^{-4}(\log t-\ell)$ is
$-t^{-3}[(\log t-\ell)/3+1/9]$, the logarithmic part of $X$ at $t^{-2}$
is $(8/27)Ct^{-2}(\ell-\log t)$. The homogeneous fast term $P\ell H_2$
alone contributes $10C(4/135)\ell\,t^{-2}$.
Restoring $k^3$, the radial logarithm supplies a second equal $\log k$ term
through $\ell-\log(u/k)=2\log k-\log u$. Hence
\begin{equation}
 [\delta x]_{k^5\log k}=\frac{16}{27}\Gamma_A^*P\,u^{-2}+\cdots .
 \label{eq:matching-unstable}
\end{equation}

\subsection{Lemma L6: Conversion to physical temperature}\label{lem:L6}
\emph{Hypotheses.} Use the constrained hierarchy and L4--L5, keeping the physical temperature fixed.
\emph{Conclusion.} Since $\vartheta_0(k)=k+O(k^2)$, the first non-analytic shift of $k$ has degree five. The source-work covector equals its extremal value plus $O(k^2)$ before its first logarithm, so this shift affects the response only from degree six.

\emph{Proof.} Define $D_*(\xi)$ by the extremal outer expansion $g(u,0,\xi)=1+D_*(\xi)u+O(u^2)$; its constant term at RN vanishes. The first inner radius blocks are
\begin{equation}
\begin{aligned}
g={}&1+kD_*t+k^2\bigl[-\tfrac12\|A\|^2t^2+Et+\tfrac12(D_*^2+\|A\|^2)\bigr]+\cdots.
\end{aligned}
\end{equation}
Degree-three focusing is polynomial. Its homogeneous logarithms would be earlier outer stable logarithms, already excluded. Horizon metric logarithms first enter at degree four through $x_1'x_3'$ or $y_2'^2$. Since $\vartheta_0(k)=k+O(k^2)$, Eq.~\eqref{fnd:physicalT} changes $k$ non-analytically only at degree five. Independently the outer term linear in $k$ solves the zero-energy homogeneous Jacobi problem and has no available unstable tail, so it vanishes. The source-work covector is consequently its extremal value plus $O(k^2)$ before its first logarithm. A shift $\delta k=O(k^5\log^p k)$ changes its non-constant analytic part only at degree six. A source-independent constant covector does not change under a $k$ shift. The analytic reparametrization has unit leading coefficient because $g_h\to1$, and the induced non-analytic energy change likewise starts at degree six.

\subsection{Lemma L7: Symplectic pairing}\label{lem:L7}
\emph{Hypotheses.} Use the fixed-charge Routhian, scalar boundary work, L5--L6, and the Lagrangian stable graph.
\emph{Conclusion.} In every finite source quotient,
\begin{equation*}
dC=d\!\left[-\frac{4Q}{9}\|\Gamma(A,A)\|^2\right].
\end{equation*}
The vanishing RN value fixes $C=-4Q\|\Gamma(A,A)\|^2/9$.

\emph{Proof.} The fixed-charge Routhian and the scalar surface work in Eq.~\eqref{eq:sourcework} fix the flux normalization. In particular, Maxwell variations do not contribute independent boundary work at fixed Gauss charges. With one half of the canonical radial flux and the scalar-work orientation, write
\begin{align}
\Omega={}&\delta_{\log}U\,\delta_{\rm ext}\dot U-\delta_{\rm ext}U\,\delta_{\log}\dot U+G(\delta_{\log}\phi,D_\tau\delta_{\rm ext}\phi)-G(\delta_{\rm ext}\phi,D_\tau\delta_{\log}\phi).\label{eq:matching-flux-normalization}
\end{align}
At infinity it is $-G_\infty(\delta_{\log}\Sigma,d\xi)=dC$. Near the horizon, $du/d\tau=-Qu^2$, $\delta_{\rm ext}x=u\,dA+\cdots$, and Eq.~\eqref{eq:matching-unstable} give
\begin{equation}
\Omega_\phi=-\frac{16Q}{9}\langle P,\Gamma(A,dA)\rangle=d\!\left[-\frac{4Q}{9}\|P\|^2\right].\label{eq:matching-flux-final}
\end{equation}
Target-metric corrections in the flux start at $O(u^2)$. The induced radius response is at most $O(u^{-1})$ and gives a vanishing endpoint pairing with the stable $O(u)$ radius variation after the radial symplectic weight is included. With its primary unstable coefficient absent, the induced fast response is at most $O(u^{-1}\log^p u)$ and also has zero endpoint pairing with the stable $O(u^2\log^q u)$ fast variation. Stable additions pair to zero because the stable graph is Lagrangian, Eq.~\eqref{fnd:lagrangian}. At RN the logarithmic response vanishes, fixing the source-independent integration constant. This proves Eq.~\eqref{eq:main} for every finite source index, with the differentiated remainders supplied by Eq.~\eqref{eq:matching-convergent-log-germ}. The coefficient is thus derived from the unstable tail of L5 and the flux pairing; the direct quartic exchange of Sec.~\ref{Sec:Thermal}, an independent calculation, gives the same value at leading order. The sixth-order calculation of Appendix~\ref{App:Checks}, which uses none of L1--L7, checks the next order: with $A=\epsilon+a_3\epsilon^3$ for $\|\Gamma(A,A)\|^2=\gamma^2A^4$, Eq.~\eqref{eq:main} predicts $-\frac{4Q}{9}\gamma^2[\epsilon^6]A^4=-\frac{16Q}{9}\gamma^2a_3$, which is Eq.~\eqref{eq:sixthmain}.

\subsection{Hamilton--Jacobi derivation and renormalization-group form}\label{App:RGform}
The Hamiltonian constraint gives a second, shorter derivation of Eq.~\eqref{eq:main}. It uses the expansion of L1 and the extremal limit of L2, but none of L3--L7, and it also shows how the fast amplitude $B$ enters at fifth order. With $R=(\rho+M_0)g$ and Eq.~\eqref{eq:RNcoords}, the mass is $M=M_0-B_0g'(1)$, the horizon is at $\rho=b_{\rm ne}$, and Eq.~\eqref{fnd:physicalT} is the statement $2TS=b_{\rm ne}$. Let $\mathsf b=b_{\rm ne}/Q=k/\sqrt{1-k^2}$, $S_0=\pi Q^2$, and
\begin{equation}
\begin{gathered}
\Psi(\xi,b_{\rm ne})=M-\tfrac12\,b_{\rm ne}\log(S/S_0),\qquad d\Psi=-G_{ij}\Sigma^i d\xi^j-\tfrac12\log(S/S_0)\,db_{\rm ne},
\end{gathered}
\label{eq:thermalPsi}
\end{equation}
where the second line follows from Eq.~\eqref{eq:sourcework}. Thus $\Psi$ is the potential of the scalar work at fixed $k$, $M=\Psi-b_{\rm ne}\partial_{b_{\rm ne}}\Psi$, and the constraint in Eq.~\eqref{eq:compactfull} at $u=1$, which reads $M^2+|\Sigma|_G^2-b_{\rm ne}^2=V(\xi)$, becomes the Hamilton--Jacobi equation
\begin{equation}
\big(\Psi-b_{\rm ne}\partial_{b_{\rm ne}}\Psi\big)^2+|\nabla_\xi\Psi|_G^2=V(\xi)+b_{\rm ne}^2 .
\label{eq:thermalHJ}
\end{equation}
At $b_{\rm ne}=0$ it reduces to Eq.~\eqref{fnd:W}. In a finite source quotient, L1 and L2 give $\Psi=Q\sum_{n\ge0}\psi_n(\xi,\ell)\,\mathsf b^n$ with $\ell=\log\mathsf b$, $\psi_0=w=W/Q$, and coefficients polynomial in $\ell$. At order $\mathsf b^n$, Eq.~\eqref{eq:thermalHJ} reads
\begin{equation}
\begin{aligned}
\big[\mathcal Z-\partial_\ell-(n-1)\big]\psi_n&=\frac1{2w}\Big[\delta_{n2}-\!\!\sum_{\substack{i+j=n\\ i,j\ge1}}\!\!\big(\mu_i\mu_j+\langle\nabla\psi_i,\nabla\psi_j\rangle_G\big)\Big],\\
\mu_i&=(1-i)\psi_i-\partial_\ell\psi_i ,
\end{aligned}
\label{eq:thermalCS}
\end{equation}
where $\mathcal Z=\langle\operatorname{grad}\log w,\nabla\rangle$ generates the extremal flow of Appendix~\ref{App:Branches}. In the normal-form coordinates of Eq.~\eqref{eq:normalform}, $\mathcal ZX=X$ and $\mathcal ZY=2Y+\Gamma(X,X)$, the beta functions of Eq.~\eqref{eq:IRdimensionlessRG}, so Eq.~\eqref{eq:thermalCS} is a Callan--Symanzik equation. With $\hat Y=Y+\Gamma(X,X)\,\ell$ one has $\mathcal Z-\partial_\ell=X\partial_X+2\hat Y\partial_{\hat Y}-\partial_\ell$ in the variables $(X,\hat Y,\ell)$. Give $X$ and $\hat Y$ the weights one and two. The kernel of the operator in Eq.~\eqref{eq:thermalCS} then consists of the polynomials in $(X,\hat Y)$ with constant coefficients and weight $n-1$; a forcing term of any other weight is solved without a new logarithm, and a forcing term $f$ of weight $n-1$ produces $-\ell f$. At the boundary point $X=A(\xi)$, and $\hat Y=B(\xi)+\Gamma(A,A)\log\mathsf b$ is the running coupling of Eq.~\eqref{eq:IRRG} at the scale $\mathsf b$. The constants in the homogeneous solutions are not fixed by Eq.~\eqref{eq:thermalHJ}; they encode horizon regularity, and each is read off at the lowest source order at which it appears, from the expansion around Reissner--Nordstr\"om at fixed $k$. There $\Psi$ has no term linear in the sources, and
\begin{equation}
\begin{gathered}
\Psi^{(2)}=\frac{B_0}{2}\Big[|\xi_1|^2+\frac{6}{3-k^2}|\xi_2|^2\Big],\qquad \Psi^{(3)}=\frac{M_0}{2}\int_k^1 v_3(\phi_1)\,du ,
\end{gathered}
\label{eq:thermalHJinputs}
\end{equation}
by Eq.~\eqref{eq:linearprofiles} and a Green identity against the linear profiles, as in Appendix~\ref{App:Exchange}. Both are analytic at $k=0$, and rational in $k$ once the factors $B_0$ and $M_0$ are removed. The quadratic term is even in $k$, and the only term of $\Psi^{(3)}$ odd in $k$ below $k^7$ comes from $5M_0\langle\xi_2,\Gamma(\xi_1,\xi_1)\rangle\int_k^1u^2h_2\,du$, with
\begin{equation}
\int_k^1u^2h_2\,du=\frac{\frac35-\frac13k^2-\frac4{15}k^5}{3-k^2},
\label{eq:thermalHJvertex}
\end{equation}
so that $[\Psi^{(3)}]_{\mathsf b^5}=-\frac{4Q}{9}\langle\Gamma(\xi_1,\xi_1),\xi_2\rangle$.

The recursion now closes. At $n=1$ there is no forcing and $\psi_1$ is a constant, which vanishes because the extremal entropy is $S_0$ for every $\xi$; all terms containing $\psi_1$ drop out of Eq.~\eqref{eq:thermalCS}. At $n=2$ the forcing $1/(2w)$ has no component of weight one, so $\psi_2$ is free of logarithms; it is unique because the homogeneous term $b_aX^a$ vanishes, $\Psi$ having no term linear in the sources. At $n=3$ the forcing vanishes and $\psi_3=q_{ab}X^aX^b+n_i\hat Y^i$; here $n_i=0$ because $\Psi$ has no linear term and $q_{ab}=0$ because $\Psi^{(2)}$ is even, so $\psi_3=0$. At $n=4$ a logarithm could multiply only the monomials $XY$ and $X^3$; it is absent because $\Psi^{(2)}$ is diagonal and $\Psi^{(3)}$ is rational, so $\psi_4$ is free of logarithms. At $n=5$ the forcing contains only the pairs $(1,4)$ and $(2,3)$ and vanishes identically. Therefore $\psi_5$ is a polynomial of weight four, $\psi_5=q_4(X)+C_{iab}X^aX^b\hat Y^i+M_{ij}\hat Y^i\hat Y^j$, where $M=0$ because $\Psi^{(2)}$ is even and $C=-\frac49\Gamma$ by Eq.~\eqref{eq:thermalHJvertex}. In every finite source quotient,
\begin{equation}
\begin{aligned}
\Psi\big|_{\mathsf b^5}={}&Q\Big[q_4(A)-\frac49\big\langle\Gamma(A,A),\,B+\Gamma(A,A)\log\mathsf b\big\rangle\Big],
\end{aligned}
\label{eq:RGform}
\end{equation}
with $A=A(\xi)$ and $B=B(\xi)$ the amplitudes of Eq.~\eqref{eq:profiles} and $q_4$ a quartic form on $E_1$ fixed by the four-jet of $G$ and $V$ at $p$. The complete coefficient of $\mathsf b^5$ thus depends on the extremal exterior only through the slow amplitude and the dimension-three coupling at the scale $\mathsf b$. The nonlinear amplitudes appear because the homogeneous solutions of Eq.~\eqref{eq:thermalCS} are functions of the normal-form coordinates, the eigen-coordinates of $\mathcal Z$. There is no squared logarithm because the coefficient of $\langle\hat Y,\hat Y\rangle$ is the linear static response of the fast mode, $-\frac52\mathcal R(2)=0$ by Eq.~\eqref{eq:IRintegerkernel}; the single logarithm is the running of $g_3$ in the cross term. In the detuned language of Appendix~\ref{App:Detuning}, with $\nu=2+\delta$ the amplitude of the $z^\nu$ mode is $B+\Gamma(A,A)/\delta$, and $-\frac Q2(2\nu+1)\mathcal R(\nu)\|B+\Gamma(A,A)/\delta\|^2\mathsf b^{2\nu+1}$ is the sum of a pole, cancelled by the analytic $\mathsf b^5$ term as in Eq.~\eqref{eq:polepair}, the bracket of Eq.~\eqref{eq:RGform} without $q_4$, and terms of order $\delta$.

\paragraph{Fixed temperature.} The free energy is $F=\Psi-b_{\rm ne}\partial_{b_{\rm ne}}\Psi-\frac12b_{\rm ne}$, with $\mathsf b=\vartheta S/S_0$. A small change of $\Psi$ at fixed $b_{\rm ne}$ changes $F$ at fixed $T$ by the same amount to first order, and $\mathsf b=\vartheta[1+O(\vartheta)]$ without logarithms to the order needed, so that
\begin{equation}
\begin{aligned}
F\big|_{\vartheta^5}={}&Q\Big[q_4(A)-\frac49\big\langle\Gamma(A,A),\,B+\Gamma(A,A)\log\vartheta\big\rangle\Big]-Q\Big(\frac{400}{3}\psi_2^4+16\,\psi_2\psi_4\Big).
\end{aligned}
\label{eq:RGformF}
\end{equation}
The last term is the image of the even part $Q(w+\psi_2\mathsf b^2+\psi_4\mathsf b^4)$ of $\Psi$ under the change of ensemble. It is analytic and free of logarithms, but it depends on the boundary data and not only on $A$ and the running coupling. For example, for a pure fast source at $A=0$ the quadratic response is $F_2=Q\|B\|^2\sqrt{1-k^2}/(1-k^2/3)$ at the temperature $\vartheta_0(k)$, and
\begin{equation}
\big[F_2\big]_{\vartheta^5}=-\frac{100Q}{9}\,\|B\|^2 ,
\label{eq:RGcounterexample}
\end{equation}
although the bracket of Eq.~\eqref{eq:RGformF} vanishes there. The renormalization-group form therefore holds for the complete fifth-order coefficient at fixed $k$, Eq.~\eqref{eq:RGform}, and at fixed physical temperature for the logarithm and for the bracket of Eq.~\eqref{eq:RGformF}. In particular, Eq.~\eqref{eq:RGformF} contains Eq.~\eqref{eq:main}. Equivalently, on variations with $dA=0$ the analytic $k^5$ coefficient of the scalar-work one-form $-G_{ij}\Sigma^id\xi^j$ is $-\frac{4Q}{9}\langle\Gamma(A,A),dB\rangle$. The matched expansion gives the same value: the coefficient of $t^{-3}$ in Eq.~\eqref{eq:matching-fast-core} is $4\Gamma(A,A)/45$, and its pairing with the stable variation $u^2dB$ in Eq.~\eqref{eq:matching-flux-normalization} has the Wronskian $5u^{-2}$.

A redefinition $Y\mapsto Y+C(X,X)$ shifts $q_4(A)$ by $\frac49\langle\Gamma(A,A),C(A,A)\rangle$, and a change of the reference scale of the logarithm shifts it by a multiple of $\|\Gamma(A,A)\|^2$; the bracket is unchanged. The quartic $q_4$ cannot in general be absorbed in $B$, since it contains contact terms. We have evaluated it only in two-field examples.

\paragraph{Checks.} For the potential~\eqref{eq:potential6} with sources $(\epsilon,\eta)$, Eqs.~\eqref{eq:mixedfifth} and~\eqref{eq:sixthmain} follow from Eq.~\eqref{eq:RGformF} with Eqs.~\eqref{eq:Amixed} and~\eqref{eq:a3}. Equation~\eqref{eq:RGform} also predicts the coefficient of $Q\epsilon^2\eta^2\mathsf b^5$ in $\Psi$,
\begin{equation}
-\frac49\gamma\,[A^2B]_{\epsilon^2\eta^2}=\frac89\gamma^2+\frac1{21}\gamma v_{03},
\end{equation}
with no logarithm, and a direct fourth-order calculation at fixed $k$, which includes the metric backreaction and uses none of the matching lemmas, gives the same value. The recursion applied to the single-field potential~\eqref{eq:even_exponent_example}, with the cubic-order input $\psi_7=-\frac{8\eta}{945}Y^3$, reproduces Eq.~\eqref{eq:smooth_thermal_log} from the resonant forcing at $n=9$. Flat directions need a separate argument, since a coordinate of exponent zero has weight zero and the weights no longer grade the monomials; Appendix~\ref{App:flatthermal} gives it for manifolds of attractors.

\subsection{Manifolds of attractors}\label{App:flatthermal}
The attractors of Appendix~\ref{special:z0} are not isolated. Let $\mathcal M$ be a connected Morse--Bott manifold of minima of $V$, with $G$ and $V$ analytic, such that at every point of $\mathcal M$ the normal exponents are exactly $1$ and $2$; their multiplicities are then constant, and the directions of $T_p\mathcal M$ have $c=0$. Differentiating $\mathsf H\Pi_i=i(i+1)\Pi_i$ along $v\in T_p\mathcal M$, with $\Pi_i$ the spectral projections, gives $\Pi_i(\nabla_v\mathsf H)\Pi_i=0$, so the flat--slow--slow and flat--fast--fast components of $\nabla^3V$ vanish. The eigenspaces, the flat--slow--fast components and $\Gamma_p$ may still vary along $\mathcal M$, and $\mathcal M$ need not be totally geodesic. An integer spectrum at the endpoint alone is not enough, as the control below shows.

\paragraph{Horizon.} Fix the endpoint $p$ in the stable-graph construction of Appendix~\ref{App:Branches}: in Eq.~\eqref{fnd:LP} the flat coordinates acquire the integral from infinity of their quadratic forcing, and the contraction is unchanged and analytic in $p$. The extremal solutions that end at $p$ form a leaf. On it the flat coordinates are an analytic graph over the normal ones without a linear term, the decaying rates are $-Q$ and $-2Q$, and the only resonance is $1+1=2$. The conjugacy~\eqref{fnd:conjugacy} then gives the profiles~\eqref{eq:profiles} with $\Gamma=\Gamma_p$, and the flat graph is $O(z^2)$ without a $z^2\log z$ term. The horizon is therefore smooth if and only if $\Gamma_p(A,A)=0$. At the constant solution the leaf chart $(p,A,B)$ has the identity as its linearization in $\phi_\infty$, so $\phi_\infty\mapsto(p,A)$ is a submersion; if $\Gamma$ does not vanish identically on $\mathcal M$, the logarithm is the first non-analytic term for an open dense set of nearby asymptotic moduli.

\paragraph{Thermodynamics.} The thermal argument of Appendices~\ref{App:Matching} and~\ref{App:RGform} goes through with three changes.

First, the flat directions produce no new logarithm through degree five. The regular response of $L_c$ to a forcing $u^n$ contains a multiple of $k^nP_c(u/k)\log(u+k)$ only if $n\ge c$ and $n\equiv c\pmod 2$, so an exchange of the forcings $u^n$ and $u^m$ in the channel $c$ gives $k^{n+m+1}\log k$ only if both resonate. Degree five requires $n+m=4$. In the fast channel this is the slow--slow--fast vertex, whose residue gives Eq.~\eqref{eq:main}. In the flat channel it requires a flat--slow--slow vertex, and in the slow channel a forcing linear in a flat displacement; both are derivatives of the slow mass along $\mathcal M$ and vanish. The remaining flat forcings, slow$\times$fast and slow$^3$ of odd degree and slow$^2\times$fast, contribute first at $k^9$, and a flat source only moves the base point.

Second, the horizon value of the flat coordinates moves with the temperature. In tubular coordinates $(s,N)$, with $N$ in the normal eigenframe that rotates along $\mathcal M$, the constancy of the spectrum gives the normal force $\mathrm{diag}(2,6)N+O(|N|^2)$ uniformly in $s$, while the tangential force is $O(|N|^2)$; in a fixed frame a flat$\times$slow source would make the fast field $O(u)$ instead of $O(u^2)$. Under the bounds of Lemma~\ref{lem:L2} the flat forcing is $f_0=O(u^2\Lambda_u^m)$, $\Lambda_u=1+|\log u|$, and its regular inverse obeys
\begin{equation}
\begin{gathered}
s'(u)=\frac{1}{u^2-k^2}\int_k^uf_0(v)\,dv=O(u\Lambda_u^m),\qquad s(u)-s_h=O(u^2\Lambda_u^m).
\end{gathered}
\end{equation}
The horizon value $s_h$ differs from the endpoint $p(\xi)$ of the extremal solution with the same data by analytic terms of order $k^2$, with a first logarithm $k^3\log k$ at third order in the sources, and Lemma~\ref{lem:L2} identifies the same extremal limit.

Third, the flat coordinates enter Eq.~\eqref{eq:thermalCS} as parameters. In coordinates $(p,X,\hat Y)$ with $\mathcal Zp=0$, the kernel at order $n$ consists of the polynomials in $(X,\hat Y)$ of weight $n-1$ with coefficients analytic in $p$. The gradient products in the forcing now contain derivatives along $p$, which do not lower the weight; through $n=5$ this is harmless, because $\psi_1=\psi_3=0$ identically in $p$ and the forcing at $n=4$ has no logarithm. The inputs~\eqref{eq:thermalHJinputs} hold at every $p$: $\Psi^{(2)}$ has no flat block, and $\Psi^{(3)}$ acquires only a flat--slow--fast term proportional to $\int_k^1uh_2\,du=(1-k^2)(3+k^2)/[4(3-k^2)]$ and, if $\mathcal M$ is curved, two-flat--one-normal terms, all even in $k$. The recursion of Appendix~\ref{App:RGform} then gives $\psi_5=q_{4,p}(X)-\frac49\langle\Gamma_p(X,X),\,Y+\Gamma_p(X,X)\log\mathsf b\rangle$, and since the first logarithm of $\log(S/S_0)$ appears at $k^4\log k$, the change to fixed temperature keeps its coefficient. Order by order in the sources, the only non-analytic term of the free energy through degree five is therefore
\begin{equation}
F\supset-\frac{4Q}{9}\,\big\|\Gamma_{p(\xi)}\big(A(\xi),A(\xi)\big)\big\|^2\,\vartheta^5\log\vartheta,
\label{eq:flatthermal}
\end{equation}
where $p(\xi)$ and $A(\xi)$ are the endpoint and the slow amplitude of the extremal solution with the same asymptotic data. The variation of $\Gamma_p$ along $\mathcal M$ enters the scalar work as the derivative of this coefficient and adds no independent logarithm. The matching route of Appendix~\ref{App:Matching} gives the same coefficient: an unstable flat tail $k^ju^{-1}$ comes from inner degree $j-1$, no flat logarithmic tail arises through outer order four, and at fixed endpoint the slow Wronskian of Lemma~\ref{lem:L7} is the only finite contribution; variations of the endpoint pair with a flat tail at $k^5\log k$ and give the derivative of the coefficient along $\mathcal M$.

\paragraph{Checks.} Take a flat target metric and $v=1+2X_s^2+6Y_s^2+10\gamma(s)X_s^2Y_s$, where $(X_s,Y_s)$ is $(x,y)$ rotated by the angle $\omega s$ and $\gamma(s)=\gamma+g's$, so that the flat--slow--fast coupling $v_{sxy}=-8\omega$ is active and $\Gamma$ varies along a flat direction that is not an isometry. For the slow source $(s,x,y)=(0,a,0)$ the polynomial Green inverses give $-\frac{4Q}{9}\gamma^2a^4\vartheta^5\log\vartheta$. With a fast source as well, the rotation moves the horizon value of $s$, but its exchange has no logarithmic residue at this order, and for $g'\neq0$ the flat boundary work reproduces the endpoint derivative $-\frac{8Q}{9}\gamma g'a^4$ of the predicted coefficient. An order-by-order solution of the complete thermal equations, in models of this type with an added cubic slow term and with $\mathcal M$ a circle that is not totally geodesic, reproduces Eq.~\eqref{eq:flatthermal} at orders $\epsilon^4$, $\epsilon^5$ and $\epsilon^6$ in the sources to better than $10^{-7}$; with $\Gamma$ evaluated at the boundary point instead of the endpoint, the $\epsilon^6$ coefficients would be off by 7--11\%. The constancy of the spectrum is needed: if the slow mass varies along $\mathcal M$, as for $v\supset2Ksx^2$, the exchange of the flat mode adds $-QK^2a^4\vartheta^5\log\vartheta/9$, and the slow exponent itself depends on the endpoint.

\section{Polynomial exchange and fixed-order checks}\label{App:Exchange}
The calculations of this appendix are done at fixed order in the sources and do not use the matching theorem of Appendix~\ref{App:Matching}.
\subsection{Fixed-temperature scalar exchange}
Use the notation and boundary work of Appendix~\ref{App:Branches}. In target normal coordinates let $v=1+v_2+v_3+\cdots$ and $J_c=\tfrac12\Pi_c\nabla v_3(\phi_1)$. The Green identity and source integration at fixed temperature give, modulo analytic terms,
\begin{equation}
(1-k^2)\langle a+b,\phi_3'(1)\rangle_G
=2\sum_c\int_k^1\langle J_c,\phi_{2,c}\rangle_G\,du.
\label{eq:quartic_green_work}
\end{equation}
Since $\phi_{2,c}=L_c^{-1}J_c$ with zero outer data, integrating the scalar work~\eqref{eq:sourcework} along the source at fixed temperature gives
\begin{equation}
F_4\big|_{\rm nonanalytic}=\frac{M_0}{2}\sum_{c=1,2}\int_k^1 J_cL_c^{-1}J_c\,du\bigg|_{\rm nonanalytic}.\label{eq:exchange}
\end{equation}
The fixed-charge spherical Routhian, including the electric Legendre transform, is
\begin{equation}
\begin{aligned}
I_2={}&\frac14\int\sqrt{-g_2}\biggl[R^2\mathcal R_2+2(\nabla R)^2+2-2R^2|\nabla\phi|_G^2-\frac{2V}{R^2}\biggr]+I_\partial.
\end{aligned}\label{eq:Routhian}
\end{equation}
The boundary contribution $I_\partial$ implements the gravitational boundary terms and the fixed-charge ensemble. The leading throat sector has the familiar two-dimensional dilaton-gravity structure \cite{Teitelboim1983,Jackiw1985,AlmheiriPolchinski2015,Jensen2016,EngelsoyMertensVerlinde2016,MertensTuriaci2023}. Stationarity of this functional is an alternative to the scalar-work derivation around Eq.~\eqref{eq:quartic_green_work}; it does not hold the radius fixed or discard its constraint.

For positive integer exponents, $\phi_1=\sum_m v_mh_m$ and its derivatives are polynomial in $u$, with coefficients analytic in $k$. Focusing, $g_2''=-|\phi_1'|^2$, and the outer constraint make $g_2$ polynomial. The cubic scalar equation contains the exchange term $\tfrac12\nabla^2v_3(\phi_1)\phi_2$; first-order fields, quadratic geometry, quartic contacts, target curvature, and the quadratic temperature-preserving shift contribute only analytic terms. Thus Eq.~\eqref{eq:exchange}, with its $M_0/2$ factor, extends to all integer eigenspaces. Equivalently, stationary-action elimination of the quadratic scalar response removes third-order variations. This proves the complete quartic reduction independently of the all-order $(1,2)$ theorem.

An analytic single-field model with only exponent two is
\begin{equation}
v(\phi)=1+6\phi^2+\eta\phi^3+O(\phi^4),\qquad \eta\ne0.
\label{eq:even_exponent_example}
\end{equation}
It has no extremal quadratic resonance, although its thermal exchange gives
\begin{equation}
\left[F_4\right]_{\vartheta^9\log\vartheta}
=-\frac{16Q}{11025}\eta^2\ne0.
\label{eq:smooth_thermal_log}
\end{equation}
For example, $v=\cosh(\sqrt{12}\,\phi)+\eta\phi^3e^{-\phi^2}$ is positive for sufficiently small $|\eta|$ and realizes this jet. This example also allows an independent check by scalar work, without relying on the on-shell action. Write $\phi=\epsilon h_2+\epsilon^2\phi_2+\epsilon^3\phi_3+\cdots$. Then $L_2\phi_2=(3\eta/2)h_2^2$, while the only term in $L_2\phi_3$ containing the non-polynomial $\phi_2$ is $3\eta h_2\phi_2$. Every remaining forcing term is polynomial because $g_2$ is polynomial. The Green identity against $h_2$ consequently gives
\begin{equation}
\left[\phi_3'(1)\right]_{\log k}
=\frac{3\eta}{1-k^2}
\left[\int_k^1h_2^2\phi_2\,du\right]_{\log k}.
\label{eq:even_exponent_green_check}
\end{equation}
Since $dF/d\epsilon=B_0\phi'(1)$ at fixed physical temperature, and the order-two shift from $k$ to that temperature is analytic, integration in the source gives
\begin{equation*}
[F_4]_{\log k}=\frac{M_0}{2}\left(\frac{3\eta}{2}\right)^2
\left[\int_k^1 h_2^2L_2^{-1}h_2^2\,du\right]_{\log k}.
\end{equation*}
This reproduces Eq.~\eqref{eq:smooth_thermal_log} independently of the stationary-action elimination.

\subsection{Exact polynomial inverse and rank-one identity}
For the forcing $5\Gamma(a,a)u^2$, the response is $(5/2)\Gamma(a,a)\mathcal Y$, where
\begin{align}
\mathcal Y={}&\frac{2}{15}(3u^2-k^2)\log\frac{u+k}{1+k}-\frac{k(u-1)(6k^2+ku+k+18u)}{15(k^2-3)},\qquad L_2\mathcal Y=2u^2.\label{eq:Yexact}
\end{align}
With $\ell=\log[2k/(1+k)]$,
\begin{equation}
\left[\int_k^1u^2\mathcal Y\,du\right]_\ell=-\frac{16}{225}k^5.\label{eq:residue}
\end{equation}
The coefficient of $\ell$ equals that of $\log k$, since their difference is analytic at $k=0$.
Define the monic Legendre polynomial $H_m=k^mP_m(u/k)/a_m$, $a_m=2^{-m}\binom{2m}{m}$. Then $L_cH_m=[m(m+1)-c(c+1)]H_m$. The only resonant inverse needed for a polynomial forcing is
\begin{equation}
\begin{aligned}
R_c=\frac1{2c+1}\biggl[&H_c\log\frac{u+k}{1+k}+q_c-\frac{H_c(u,k)}{H_c(1,k)}q_c(1,k)\biggr],
\end{aligned}\label{eq:R-c-exact}
\end{equation}
where, for general $c\ge1$, $q_c$ is any polynomial solving
\begin{equation}
L_cq_c=2cH_c-2(u-k)\partial_uH_c.
\label{eq:qc_general}
\end{equation}
The right-hand side has degree at most $c-1$, and $L_c$ is invertible on this subspace because its Legendre eigenvalues $j(j+1)-c(c+1)$ are nonzero for $j<c$. Thus there is a unique solution of degree at most $c-1$. Adding a multiple of $H_c$ changes neither $L_cq_c$ nor $R_c$, so the convenient alternative choices $q_1=u-k$ and $q_2=7u^2/6-ku-k^2/6$ give the same inverse. Direct substitution yields $L_cR_c=H_c$, horizon regularity, and $R_c(1)=0$ for every integer $c\ge1$. Decompose a homogeneous forcing into $k^{d-m}H_m$. If $\beta_{ec}H_c$ is the resonant component of $f_e$, integration by parts yields
\begin{equation}
\begin{gathered}
\left[\int_k^1f_dL_c^{-1}f_e\,du\right]_\ell=\beta_{ec}[J_{dc}(-k)-J_{dc}(k)],\qquad J_{dc}^\prime=\frac{f_dH_c}{2c+1}.
\end{gathered}\label{eq:polynomial-residue}
\end{equation}
Both endpoints are included. The joint degree is $d+e+1$. For $f_2=u^2$, $f_3=uh_2$, $f_4=h_2^2$, the nonzero symmetric exchange entries are
\begin{align}
\mathcal K^{(1)}_{33}&=-\frac{32k^7}{225(3-k^2)^2},\label{eq:K-one}\\
\mathcal K^{(2)}_{22}&=-\frac{8k^5}{225},\quad \mathcal K^{(2)}_{24}=-\frac{32k^7}{525(3-k^2)^2},\nonumber\\
\mathcal K^{(2)}_{44}&=-\frac{128k^9}{1225(3-k^2)^4}.\label{eq:K-two}
\end{align}
The endpoint formula implies the exact rank-one identity
\begin{equation}
\begin{aligned}
&\left[\int_k^1J L_c^{-1}K\,du\right]_{\log k}=-\frac{a_c^2}{2k^{2c+1}}
\left(\int_{-k}^{k}JH_c\,du\right)
\left(\int_{-k}^{k}KH_c\,du\right).
\end{aligned}
\label{eq:rank_one_exchange}
\end{equation}
The auxiliary interval $[-k,k]$ extracts a polynomial moment; it is not part of the exterior radial domain. For any polynomial $f$, expand $f=\sum_j\beta_jH_j$ at fixed $k>0$. Legendre orthogonality on the auxiliary interval $[-k,k]$ gives
\begin{equation}
\beta_c=\frac{(2c+1)a_c^2}{2k^{2c+1}}
\int_{-k}^{k}f(u,k)H_c(u,k)\,du.
\label{eq:rank_one_projection}
\end{equation}
Choose the primitive of $gH_c/(2c+1)$ in Eq.~\eqref{eq:polynomial-residue} to vanish at $-k$. Its endpoint difference is $-\int_{-k}^kgH_c/(2c+1)\,du$; Eq.~\eqref{eq:rank_one_projection} then proves Eq.~\eqref{eq:rank_one_exchange}. Linearity includes finite sums with coefficients analytic in $k$. Non-resonant components and their outer-data corrections have polynomial inverses, so the negative sign is exact.

For any complete positive integer spectrum without zero modes, the quartic coefficient is therefore
\begin{equation}
\begin{aligned}
\left[F_4(k)\right]_{\log k}
&=-\frac{M_0(k)}4\sum_{c\ge1}\frac{a_c^2}{k^{2c+1}}\left\|\int_{-k}^k J_cH_c\,du\right\|_G^2.
\end{aligned}
\label{eq:integer_quartic_free_energy}
\end{equation}

\subsection{Gaunt selection rules}
The rank-one identity gives selection rules for general integer exponents. Let $m,n,p,q\ge1$ be input exponents, $h_j=H_j/H_j(1,k)$, and $g_{mnc}=\int_{-1}^1P_m(x)P_n(x)P_c(x)\,dx$. For the mixed product $f=H_mH_n$ one has
\begin{equation}
\int_{-k}^k H_mH_nH_c\,du
=\frac{k^{m+n+c+1}}{a_ma_na_c}g_{mnc}.
\label{eq:mixed_gaunt_projection}
\end{equation}
Applying \eqref{eq:rank_one_exchange} twice, and dividing by the four outer normalization factors, gives the mixed-mode kernel
\begin{equation}
\begin{aligned}
&\left[\int_k^1 h_mh_n L_c^{-1}(h_ph_q)\,du\right]_{\log k}=-\frac{k^{m+n+p+q+1}g_{mnc}g_{pqc}}{
2a_ma_na_pa_q H_m(1,k)H_n(1,k)H_p(1,k)H_q(1,k)}.
\end{aligned}
\label{eq:gaunt_mixed}
\end{equation}
The Gaunt coefficient has a factorial form. When $m+n+c$ is even and $|m-n|\le c\le m+n$, let $s=(m+n+c)/2$. Then
\begin{equation}
g_{mnc}=\frac{2(s!)^2}{(2s+1)!}
\prod_{j\in\{m,n,c\}}
\frac{(2s-2j)!}{[(s-j)!]^2};
\label{eq:mixed_gaunt_factorial}
\end{equation}
otherwise $g_{mnc}=0$: a Gaunt factor vanishes unless its indices satisfy the triangle inequalities and have even sum, as follows from the Legendre product decomposition. The mixed-mode selection rule is therefore entirely algebraic. Thus $J_c=\sum_{m,n}C_{c;mn}h_mh_n$ contributes the negative square in Eq.~\eqref{eq:rank_one_exchange}, including cross terms. Metric, contact, and temperature-conversion effects have already been included at quartic source order. Higher source orders require additional matching estimates.

For two identical inputs of exponent $m$, Eq.~\eqref{eq:gaunt_mixed} reduces to
\begin{equation}
\begin{aligned}
\mathcal K_{m,c}(k)&\equiv
\left[\int_k^1h_m^2L_c^{-1}h_m^2\,du\right]_{\log k}=-\frac{k^{4m+1}}{2H_m(1,k)^4}
\left(\frac{g_{mmc}}{a_m^2}\right)^2,\\
g_{mmc}&=\int_{-1}^1P_m(x)^2P_c(x)\,dx.
\end{aligned}
\label{eq:gaunt_exchange}
\end{equation}
The single-mode Gaunt integral vanishes unless $c\le2m$ is even: for $c>2m$ the resonant projection vanishes, while parity makes $g_{mmc}=0$ for odd $c$. Every nonzero coefficient has the closed form
\begin{equation}
\begin{aligned}
g_{mm,2\ell}&=\frac{2[(2\ell)!]^2(2m-2\ell)!}{(2m+2\ell+1)!}\left[\frac{(m+\ell)!}{(\ell!)^2(m-\ell)!}\right]^2,
\quad 0\le\ell\le m.
\end{aligned}\label{eq:gaunt_factorial}
\end{equation}
In the maximal channel,
\begin{equation}
\begin{aligned}
g_{mm,2m}&=\frac{2[(2m)!]^4}{(4m+1)!(m!)^4},\\
\mathcal K_{m,2m}&=-\frac{2^{4m+1}[(2m)!]^4}{[(4m+1)!]^2}
\frac{k^{4m+1}}{H_m(1,k)^4}.
\end{aligned}
\label{eq:gaunt_maximal}
\end{equation}
These expressions retain the full $k$-dependent mode normalization. Equations~\eqref{eq:gaunt_exchange}--\eqref{eq:gaunt_maximal} give $-8k^5/225$ for $(m,c)=(1,2)$; for $(m,c)=(2,2)$ the general channel gives $-128k^9/[1225(3-k^2)^4]$, reproducing two independent entries above. Equation~\eqref{eq:gaunt_exchange} can also be verified directly. The same orthogonality gives the exact decomposition
\begin{equation}
H_m(u,k)^2=\sum_{c=0}^{2m}
\frac{(2c+1)a_c}{2a_m^2}\,g_{mmc}\,k^{2m-c}H_c(u,k),
\label{eq:gaunt_decomposition}
\end{equation}
and the resonant coefficient and endpoint difference in Eq.~\eqref{eq:polynomial-residue} are
\begin{equation*}
\begin{gathered}
\beta_c=\frac{(2c+1)a_c g_{mmc}}{2a_m^2}k^{2m-c},\qquad J_{2m,c}(-k)-J_{2m,c}(k)
=-\frac{k^{2m+c+1}g_{mmc}}{(2c+1)a_m^2a_c}.
\end{gathered}
\end{equation*}
Their product is $-k^{4m+1}g_{mmc}^2/(2a_m^4)$, and dividing by $H_m(1,k)^4$ reproduces \eqref{eq:gaunt_exchange}.

For a slow mode of exponent $m$ quadratically sourcing a fast mode of exponent $2m$, the local superpotential has $W/Q=1+\frac m2|x|^2+m|y|^2+\langle y,\Gamma_m(x,x)\rangle+\cdots$ along the selected directions, where $\Gamma_m$ is the normal-form coupling in the conventions of Eq.~\eqref{eq:normalform}. The identity $V=W^2+|\nabla W|^2$ fixes the fast quadratic forcing to $(4m+1)\Gamma_m(a,a)h_m^2$, and multiplication of \eqref{eq:gaunt_maximal} by $Q(4m+1)^2/2$ gives its quartic \emph{fast-channel} contribution,
\begin{equation}
\left[F_{4,\mathrm{fast}}^{\mathrm{ex}}\right]_{\vartheta^{4m+1}\log\vartheta}
=-Q\left[\frac{2^{2m}[(2m)!]^2}{(4m)!}\right]^2\|\Gamma_m(a,a)\|^2.
\label{eq:general_fast_coefficient}
\end{equation}
At $m=1$ the coefficient is $-4Q\|\Gamma(a,a)\|^2/9$. For even $m$, a slow cubic self-coupling has a nonzero $c=m$ Gaunt integral and can contribute at the same thermal order without an extremal quadratic resonance. This is a channel identity for the scalar exchange; the all-order theorem remains restricted to $(1,2)$.

For an arbitrary tangent, the cubic coefficients are $C_c^{11}=\frac14\Pi_c(\nabla^3v)(a,a)$, $C_c^{12}=\frac12\Pi_c(\nabla^3v)(a,b)$, and $C_c^{22}=\frac14\Pi_c(\nabla^3v)(b,b)$, with raised output index; $C_2^{11}=5\Gamma(a,a)$. The order-two geometry is polynomial:
\begin{equation}
\begin{aligned}
g_2={}&\frac{|a|^2}{2}(1-u)(u+k^2)+\frac{|b|^2}{3(1-k^2/3)^2}[(1-k^4)u+k^4-u^4].
\end{aligned}
\end{equation}
Its constraint, the analytic temperature shift, quartic contacts, and target curvature give polynomial Green moments. A bent source path only pulls back analytic quadratic and cubic free energies. These observations account for every quartic contribution.

\subsection{Mixed fifth order and the exponential model}\label{App:Checks}
For canonical fields with $x\mapsto-x$ symmetry, retain the weighted potential jet
\begin{align}
v={}&1+2x^2+6y^2+10\gamma x^2y+v_4x^4+v_{22}x^2y^2+v_{41}x^4y+v_{03}y^3+v_6x^6+O_{\operatorname{wt}\ge8},\label{eq:potential6}
\end{align}
where $\operatorname{wt}(x)=1$ and $\operatorname{wt}(y)=2$. For physical sources $(x_\infty,y_\infty)=(\epsilon,\eta)$, the extremal slow amplitude has the expansion
\begin{align}
A(\epsilon,\eta)&=\epsilon-\gamma\epsilon\eta+O(\|(\epsilon,\eta)\|^3),\label{eq:Amixed}\\
A(\epsilon,0)&=\epsilon+a_3\epsilon^3+O(\epsilon^5),\nonumber\\
a_3&=\frac{3-2v_4+7\gamma^2}{10}.\label{eq:a3}
\end{align}
Direct thermal calculations give
\begin{align}
\left[[\epsilon^4\eta]F\right]_{\vartheta^5\log\vartheta}&=\frac{16Q}{9}\gamma^3,\label{eq:mixedfifth}\\
[F_6]_{\vartheta^5\log\vartheta}&=-\frac{16Q}{9}\gamma^2a_3,\nonumber\\
[F_6]_{\vartheta^5\log^2\vartheta}&=0.\label{eq:sixthmain}
\end{align}
The coefficients $v_{22},v_{41},v_{03},v_6$ enter the full response but cancel from these residues.

One positive analytic model realizing the jet is
\begin{equation}
v(x,y)=\tfrac12[e^{2\sqrt3y}\cosh(2\sqrt2x)+e^{-2\sqrt3y}],\label{eq:exponentialmodel}
\end{equation}
with one Maxwell field, $f=1/v$, and charge $Q$. It has $\gamma=2\sqrt3/5$, $v_4=4/3$, $v_{22}=12$, $v_{41}=8\sqrt3/3$, $v_{03}=0$, $v_6=16/45$, and $a_3=277/750$. The exact logarithmic coefficients divided by $Q$ are $-16/75$, $128\sqrt3/375$, and $-8864/28125$ for $\epsilon^4$, $\epsilon^4\eta$, and $\epsilon^6$, respectively.

For the calculation, set $x=\epsilon u+\epsilon^3Z+\epsilon^5W_5+\cdots$, $y=\epsilon^2Y+\epsilon^4H+\cdots$, and $g=1+\epsilon^2g_2+\epsilon^4g_4+\cdots$. Here $Y=(5\gamma/2)\mathcal Y$ and $g_2=(1-u)(u+k^2)/2$. The lower response equations are
\begin{align}
L_1Z&=S_3=10\gamma uY+2v_4u^3-4g_2u,\label{eq:S3}\\
L_2H&=10\gamma uZ+v_{22}u^2Y+\tfrac12v_{41}u^4+\tfrac32v_{03}Y^2-12g_2Y-10\gamma g_2u^2.\label{eq:S4fast}
\end{align}
All response fields have zero outer data. For the mixed coefficient, put $J=\int_k^u vS_3(v)dv$ and
\begin{equation}
K_H=\frac{10\gamma}{3-k^2}\left[\frac35(u^5-k^5)-\frac{k^2}{3}(u^3-k^3)\right].
\end{equation}
Reduction of order and the Green identity against $h_2$ give
\begin{equation}
(1-k^2)H'(1)=-\int_k^1\frac{JK_H}{u^2D}\,du+\int_k^1h_2\mathcal R_Hdu,\label{eq:mixedGreen}
\end{equation}
where $\mathcal R_H=L_2H-10\gamma uZ$. The primitive has $J=J_{\rm reg}-(16\gamma^2/9)k^5\ell$, while
\begin{equation}
\begin{aligned}
\int_k^1\frac{K_H}{u^2D}\,du&=\frac{\gamma[9-k^2-8k^3-8k^2\ell]}{3(3-k^2)}=\gamma+O(k^2|\log k|).
\end{aligned}
\end{equation}
Here $K_H(k)=0$, so the separated moment is regular. After this cancellation, the remaining non-analytic integrals have the form $c(k)k^a\int_k^1u^b(u+k)^{-m}\log^n[(u+k)/(1+k)]du$ with $a\ge0$ and $a+b-m+1\ge6$. Their fifth derivatives are integrable, and their sixth derivatives have at most finite logarithmic growth. Thus their remainder is $O(k^6[1+|\log k|^p])$. Analytic lower-primitive terms start at degree four and multiply a moment whose first logarithm has degree two. Scalar work gives $[\epsilon^4\eta]F/Q=\sqrt{1-k^2}H'(1)+\kappa_2\partial_k[\sqrt{1-k^2}Y'(1)]$, where $\kappa_2$ is the order-two shift of $k$ at fixed temperature. Both $Y'(1)$ and $\kappa_2$ are analytic, establishing Eq.~\eqref{eq:mixedfifth}.

\subsection{Sixth-order response and endpoint control}
The full equations further give
\begin{align}
L_1W_5={}&10\gamma(uH+ZY)+6v_4u^2Z+v_{22}uY^2+2v_{41}u^3Y+3v_6u^5\nonumber\\&-4g_2Z-20\gamma g_2uY-4v_4g_2u^3+6g_2^2u-4g_4u,\label{eq:S5full}\\
g_4''={}&-2Z'-Y'^2-g_2.\label{eq:g4full}
\end{align}
All higher-order fields vanish at $u=1$. With $h_2^g=g_2'(1)=-(1+k^2)/2$, the boundary constraint fixes
\begin{align}
h_4^g:=g_4'(1)={}&\frac{1-k^2}{2}\bigl\{(h_2^g)^2+2Z'(1)+Y'(1)^2\bigr\}-\frac{v_4}{2}.\label{eq:g4slope}
\end{align}
Define
\begin{align}
q(u)&=\frac{u^4-4k^3u+3k^4}{12},\nonumber\\
w(u)&=30\gamma uY+6v_4u^3-4ug_2-8q'(u),\nonumber\\
J(u)&=\int_k^u vS_3(v)\,dv,\qquad K(u)=\int_k^u vw(v)\,dv.\label{eq:sixthJK}
\end{align}
Self-adjointness gives $5\gamma\int u^2H=\int YL_2H$. The focusing identity, using $q''=u^2$ and $q(k)=q'(k)=0$, is
\begin{align}
\int_k^1u^2g_4\,du={}&-q(1)h_4^g+2\int_k^1q'Z\,du-\int_k^1q(Y'^2+g_2)\,du.
\end{align}
Finally $(Z/u)'=J/(u^2D)$ implies $\int wZ=-\int JK/(u^2D)$. Combining these identities yields
\begin{align}
(1-k^2)W_5'(1)={}&-\int_k^1\frac{JK}{u^2D}\,du+\int_k^1\mathcal R_6\,du+4q(1)h_4^g,\label{eq:sixthGreen}
\end{align}
where
\begin{align}
\mathcal R_6={}&3v_{22}u^2Y^2+3v_{41}u^4Y+3v_{03}Y^3-24g_2Y^2-40\gamma g_2u^2Y+3v_6u^6\nonumber\\
&-4v_4g_2u^4+6g_2^2u^2+4q(Y'^2+g_2).\label{eq:sixthR}
\end{align}
Equation~\eqref{eq:sixthGreen} retains the boundary-determined slow homogeneous component, fast response, and metric slope. No amplitude-matching formula enters it.

The primitives defining $J,K$ are polynomials times $L=\log[(u+k)/(1+k)]$, plus polynomials with coefficients analytic in $k$. Their lower logarithms are
\begin{align}
J&=J_{\rm reg}-C_Jk^5\ell,\nonumber\\
K&=K_{\rm reg}-3C_Jk^5\ell,\qquad C_J=\frac{16\gamma^2}{9}.\label{eq:sixthlower}
\end{align}
At $k=0$ the complete functions reduce to
\begin{align}
J_0&=-\tfrac12u^4+u^5\left[2\gamma^2\log u+\frac{2v_4+2-2\gamma^2}{5}\right],\nonumber\\
K_0&=-\tfrac12u^4+u^5\left[6\gamma^2\log u+\frac{6v_4-2/3-6\gamma^2}{5}\right].\label{eq:sixthlimits}
\end{align}

Since $J_{\rm reg}$ and $K_{\rm reg}$ do not separately vanish at $u=k$, their separated products cannot be integrated through the horizon. Split the integral into $[k,2k]$ and $[2k,1]$. In the first interval keep the complete $J,K$ and set $u=kt$. Both lowest-degree terms are $-(u^4-k^4)/2$, and
\begin{equation}
\frac{(u^4-k^4)^2}{4u^2(u^2-k^2)}=\frac14\left(u^4+k^2u^2-k^4-\frac{k^6}{u^2}\right).\label{eq:sixthlowest}
\end{equation}
Its integral has no logarithm. All other combined inner terms have integrated degree at least six and are $O(k^6[1+|\log k|^p])$ for a finite $p$. Their horizon zeros remain explicit.

On $[2k,1]$ the separation \eqref{eq:sixthlower} is nonsingular. Beyond \eqref{eq:sixthlowest}, a term of joint integrand degree $d\ge5$ has a fifth $k$ derivative bounded by $C u^{d-5}(1+|\log u|^p)$. This is integrable at zero. Taylor expansion through degree five and one more integral-remainder estimate leave $O(k^6[1+|\log k|^q])$. In particular such terms supply no $k^5/u$ residue. The same reasoning applies to \eqref{eq:sixthR}: its sole joint-degree-four contribution is polynomial, and every logarithmic term has degree at least five. Terms containing $v_{22},v_{41},v_{03},v_6$ have degree at least six in the integrand.

The single lower logarithms multiply an outer moment with integrable limit $(K_0+3J_0)/u^4$. Extending it to zero discards only $O(k)$ times finite logarithms. The product of the two lower logarithms is $k^{10}\ell^2$ times an outer integral of order $k^{-3}$, and therefore starts at $k^7\ell^2$. This separation is used only on the outer interval; the complete inner expression was controlled above.

The boundary constraint gives $[h_4^g]_{k^5\ell}=-C_J$, since $Y'(1)$ is rational in $k$. Hence
\begin{align}
[W_5'(1)]_{k^5\ell}&=C_J\left[\int_0^1\frac{K_0+3J_0}{u^4}\,du-\frac13\right],\nonumber\\
\int_0^1\frac{K_0+3J_0}{u^4}\,du-\frac13&=\frac{6v_4-21\gamma^2-9}{5}=-6a_3.\label{eq:sixthmoment}
\end{align}
Thus the scalar boundary coefficient is $-(32/3)\gamma^2a_3$. There is no $k^5\ell^2$ term.

\subsection{Fixed-temperature conversion}
To complete the ensemble conversion, put $l=\log\vartheta_0(k)$, $s_2=[\epsilon^2]\log S=k(1-k^2)$ and $s_4=[\epsilon^4]\log S$. At fixed $T$,
\begin{align}
k(\epsilon)&=k+\epsilon^2\kappa_2+\epsilon^4\kappa_4+\cdots,\nonumber\\
\kappa_2&=\frac{s_2}{l'},\qquad \kappa_4=\frac{s_4+\kappa_2s_2'-\tfrac12l''\kappa_2^2}{l'}.\label{eq:kappas}
\end{align}
Source work gives, with $\beta_k=\sqrt{1-k^2}$,
\begin{align}
\frac{6F_6}{Q}={}&\beta_k W_5'(1)+\kappa_2(\beta_k Z'(1))'+\kappa_4\beta_k'+\tfrac12\kappa_2^2\beta_k''.\label{eq:sixthfixedT}
\end{align}
The two notions of prime in this formula are unambiguous: scalar primes at $(1)$ are $u$ derivatives; primes outside parentheses and on $\beta_k$ are $k$ derivatives. One has $\kappa_2=O(k^2)$ and $l'=(1-2k)/[k(1-k^2)]$. The direct focusing calculation gives $[s_4]_{\log}=O(k^4\log^p k)$, hence $[\kappa_4]_{\log}=O(k^5\log^p k)$. To see the first bound without assuming the answer, in the horizon integral of \eqref{eq:g4full} the only new singular factor is $Y'^2=O((u+k)^2[1+|\log(u+k)|^2])$; its additional focusing moment has integrated degree four. The $Z'$ term has the same or higher degree after its lower moment is kept combined. Thus no lower logarithm is present in $g_4(k)$ or $s_4=2g_4(k)-g_2(k)^2$.

The non-analytic parts of every conversion term in \eqref{eq:sixthfixedT} start at $k^6\log^p k$ or higher, since $\beta_k'=O(k)$. Using $k=\vartheta+O(\vartheta^2)$ proves \eqref{eq:sixthmain}.

\subsection{Delayed obstruction}
For three canonical scalars, consider
\begin{align}
W/Q={}&1+\tfrac12(x_1^2+x_2^2)+y^2+\gamma y(x_1^2-x_2^2)+\lambda x_1^3,\nonumber\\
V={}&W^2+|\nabla W|^2,\label{eq:countermodel}
\end{align}
on the straight physical source $(\epsilon,\epsilon,0)$. Its nonlinear amplitudes satisfy
\begin{align}
A_1&=\epsilon-3\lambda\epsilon^2+O(\epsilon^3),\qquad A_2=\epsilon+O(\epsilon^3),\nonumber\\
\Gamma(A,A)&=\gamma(A_1^2-A_2^2)=-6\gamma\lambda\epsilon^3+O(\epsilon^4).\label{eq:counterA}
\end{align}
Although $\Gamma((1,1),(1,1))=0$, the predicted sixth-order coefficient is
\begin{equation}
\left[[\epsilon^6](F_\gamma-F_0)\right]_{\vartheta^5\log\vartheta}=-16Q\gamma^2\lambda^2.\label{eq:delayedresult}
\end{equation}
The direct thermal calculation provides a separate check. The second-order fast field vanishes, while
\begin{equation}
x_{1,2}=3\lambda\{u^2-(1-k^2)u-k^2\},\qquad x_{2,2}=0.
\end{equation}
The first fast response satisfies
\begin{align}
L_2y_3&=J_3=\gamma\lambda\{36u^3-30(1-k^2)u^2-30k^2u\},\nonumber\\
y_3&=\gamma\lambda\left[-15(1-k^2)\mathcal Y+6u^3-\tfrac32k^2u-(6-\tfrac32k^2)h_2\right].\label{eq:delayedfield}
\end{align}
The $36u^3$ term includes the quartic contact $12\gamma\lambda yx_1^3$ in the potential; omitting it would be inconsistent with \eqref{eq:countermodel}.

At sixth source order, fixed-temperature stationarity gives the entire $\gamma$-dependent contribution:
\begin{equation}
[\epsilon^6](F_\gamma-F_0)=\frac{M_0}{2}\int_k^1J_3y_3\,du+\text{terms analytic in }k.\label{eq:delayedaction}
\end{equation}
The additional pure-slow contact is polynomial. The logarithmic moments of $u\mathcal Y$ and $u^3\mathcal Y$ vanish by parity; \eqref{eq:residue} gives the remaining moment. Therefore
\begin{equation}
\left[[\epsilon^6](F_\gamma-F_0)\right]_\ell=-16Q\gamma^2\lambda^2(1-k^2)^{3/2}k^5.
\end{equation}
This gives Eq.~\eqref{eq:delayedresult} at the same charges, sources, and physical temperature.

\section{Special geometry and selection rules}\label{App:Special}\label{App:Models}
\subsection{The aligned cubic branch}
Let
\begin{equation}
\begin{gathered}
F=-\frac{d_{abc}X^aX^bX^c}{6X^0},\qquad D(y)=\frac16d_{abc}y^ay^by^c,\qquad t^a=x^a+iy^a,
\end{gathered}
\label{special:cubic}
\end{equation}
with real symmetric $d_{abc}$. Work at an interior point with $0<D<\infty$,
finite positive-definite $g_{a\bar b}=\partial_a\partial_{\bar b}K$,
$K=-\log(8D)$, and $0<Q^2=V_*<\infty$. The real metric is $G=g\oplus g$.
Write $f_a=\mathcal D_a Z$, $\bar f^a=g^{a\bar b}\bar f_{\bar b}$,
$C_{abc}=e^Kd_{abc}$, and
\begin{equation}
L_a=\frac{D_a}{2D},\qquad
 g^{a\bar b}L_b=2y^a,\qquad L_ag^{a\bar b}L_b=3.
\label{special:noscale}
\end{equation}
Here $\mathcal D$ is the K\"ahler and Levi-Civita derivative, and $D_a$
is a derivative of the cubic volume. These conventions give
\begin{equation}
\mathcal D_af_b=-iC_{abc}\bar f^c,\qquad
\partial_aV=2\bar Zf_a-iC_{abc}\bar f^b\bar f^c.
\label{special:critical}
\end{equation}
A non-BPS point with $Z\ne0$ is \emph{aligned} if
\begin{equation}
 f_a=\zeta L_a,\quad \zeta\ne0,
\qquad
\left(\delta^a{}_b-\frac23y^aL_b\right)\bar f^b=0.
\label{special:alignment}
\end{equation}
The last equation is an equivalent condition: the matter charge has no
component tangent to the surface of constant volume. Alignment is K\"ahler-phase independent and covariant
under coordinate changes carrying the specified cubic structure; it does
not identify an arbitrary-gauge $K_a$ with an invariant covector.
In particular, it is not preserved by the additional dualities of the
symmetric models. For example, the $st^2$-type cubic
$D=\frac23y_1^3+y_1^2y_2$ with charges
$(q_0,q_1,q_2;p^0,p^1,p^2)=(3,2,3;3,0,3)$ has a regular non-BPS critical
point at $(x;y)=(0,0;1,\tfrac13)$ with $V_*=18$, $|Z|^2/V_*=\tfrac14$ and
spectrum $\{0,1,1,2\}$, at which $\partial_a\mathcal W=(10+6i,6)$ is not parallel to
$L=(\tfrac43,\tfrac12)$. Such points are dual to aligned ones~\cite{BellucciEtAl2006}, and the
vanishing of the tensor $\Gamma$ is transported to them.
Since $C_{abc}y^by^c=L_a/2$, aligned criticality is equivalent to
\begin{equation}
\begin{gathered}
\bar Z\zeta=i\bar\zeta^2,\qquad Z=-i\frac{\zeta^2}{\bar\zeta},\qquad |f|^2=3|Z|^2=3|\zeta|^2,\qquad Q^2=4|\zeta|^2.
\end{gathered}
\label{special:alignedcritical}
\end{equation}

\paragraph{Aligned-branch theorem.}
Under these regularity and criticality hypotheses, for every $n\ge1$, put
\begin{equation}
\begin{gathered}
\zeta=i\rho e^{-i\phi},\quad Z=-\rho e^{-3i\phi},\quad \rho>0,\qquad s_\phi=\sin(2\phi),\qquad c_\phi=\cos(2\phi).
\end{gathered}
\end{equation}
For a real tangent vector $\delta t=(a,b)=(\delta x,\delta y)$, define
$\hat a=c_\phi a-s_\phi b$, $\hat b=s_\phi a+c_\phi b$. Then
\begin{align}
(\nabla^2V)(\delta t,\delta t)
 &=4Q^2\bigl[(L\cdot\hat a)^2+g(\hat b,\hat b)\bigr],\label{special:alignedhessian}\\
(\nabla^3V)(\delta t,\delta t,\delta t)
 &=-8Q^2C_{abc}\hat b^a\hat b^b\hat b^c=-\frac{Q^2}{D}d_{abc}\hat b^a\hat b^b\hat b^c.
\label{special:alignedcubic}
\end{align}
The eigenspaces of $\mathsf H=(2Q^2)^{-1}G^{-1}\nabla^2V$ are
\begin{align}
E_1&=\{\hat a=0\},& \dim E_1&=n,\nonumber\\
E_2&=\{\hat b=0,\ \hat a\parallel y\},& \dim E_2&=1,\nonumber\\
E_0&=\{\hat b=0,\ L\cdot\hat a=0\},& \dim E_0&=n-1,
\label{special:eigenspaces}
\end{align}
with eigenvalues $2,6,0$, respectively. Polarization gives
\begin{equation}
\nabla^3V(E_0\oplus E_2,T_p\mathcal M,T_p\mathcal M)=0,
\qquad \Gamma(E_1,E_1)=0.
\label{special:selection}
\end{equation}
Stability in the zero directions requires the higher-order analysis below.

\paragraph{Proof.}
Denote the holomorphic connection by $\mathsf C^a{}_{ib}$, to distinguish it
from the resonance map. Cubic homogeneity implies
\begin{equation}
\begin{gathered}
\mathsf C^a{}_{ib}y^b=i\delta_i^a,\qquad
\mathcal D_i y^a=\frac i2\delta_i^a,\qquad R_{ij}:=L_iL_j-g_{ij}=2C_{ijk}y^k.
\end{gathered}
\label{special:homogeneity}
\end{equation}
Indeed $y^a\partial_{y^a}g=-2g$, and the symmetry of its imaginary-coordinate
derivatives gives $y^a\partial_a g=ig$ for the holomorphic derivative.
For the K\"ahler-weight-two cubic tensor,
\begin{equation}
\mathcal D_iC_{jkl}=2K_iC_{jkl}
-\mathsf C^m{}_{ij}C_{mkl}
-\mathsf C^m{}_{ik}C_{jml}
-\mathsf C^m{}_{il}C_{jkm}.
\end{equation}
Using $K_i=iL_i$ and $y^iL_i=3/2$ yields
\begin{equation}
 y^i\mathcal D_iC_{jkl}=0,\qquad
 (\mathcal D_i\mathcal D_jC_{klm})y^ly^m=0.
\label{special:radialidentities}
\end{equation}
For the second identity, use the full symmetry of $\mathcal D_iC_{jkl}$
and differentiate its double radial contraction with
$\mathcal D_i y^a=i\delta_i^a/2$. These radial identities do not require
$\mathcal DC=0$.
The independent complex Hessian components are
\begin{align}
M_{ij}:=\nabla_i\partial_jV
 &=-4i\bar ZC_{ijk}\bar f^k-i(\mathcal D_iC_{jkl})\bar f^k\bar f^l,\nonumber\\
N_{i\bar j}:=\nabla_i\partial_{\bar j}V
 &=2g_{i\bar j}|Z|^2+2f_i\bar f_{\bar j}
 +2C_{ikl}\bar C_{\bar j}{}^{km}\bar f^lf_m.
\label{special:complexhessian}
\end{align}
Substituting $\bar f^a=2\bar\zeta y^a$,
Eq.~\eqref{special:radialidentities}, and $Rg^{-1}R=g+LL$ gives
\begin{equation}
 M=Q^2e^{4i\phi}R,\qquad N=Q^2(g+LL).
\end{equation}
Since $e^{2i\phi}(a+ib)=A+iB$, these expressions give
Eq.~\eqref{special:alignedhessian}; the radial projector
$P_y v=(2/3)y(L\cdot v)$ gives Eq.~\eqref{special:eigenspaces}.
Differentiate Eq.~\eqref{special:complexhessian} at fixed charges before
imposing alignment. Equation~\eqref{special:radialidentities} removes the
derivative-of-$C$ terms, leaving
\begin{equation}
\mathcal T_{ijk}=-4i\bar Z^2C_{ijk},\qquad
\mathcal T_{ij\bar k}=-4i\bar\zeta^2C_{ijk},\qquad \mathcal T=\nabla^3V.
\label{special:complextensor}
\end{equation}
The real cubic frame identifies the components on the right of the mixed
equation. Its simplification uses
\begin{align}
R_i{}^aR_j{}^bC_{kab}&=C_{ijk}+L_ig_{jk}+L_jg_{ik},\nonumber\\
C_{ija}R_k{}^a&=L_kR_{ij}-C_{ijk},
\end{align}
and $\bar Z\zeta=i\bar\zeta^2$; the terms proportional to $Lg$ and $LR$
cancel. With $\xi=A+iB$, adding the components and their conjugates gives
\begin{equation}
\mathcal T(\Delta^3)
=2Q^2\operatorname{Re}\!\left\{C[-i\xi^3+3i\xi^2\bar\xi]\right\}
=-8Q^2C(B,B,B).
\end{equation}
The $A^2B$ terms cancel, proving Eq.~\eqref{special:alignedcubic} and, by
polarization, Eq.~\eqref{special:selection}. $\square$

In the D0--D4 phase, axion reflection gives a shorter proof of the resonant
zero. The homogeneity proof also applies to non-symmetric cubics on their aligned
branch, without a global orbit classification.

\subsection{The zero-mode quartic}
At a positive-metric point, a real linear basis and a positive volume
normalization put the cubic exactly in the form
\begin{equation}
\begin{gathered}
\widehat D(r,z)=r^3-r|z|^2+\mathcal C(z),\qquad g_* =\operatorname{diag}(3/4,1/2,\ldots,1/2),
\end{gathered}
\label{special:localcubic}
\end{equation}
where $\mathcal C$ is an arbitrary homogeneous shape cubic. The zero vector
and its norm are
\begin{equation*}
\begin{gathered}
N(a)=(c_\phi(0,a),-s_\phi(0,a)),\qquad \|N(a)\|_G^2=|a|^2/2.
\end{gathered}
\end{equation*}
Eliminating the massive coordinates by
$\partial_{\rm massive}V=0$ gives
\begin{align}
 V_{\rm red}(a)&=Q^2+\kappa(a)+O(|a|^5),\nonumber\\
 \kappa(a)&=Q^2c_\phi^3
 \left[\frac{|a|^4}{12}-\frac{|\nabla\mathcal C(a)|^2}{16}\right].
\label{special:quartic}
\end{align}
There is no cubic term. The coefficient includes massive-field
relaxation~\cite{NTT2007}; a straight null-line restriction alone does not
determine stability. If $M_{IJ}=V_{IJ}$ is the massive Hessian in fixed
local coordinates, then
\begin{equation}
\kappa(a)=\frac{V_{NNNN}}{24}
-\frac18V_{INN}(M^{-1})^{IJ}V_{JNN}.
\label{special:quarticexchange}
\end{equation}
All derivatives here are ordinary coordinate derivatives. With
$A_2=|a|^2$, $B_4=|\nabla\mathcal C(a)|^2$, the two terms are
\begin{equation}
\begin{gathered}
\begin{aligned}
\frac{V_{NNNN}}{24Q^2}
 &=\left(\frac{s_\phi^2}{2}+\frac{c_\phi^2}{6}+\frac{c_\phi^3}{12}\right)A_2^2+\frac{c_\phi^2(1-c_\phi)}{16}B_4,
\end{aligned}\\
\frac{V_{INN}(M^{-1})^{IJ}V_{JNN}}{8Q^2} =\left(\frac{s_\phi^2}{2}+\frac{c_\phi^2}{6}\right)A_2^2
 +\frac{c_\phi^2}{16}B_4.
\end{gathered}
\label{special:quarticproof}
\end{equation}
Expansion of the cubic and inverse metric gives these expressions using
$\mathcal C''(a)a=2\nabla\mathcal C(a)$ and
$a\cdot\nabla\mathcal C(a)=3\mathcal C(a)$. Their difference proves
Eq.~\eqref{special:quartic}. Equivalently, for a real primitive vector $f$
with $D_af^a=0$,
\begin{equation}
\kappa(f)=\frac{Q^2c_\phi^3}{2}
\left[(g_{a\bar b}f^af^b)^2
-g^{i\bar j}C_{iab}\bar C_{\bar j\bar c\bar d}f^af^bf^cf^d\right].
\label{special:quarticinvariant}
\end{equation}
A positive-definite quartic gives a strict local minimum; a negative value
gives nonlinear instability. A zero quartic alone does not prove an exact
modulus. At $c_\phi=0$, however, the charges are D0--D6 and
\begin{equation}
V(x=0,y)=\frac{q_0^2}{2D(y)}+\frac{(p^0)^2D(y)}2.
\label{special:d0d6flat}
\end{equation}
The locus $D(y)=|q_0/p^0|$ consists of exact critical points with positive
normal Hessian and $n-1$ local moduli. Static elimination determines the
leading center potential; it does not identify the full invariant center
manifold of the coupled radial equations.

\subsection{STU reduction and the four-charge family}
The STU model illustrates the selection rule~\eqref{special:selection} and allows two further checks of it. Use the electromagnetic frame of Ref.~\cite{CastroVerheijden2021}, with vanishing gauge coupling, charges $Q_1=P^1=q_{\rm d}>0$, and all other charges zero. The extremal radius is $Q^2=q_{\rm d}^2/2$. Its physical scalars are $(\varphi_i,\chi_i)$, $i=1,2,3$, with target metric $\frac14\sum_i[d\varphi_i^2+e^{2\varphi_i}d\chi_i^2]$. The origin has positive exponents $(1,1,1,2)$ and two flat directions.

The fixed-charge Maxwell Routhian, including the electric Legendre transform, gives
\begin{equation}
V=\frac14\left[\mathbf P^T\mathcal K\mathbf P+(\mathbf Q-2\mathcal H\mathbf P)^T\mathcal K^{-1}(\mathbf Q-2\mathcal H\mathbf P)\right],\label{stu:routh}
\end{equation}
where $\mathcal K,\mathcal H$ denote the electromagnetic matrices in Appendix B of Ref.~\cite{CastroVerheijden2021}. The factor $1/4$ converts its Maxwell normalization to ours. With $v=V/Q^2$ and $Q^2=q_{\rm d}^2/2$, the evaluated six-scalar potential is
\begin{align}
2v={}&e^{-\varphi_1+\varphi_2+\varphi_3}+e^{\varphi_1-\varphi_2-\varphi_3}+\chi_3^2e^{\varphi_1-\varphi_2+\varphi_3}+\chi_2^2e^{\varphi_1+\varphi_2-\varphi_3}+(\chi_1-\chi_2\chi_3)^2e^{\varphi_1+\varphi_2+\varphi_3}.\label{stu:potential}
\end{align}
The quartic charge invariant, Eq.~(4.1) of Ref.~\cite{CastroVerheijden2021}, evaluates to $-q_{\rm d}^4/16<0$ on these charges, so they lie on the non-BPS $Z\neq0$ orbit.

For the covariant resonance calculation, put $d_i=\varphi_i/2$ and $\alpha_i=\chi_i/2$. The target metric is
\begin{equation}
ds_G^2=\sum_{i=1}^3\left(dd_i^2+e^{4d_i}d\alpha_i^2\right).\label{stu:metric}
\end{equation}
At the origin the slow directions are $\alpha_i$, and the fast unit vector is $e=(-1,1,1)/\sqrt3$ in the $d$ subspace, with $y_{\rm lin}=e\cdot d$. Two orthogonal $d$ directions are flat. The ordinary third derivative is
\begin{equation}
\partial_{y_{\rm lin}}\partial_{\alpha_i}\partial_{\alpha_j}v=\frac8{\sqrt3}\operatorname{diag}(1,-1,-1)_{ij}.\label{stu:raw}
\end{equation}
It is not the resonant tensor. The nonzero target connection components at the origin are $\mathsf C^{d_i}_{\alpha_i\alpha_i}=-2$ and $\mathsf C^{\alpha_i}_{d_i\alpha_i}=2$. In $\nabla^3v$, all three connection--Hessian terms must be retained. For each diagonal slow channel the ordinary derivative is $-8e_i$ and their combined correction is $+8e_i$; the off-diagonal channels vanish. Therefore
\begin{equation}
(\nabla^3v)_0(E_2,E_1,E_1)=0.\label{stu:covzero}
\end{equation}

This cancellation also appears directly at positive temperature. On the RN background, let $\alpha_i=\epsilon s_i u+O(\epsilon^2)$ and excite no linear fast mode. The coefficient $y_2=[\epsilon^2]y_{\rm lin}$ obeys
\begin{align}
L_2y_2&=D_s\{u^2-(u^2-k^2)\}=D_sk^2,\label{stu:forcing}\\
D_s&=\frac2{\sqrt3}(s_1^2-s_2^2-s_3^2),\nonumber\\
y_2(u)&=\frac{D_sk^2(u^2-1)}{2(3-k^2)}.\label{stu:polynomial}
\end{align}
The potential and connection supply the two terms in braces. The solution is horizon regular, vanishes at $u=1$, and is polynomial. An order-$\epsilon^2$ fast boundary source adds the polynomial homogeneous solution $P_2(u/k)/P_2(1/k)$. This check keeps the original physical source coordinates fixed.

Introduce analytic coordinates
\begin{equation}
e^{-\varphi_i}=e^{2h_i}\operatorname{sech}(2t_i),\qquad \chi_i=e^{2h_i}\tanh(2t_i).\label{stu:coordinates}
\end{equation}
Their inverse includes $e^{4h_i}=e^{-2\varphi_i}+\chi_i^2>0$. The potential and target metric above become
\begin{align}
ds_G^2&=\sum_i[dt_i^2+\cosh^2(2t_i)dh_i^2],\label{stu:adaptedmetric}\\
v&=\cosh[2(h_1-h_2-h_3)]\prod_i\cosh(2t_i)-\prod_i\sinh(2t_i).\label{stu:adaptedpotential}
\end{align}
Let $e_h=(1,-1,-1)/\sqrt3$, $y=e_h\cdot h$, and $K_h=\operatorname{diag}(\cosh^2(2t_i))$. Translations of $h$ perpendicular to $e_h$ are exact symmetries. For any $n\cdot e_h=0$, their radial Noether momenta are proportional to
\begin{equation}
J_n=(\rho^2-b_{\rm ne}^2)n\cdot K_hh',\qquad J_n'=0.\label{stu:currents}
\end{equation}
A regular non-extremal horizon forces $J_n=0$. At a controlled extremal endpoint with finite scalar limits, the same conclusion follows from the $\rho^{-2}$ velocity that a nonzero momentum would require in a nonsingular target metric. Thus this condition is selected by the horizon, rather than imposed as an extra source constraint.

At zero momenta the exact reconstruction equation is
\begin{equation}
h_i'=\frac{(e_h)_i\operatorname{sech}^2(2t_i)}{D(t)}y',\qquad D(t)=\frac13\sum_i\operatorname{sech}^2(2t_i).\label{stu:reconstruction}
\end{equation}
Substitution, equivalently zero-momentum Routh reduction, yields
\begin{align}
ds_{G,\mathrm{red}}^2&=\sum_i dt_i^2+\frac{3dy^2}{\sum_i\operatorname{sech}^2(2t_i)},\label{stu:reducedmetric}\\
v_{\mathrm{red}}&=\cosh(2\sqrt3y)\prod_i\cosh(2t_i)-\prod_i\sinh(2t_i).\label{stu:reducedpotential}
\end{align}
The removed coordinates generally vary: $n\cdot K_hh'=0$ does not imply $n\cdot h'=0$. Their integration constants retain the two original asymptotic flat-direction data. The physical source map must also be retained. In the coordinates $(d,\alpha)$ and with the fast vector $e$ introduced above, expansion gives
\begin{equation}
y=e\cdot d+\frac{\alpha_1^2-\alpha_2^2-\alpha_3^2}{\sqrt3}+O(3).\label{stu:sourcemap}
\end{equation}
Thus a straight source in $(d,\alpha)$ generally induces a quadratic fast source in the quotient. It must be held at that transformed value when comparing the two descriptions.

The quotient has an isolated positive-Hessian attractor. Its expansion is $v_{\mathrm{red}}=1+2\sum_i t_i^2+6y^2-8t_1t_2t_3+O(4)$, and the first derivatives of its metric vanish at the origin. The exact $y\mapsto-y$ symmetry forbids the only possible positive-weight resonance, namely a weight-two output driven by two weight-one inputs. The stable-branch and normal-form construction in Appendix~\ref{App:Branches} therefore has no nonlinear resonant term. The quotient fields extend analytically through the extremal horizon. Equation~\eqref{stu:reconstruction}, with $D(0)=1$, reconstructs analytic $h_i$, and \eqref{stu:coordinates} restores the original scalars. The same full-field reconstruction also controls the metric and Maxwell fields.

A stronger thermal statement follows on the consistent sector $y=0$ with $h_i=0$. Its original scalar sources lie on the curved surface $\varphi_i=\log\cosh(2t_i)$, $\chi_i=\tanh(2t_i)$. The reduced potential becomes
\begin{equation}
\begin{gathered}
v=\frac14\sum_{A=1}^4e^{2a_A\cdot t},\\
\begin{aligned}
a_1&=(1,1,-1),&a_2&=(1,-1,1),\\
a_3&=(-1,1,1),&a_4&=(-1,-1,-1).
\end{aligned}
\end{gathered}\label{stu:tetrahedron}
\end{equation}
These vectors satisfy $\sum_Aa_A=0$ and $a_A\cdot a_B=4\delta_{AB}-1$. The known four-charge solution~\cite{Cvetic:1995uj,Cvetic:1995kv}, reviewed in Ref.~\cite{ChowCompere2014}, can be written at fixed $Q$ and arbitrary finite $t_\infty$ as
\begin{align}
m_A&=\sqrt{b_{\rm ne}^2+Q^2e^{2a_A\cdot t_\infty}},\nonumber\\
p_A(\rho)&=\rho+m_A,\qquad R^2=\sqrt{\prod_Ap_A},\nonumber\\
t&=t_\infty-\frac14\sum_Aa_A\log p_A.\label{stu:solution}
\end{align}
A common reference length cancels from the scalar logarithms. 

The thermodynamic quantities are
\begin{align}
M&=\frac14\sum_Am_A,&R_h^2&=\sqrt{\prod_A(b_{\rm ne}+m_A)},\nonumber\\
S&=\pi R_h^2,&T&=\frac{b_{\rm ne}}{2\pi R_h^2},\qquad F=M-\frac{b_{\rm ne}}2.\label{stu:thermodynamics}
\end{align}
For every finite fixed $t_\infty$, all $m_A(0)>0$, $R_h^2(0)=Q^2$, and $\partial T/\partial b_{\rm ne}|_0=(2\pi Q^2)^{-1}$. The analytic inverse-function theorem proves an ordinary Taylor series for $F(T;t_\infty)$ near zero temperature. This explicit fixed-source family has no classical thermal logarithms at any source order. Its analyticity follows from a non-extremal saddle and an invertible temperature map. The specified curved source surface is essential; exact extremal solutions with more general asymptotic scalars address a different boundary problem \cite{CveticPopeSaha2021}.

For a direct check of Eq.~\eqref{stu:solution}, put $s_n=\sum_Ap_A^{-n}$. Then $R''/R=-s_2/4+s_1^2/16=-|t'|^2$, and the radial constraint is
\begin{equation}
1-\frac{\rho s_1}{2}+\frac{\rho^2-b_{\rm ne}^2}{4}s_2=\frac14\sum_A\frac{m_A^2-b_{\rm ne}^2}{p_A^2}.\label{stu:constraint}
\end{equation}
The scalar equations follow from $(\Delta/p_A)'=1-(m_A^2-b_{\rm ne}^2)/p_A^2$ and $\sum_Aa_A=0$, where $\Delta=\rho^2-b_{\rm ne}^2$. This directly verifies the finite-temperature radial equations.

\subsection{Limits of the selection rule}
\paragraph{An unaligned cubic counterexample.}
Consider the local three-modulus cubic
\begin{equation}
\begin{gathered}
\begin{aligned}
D={}&y_1^3-y_1(y_2^2+y_3^2)-\frac{\sqrt{66}}9y_2^3
+\frac{\sqrt{66}-9\sqrt2}{12}y_2y_3^2,
\end{aligned}\\
t_*=(i,0,0),\qquad (q_0,q_1,q_2,q_3;p^0,p^1,p^2,p^3) =\left(0,\frac{\sqrt{66}}8,1,0;\frac{3\sqrt2}8,0,0,0\right).
\end{gathered}
\label{special:ternary}
\end{equation}
It has $D_*=1$, $Q^2=9/16$, $|Z|^2=3(7+\sqrt{33})/128$, and
\begin{equation*}
G_*=\operatorname{diag}(3/4,1/2,1/2)\oplus
\operatorname{diag}(3/4,1/2,1/2).
\end{equation*}
Its $\mathsf H$ eigenvalues are all positive:
\begin{equation}
\begin{gathered}
\left\{2,2,6,\frac{7-\sqrt{33}}3,h_-,h_+\right\},\qquad h_\pm=\frac{15+\sqrt{33}\pm\sqrt{282+22\sqrt{33}}}{6}.
\end{gathered}
\end{equation}
Alignment fails because $f_2\ne0=L_2$. In the $G$-unit bases
\begin{equation}
 u_1=\frac2{\sqrt3}\partial_{y_1},\qquad
 u_2=\sqrt2\partial_{x_3},\qquad w=\sqrt2\partial_{y_2},
\end{equation}
the complete resonant matrix is
\begin{equation}
\langle w,\Gamma(u_i,u_j)\rangle_G
=\begin{pmatrix}0&0\\0&(9-\sqrt{33})/15\end{pmatrix}.
\label{special:ternarygamma}
\end{equation}
The nonzero entry is approximately $0.217029156897465$; its algebraic
expression is exact. Deleting the third modulus removes $u_2$ and leaves a
protected binary example. Cubic special geometry alone therefore does not
protect unaligned branches. This local example is not asserted to arise
from a compactification; no complete exterior or total thermal expansion
has been constructed for it. Its additional exponents below one prevent
application of the complete-spectrum $(1,2)$ theorem to its total free energy.

\paragraph{A non-cubic fifth-order deformation.}
For general special geometry, the holomorphic component of $\nabla^3V$
contains $-i(\mathcal D_a\mathcal D_bC_{cde})\bar f^d\bar f^e$ and cannot
be determined from $C,\mathcal DC,Z,f$ alone. For
\begin{equation}
\begin{gathered}
F=(X^0)^2\left[-t^3+\frac{i\beta}{120}(t-i)^5\right],\qquad q_0=p^1=1,\qquad \beta\in\mathbb R,
\end{gathered}
\label{special:quintic}
\end{equation}
the point $t=i$ remains critical with $Q^2=2$, $g=3/4$,
$\mathsf H=\operatorname{diag}(6,2)$, $C=3/4$, and $\mathcal DC=0$. Nevertheless,
\begin{equation}
\begin{gathered}
\mathcal D^2C=-\frac{i\beta}{8},\qquad
(\nabla^3V)_{xyy}=-\frac\beta2,\qquad \left\langle\frac2{\sqrt3}\partial_x,
\Gamma\!\left(\frac2{\sqrt3}\partial_y,\frac2{\sqrt3}\partial_y\right)
\right\rangle_G =-\frac{\beta}{30\sqrt3}.
\end{gathered}
\label{special:quinticgamma}
\end{equation}
Thus the lower jets and integer spectrum can stay fixed while the resonance
varies. This is a local analytic special-K\"ahler example.

\subsection{Unaligned critical points}\label{special:unaligned}
\paragraph{General bound.} Normalize the cubic at the point as in
Eq.~\eqref{special:localcubic}, $\widehat D=r^3-r|z|^2+\mathcal C(z)$ at
$y=(1,0,\ldots,0)$, and use a real orthonormal complex frame $(e_0,e_A)$
with $e_0$ radial. Then
\begin{equation}
\begin{gathered}
C_{000}=\frac{2}{\sqrt3},\qquad C_{00A}=0,\qquad C_{0AB}=-\frac{\delta_{AB}}{\sqrt3},\qquad C_{ABC}=T_{ABC},\qquad \mathcal D_iC_{jkl}=iE_{ijkl},
\end{gathered}
\label{unal:frame}
\end{equation}
where every component of $E$ with a radial index vanishes and
\begin{equation}
\begin{aligned}
E_{ABCD}={}&\tfrac23(\delta_{AB}\delta_{CD}+\delta_{AC}\delta_{BD}
+\delta_{AD}\delta_{BC})\\
&-\textstyle\sum_E(T_{ABE}T_{CDE}+T_{ACE}T_{BDE}+T_{ADE}T_{BCE}).
\end{aligned}
\end{equation}
Write $f=(u,v)$ with $v\in\mathbb C^{n-1}$, so that
$Q^2=V=|Z|^2+|u|^2+\|v\|^2$ and alignment is $v=0$. From
Eq.~\eqref{special:complexhessian},
\begin{equation}
N_{00}=2|Z|^2+\tfrac{14}{3}|u|^2+\tfrac23\|v\|^2,\qquad
M_{00}=-\tfrac{8i}{\sqrt3}\bar Z\bar u .
\end{equation}
For the unit complex radial vector $e^{i\alpha}e_0$ the normalized
Rayleigh quotient of $\mathsf H$ and its minimum $\varrho$ over $\alpha$ are
\begin{equation}
\frac{\operatorname{Re}(e^{2i\alpha}M_{00})+N_{00}}{V},\qquad
\varrho=\frac{N_{00}-|M_{00}|}{V}.
\end{equation} The radial component of
Eq.~\eqref{special:critical},
$2\bar Zu=\frac{i}{\sqrt3}(2\bar u^2-\bar v\cdot\bar v)$, gives
$|2u^2-v\cdot v|=2\sqrt3\,|Zu|$, hence
\begin{equation}
\varrho-2=\frac{4}{3V}\Big(2|u|^2-\|v\|^2-\big|2u^2-v\cdot v\big|\Big)\le0,
\label{unal:radialbound}
\end{equation}
because $|2u^2-v\cdot v|\ge2|u|^2-|v\cdot v|\ge2|u|^2-\|v\|^2$. If $\varrho<2$,
the min--max principle gives $h_{\min}<2$. If $\varrho=2$ and $v\neq0$, both
inequalities are saturated. This forces $f=e^{i\theta}(U,v_0)$ with
$U>0$, $v_0\in\mathbb R^{n-1}$ and $0<\|v_0\|^2<2U^2$, the upper bound
because $Z\neq0$. After a rotation $v_0=(v_1,0,\ldots,0)$, the transverse
components of Eq.~\eqref{special:critical} give
\begin{equation}
T_{\alpha11}=0\quad(\alpha\ge2),\qquad
T_{111}=\frac{4U^2-v_1^2}{\sqrt3\,Uv_1}.
\end{equation}
No other transverse component then enters $M$ or $N$ restricted to
$\operatorname{span}(e_0,e_1)$, and this restriction is exactly the
two-modulus Hessian below with $a=(2-s)/(4+s)>0$, $s=v_1^2/U^2\in(0,2)$.
Equation~\eqref{unal:p2} gives an eigenvalue below $2$ on this subspace,
hence $h_{\min}<2$. This proves the theorem of Sec.~\ref{Sec:Selection}.

\paragraph{Two moduli in closed form.} For $n=2$ take
$D=r^3-rz^2+bz^3$ at $(x;y)=(0,0;1,0)$, with $27b^2\neq4$. Then
$g=\operatorname{diag}(\tfrac34,\tfrac12)$, the only independent nonzero
component of $\mathcal DC$ is $\mathcal D_2C_{222}=\frac i8(4-27b^2)$, and
the unaligned non-BPS $Z\neq0$ critical points are, up to the charge
scale $\rho>0$,
\begin{equation}
\begin{gathered}
f=\rho e^{i\theta}(\xi,1),\qquad Z=-\frac{i\rho k}{2}e^{3i\theta},\qquad 8\xi^2-9b\xi-3=0,\qquad k=3b-\tfrac43\xi\neq0 .
\end{gathered}
\label{unal:family}
\end{equation}
These are the solutions, at a fixed regular point, of the equation for the
non-BPS $Z\ne0$ charge directions,
\begin{equation}
\begin{gathered}
C_{abc}z^bz^c=\lambda g_{a\bar b}\bar z^b,\qquad \lambda\ne0,\qquad \bar Z=\frac i2\lambda,\qquad z^a=\bar f^a,
\end{gathered}
\label{cy:branchequation}
\end{equation}
which for this normal form has one aligned and two unaligned projective
directions; the latter are the two roots $\xi=f_1/f_2$. This counts charge
directions at a fixed point, not stable attractors for a fixed charge vector.
A complex ratio $f_1/f_2$ is possible only for the degenerate cubic
$27b^2=4$, and every $(Z,f)$ corresponds to a unique real charge vector.
After the reflection $z\to-z$, which makes $\xi>0$, put
\begin{equation}
a=\frac{4\xi^2-3}{8\xi^2+3}\in\Big(-1,\frac12\Big),\qquad
\chi=\cos2\theta .
\end{equation}
Then $|Z|^2/V=a^2$ and $b=2(1+4a)/[3\sqrt3\sqrt{(1+a)(1-2a)}]$; the two
roots at fixed shape are related by $a\mapsto-a-\frac12$, $a=0$ is the
$Z=0$ limit and $a=-\frac12$ the degenerate cubic. The spectrum of
$\mathsf H$ consists of $h=2$, with eigenvector $\delta t^a=ie^{2i\theta}y^a$,
and the eigenvalues of
\begin{equation}
S=\begin{pmatrix}A&\kappa&0\\ \kappa&2+2e+\beta\chi&-\beta\sin2\theta\\
0&-\beta\sin2\theta&2-\beta\chi\end{pmatrix},
\label{unal:S}
\end{equation}
\begin{equation}
\begin{gathered}
A=2+\tfrac{16}{3}a(1+a),\qquad e=\tfrac43a(1+7a),\\
\beta=-12a(1+2a),\qquad
\kappa=-\tfrac{8\sqrt2}{3}\,a\sqrt{(1+a)(1-2a)} .
\end{gathered}
\end{equation}
Since $A>0$ and the trace of the Schur complement of $A$ is positive, the
point is a non-degenerate local minimum exactly when $\det S=\Delta(a,\chi)>0$
(at $\Delta=0$ one eigenvalue vanishes and higher orders decide), where
\begin{equation}
\begin{aligned}
\Delta={}&-4(1+2a)\big[(1+\chi)(8a^2+4a-1)(12a^2+6a+1)\\
&+(1-\chi)(1+4a)(8a^2+4a+1)(12a^2+6a-1)\big].
\end{aligned}
\label{unal:Delta}
\end{equation}
Minima exist for $a$ in $(-1,-\frac{1+\sqrt3}{4})$,
$(-\frac{3+\sqrt{21}}{12},-\frac12)$, $(-\frac12,0)$ and
$(0,\frac{\sqrt3-1}{4})$. For $a<0$ the diagonal entry $A<2$ is a Rayleigh
quotient, while for $a>0$
\begin{equation}
\det(2I-S)=768\,a^3(1+a)(1+2a)\,[1+2a(1-\chi)]>0,
\label{unal:p2}
\end{equation}
so the three eigenvalues of $S$ cannot all be at least $2$. Hence
$h_{\min}<2$ throughout. At $\chi=1$ the eigenvalues of $S$ are
$2+12a+24a^2$ and $2-2a\pm2|a|\sqrt{17+16a}$; for
$0<a<(\sqrt3-1)/4$ the smallest decreases continuously from $2$ to $0$, so
$c_{\min}$ takes every value in $(0,1)$. On the minimum region the
eigenvalue $h=2$ is simple and $1<c_{\max}<4$. For the degenerate cubic all
non-BPS $Z\neq0$ directions, including the complex ones, give the
spectrum $\{0,2,2,6\}$.

\paragraph{Soft minima in every sector.} The family also gives soft minima within each compactification. A cubic of
non-degenerate type fixes the shape $b$ of each ray of its K\"ahler cone, up to sign, and the two roots at that
shape, $u=1+4a$ and $-u$ with $b^2=32u^2/[27(9-u^2)]$, are exchanged by $a\mapsto-a-\frac12$.
Equation~\eqref{unal:Delta} is linear in $\chi$, $\Delta=\frac12[(1+\chi)P+(1-\chi)M]$, with
\begin{equation}
P=\tfrac12(1+u)(3-u^2)(3u^2+1),\qquad M=\tfrac12u(1+u)(1+u^2)(7-3u^2),
\label{unal:PM}
\end{equation}
and $P(-u)=\frac{1-u}{1+u}P(u)$, $M(-u)=-\frac{1-u}{1+u}M(u)$. If $PM\neq0$, exactly one of the two roots
therefore has $P$ and $M$ of opposite signs, and on it $\Delta$ has a simple zero at
$\chi_*=-(P+M)/(P-M)\in(-1,1)$. By the positive trace of the Schur complement a single eigenvalue of $S$
vanishes there, so $c_{\min}\to0^+$ as $\chi\to\chi_*$ from the side $\Delta>0$. The product $PM$ vanishes only
for $b^2\in\{0,\frac{56}{135},\frac{16}{27}\}$, and each value of $b^2$ occurs on at most six rays, because a
real projective map that preserves the roots of $D$ is fixed by its permutation of them, so every open sector
of the K\"ahler cone contains rays of generic shape. The dilation~\eqref{cy:dilation} reaches any volume, the
implicit-function theorem extends a minimum with positive-definite Hessian to an open set of charges, and charge
scaling makes the set a cone. For every cubic of non-degenerate type, every $\epsilon>0$ and every open sector of
the K\"ahler cone there is therefore an open cone of charges with isolated unaligned non-BPS minima with $Z\neq0$,
positive-definite Hessian and $0<c_{\min}<\epsilon$ in that sector, above any given volume; since rational
directions are dense, the cone contains infinitely many integral rays.

\paragraph{Checks and a sharpness example.} The 73 sampled unaligned minima
of Appendix~\ref{App:CY}, converted to this frame, satisfy
Eq.~\eqref{unal:family} and have the eigenvalues of Eq.~\eqref{unal:S}
to numerical precision, and 112 constructed family points, converted
back to real charges, are critical with the predicted spectra. Random
cubics with $n=3$ and $4$ gave 832 strict unaligned minima, all with
$h_{\min}<2$. The bound
cannot be strengthened to $n-1$ eigenvalues below $2$: for
$D=r^3-r(z^2+w^2)+\frac59z^3-5zw^2$ with
$(q_0,q_r,q_z,q_w;p^0,p^r,p^z,p^w)=(0,-15,24,0;11,0,0,0)$ at
$(x;y)=(0;1,0,0)$ the spectrum is
$\{\frac{2}{121}(110\mp\sqrt{2233}),2,\frac{398}{121},6,\frac{3902}{121}\}$,
positive, with a single eigenvalue below $2$. For $n\ge3$ the eigenvalue
$h=2$ need not be present: three of the 832 random minima lack it.

\subsection{The D0--D4 edge and five-dimensional strings}\label{special:fived}
For D0--D4 charges, $p^0=q_a=0$, Eqs.~\eqref{special:cubic}--\eqref{special:critical}
give, for every cubic,
\begin{equation}
\begin{aligned}
V={}&\frac{\bigl[q_0+p^aD_a(x)\bigr]^2+P^2-D\,\Pi}{2D}
 +\frac{g^{ab}\hat q_a\hat q_b}{8D},
\end{aligned}
\label{fived:V}
\end{equation}
where $P=p^aD_a(y)$, $\Pi=D_{ab}p^ap^b$, $D_{ab}=d_{abc}y^c$,
$\hat q_a=d_{abc}p^bx^c$, and $g^{ab}=2y^ay^b-4D(D^{-1})^{ab}$. At $x=0$, with
$y=\mathcal V^{1/3}Y$ and $D(Y)=1$, this is the 4d/5d relation of Ref.~\cite{CFM2007},
\begin{equation}
V=\frac{q_0^2}{2\mathcal V}+\frac{\mathcal V^{1/3}}{2}\,V_5(Y),\qquad
V_5=P^2-\Pi=2G_{ab}p^ap^b ,
\label{fived:V5}
\end{equation}
in which $G_{ab}=-\frac12\partial_a\partial_b\log D$ and $V_5$ is the effective potential
of the five-dimensional string obtained from an M5-brane on the divisor class
$p$~\cite{LSVY2021,MMT2022}; $q_0$ is its momentum along the M-theory circle. Axion-free
critical points are critical points $Y_*$ of $V_5$ on $D=1$, with
\begin{equation}
\begin{gathered}
\mathcal V^{2/3}=\frac{|q_0|}{\sigma},\qquad
Q^2=2|q_0|^{1/2}\sigma^{3/2},\qquad
\sigma^2\equiv\tfrac13V_5(Y_*),
\end{gathered}
\label{fived:crit}
\end{equation}
so that $S=2\pi\sqrt{|q_0|\,c_{\rm eff}/6}$ with $c_{\rm eff}=6\sigma^3$~\cite{CFM2007}.
For $p$ inside the K\"ahler cone, $Y_*\propto p$, $c_{\rm eff}=6D(p)$~\cite{Maldacena:1997de},
and the two signs of $q_0$ give the BPS and the aligned non-BPS black
hole~\cite{TripathyTrivedi2005,DST2006}; these are the D0--D4 charges of the aligned
branch. For $p$ outside the cone the point is unaligned: it is a non-BPS string of
Refs.~\cite{LSVY2021,MMT2022} with momentum. Since $Z$ is real and $\mathcal D_aZ$
imaginary at $x=0$, these points are exactly the edge $\chi=-1$ of the family
\eqref{unal:family}. The edge $\chi=1$ consists in the same way of the D2--D6 charges, the
reductions of five-dimensional black holes. The strings, their tensions and their
recombination are the subject of Refs.~\cite{LSVY2021,MMT2022}; the four-dimensional
Hessian, including the axions, and the resulting exponents are derived here.

\paragraph{Spectrum.}
Let $s=\operatorname{sgn}q_0$ and $\tau=p^aD_a(Y_*)/\sigma$; $|\tau|/3$ is the ratio of
the central charge of the string to its tension, equal to one for a BPS string. Because
$V$ is even in $x$, $\mathsf H$ is block diagonal. The saxion block does not depend on
$s$: it consists of $h=2$ for the volume and of the spectrum of
$\frac{3}{4V_5}G^{-1}\nabla^2V_5$ on $D=1$. The axion block is $N(N+s)$, where
$N=g^{-1}(Y_*)M/(4\sigma)$ and $M_{ab}=d_{abc}p^c$. At an unaligned point, $Y_*\not\parallel p$, the plane spanned by $Y_*$ and $p$ is
invariant under $N$, with eigenvalues that solve $\mu^2+\tau\mu-(1+\tau^2)=0$. For two
moduli the exponents are therefore
\begin{equation}
\begin{aligned}
c\in\Bigl\{&1,\ \tfrac12\bigl(\sqrt{15-6\tau^2}-1\bigr),\
 \tfrac12\bigl(\sqrt{5\tau^2+4}-s\tau\bigr),
 \tfrac12\bigl(\sqrt{5\tau^2+4}+s\tau\bigr)-1\Bigr\},
\end{aligned}
\label{fived:exponents}
\end{equation}
and the point is a non-degenerate local minimum exactly when $0<s\tau<\sqrt{7/3}$ or
$-\sqrt{7/3}<s\tau<-1$: for one orientation of the momentum, $s=\operatorname{sgn}\tau$,
if $0<|\tau|<1$ (for neither at $\tau=0$), and for both if $1<|\tau|<\sqrt{7/3}$. On the family \eqref{unal:family},
$a=(s\tau-1)/4$, reversing the momentum is the map $a\mapsto-a-\frac12$, and
$\Delta(a,-1)=\frac12T(T+1)(T^2+1)(7-3T^2)$ with $T=s\tau$. Every strict minimum has an
exponent $0<c<1$, in agreement with the theorem of Sec.~\ref{Sec:Selection}.

For the proof, Eq.~\eqref{fived:V} follows from
$\mathcal W=q_0+p^aD_a(x)-P+i\hat q_ay^a$ and Eq.~\eqref{special:noscale}. At $x=0$ the
metric is
$\frac1{12}(d\mathcal V/\mathcal V)^2+\frac12G_{ab}dY^adY^b$, so the volume decouples
with $h=2$, and $V_*=2\sigma^2\mathcal V^{1/3}$. Equation~\eqref{fived:V} gives
$\partial_x\partial_xV=[q_0M+\frac14Mg^{-1}M]/D$; after the dilation \eqref{cy:dilation}
to $\mathcal V=1$, $q_0=s\sigma$, this is $2V_*\,g\,N(N+s)$. Since $g^{-1}\partial D=4Y$,
one has $4\sigma NY=4PY-4p$, and the condition for a critical point,
$2PMY-Mp=(P^2+\sigma^2)\partial D$, gives $4\sigma Np=4(P^2-\sigma^2)Y-8Pp$. Hence
$\operatorname{tr}N=-\tau$ and $\det N=-(1+\tau^2)$ on the plane. For two moduli the last
eigenvalue is $2-\beta\chi$ of Eq.~\eqref{unal:S} at $\chi=-1$ and
$a^2=(s-\tau)^2/16$, that is $\frac12(7-3\tau^2)$.

\paragraph{The two threefolds and decay.}
Table~\ref{tab:fived} lists rays $p=(\cos\psi,\sin\psi)$ with $p^1>0>p^2$. For the
bicubic, whose cone of effective divisors is the K\"ahler cone, every such ray has one
critical point~\cite{LSVY2021}, and $|\tau|<1$: for $\tau\neq0$ exactly one orientation of the momentum
is a strict minimum, and $c_{\min}$ ranges over $(0,1)$. At $\tau=0$, where $p^1=-p^2$,
an axionic eigenvalue vanishes. Relaxing the two massive saxions gives $V_{\rm eff}/V_*=1-f^4/21+O(f^6)$ for a
unit-normalized axionic displacement $f$, for every two-modulus cubic, so this point is unstable at quartic order.
For small $\tau$ a pair of index-one saddles lies next to the minimum, at $f^2=\frac{21}{2}h+O(h^2)$ with
$V_{\rm sad}/V_{\rm min}-1=\frac{21}{4}h^2+O(h^3)$, where $h\simeq c_{\min}$ is the lowest eigenvalue. For integers $N\ge2$
and $K\ge1$ the integral charges $(q_0;p^1,p^2)=(2KN;2N,2-2N)$ are those of
$\mathcal O(P_N)-\mathcal O(-P_N)$, $P_N=NJ_1-(N-1)J_2$, plus $2KN+3N(N-1)-3$ D0-branes, since
$P_N^3=-9N(N-1)$ and $c_2\cdot P_N=36$. They have $\tau=\frac{9\sqrt6}{28N}+O(N^{-2})$, hence
$c_{\min}=\frac{9\sqrt6}{56N}+O(N^{-2})$, and $R_{\rm dec}\to4\sqrt6/(9+\sqrt6)=0.8558$. For $N=4$, $K=180$ the
charges $(1440,0,0;0,8,-6)$ have the exact attractor $Y_*^2/Y_*^1=(\sqrt{17}-1)/4$, with $\tau=0.223456$,
$c=(0.142464,0.919008,1,1.417055)$ and $R_{\rm dec}=0.8588$; Appendix~\ref{App:NumericsCY} gives their thermal
solutions. For
degree 18 the effective cone is generated by $L$ and the section
$E=H-3L$~\cite{CandelasFontKatzMorrison1994,LSVY2021}. Effective classes outside the
K\"ahler cone have no critical point inside it. Non-effective classes $p=p_EE+p_LL$ have
two for $p_L/p_E<-5.176$: a saddle, and a point with $1<|\tau|<\sqrt{7/3}$ that is a
strict minimum for both orientations.

By Eq.~\eqref{fived:crit}, $\sqrt{8\mathcal V}\,Q=3\sigma\mathcal V^{2/3}+|q_0|$: the
mass at the attractor is a tension term plus a momentum term. Compare it with D0-branes,
BPS D4-branes on the generators $e_i$ of the effective cone and anti-D4-branes, with
$p=\sum_i\alpha_ie_i$. Every two- and three-body split of this kind gives
\begin{equation}
\begin{gathered}
R_{\rm dec}\equiv\frac{Q}{\sum_i|Z_i|}=\frac{4r_5}{3+r_5},\qquad
r_5=\frac{3\sigma}{\sum_i|\alpha_i|\,e_i^aD_a(Y_*)} ,
\end{gathered}
\end{equation}
with $r_5$ the tension ratio of Ref.~\cite{LSVY2021}. For the bicubic $0.8557<R_{\rm dec}<1$ on
all rays, so the attractor is bound with respect to D0- and (anti-)D4-brane
constituents. For degree 18, $1<R_{\rm dec}<1.264$ on the branch of minima, and this decay is
energetically allowed. Only these channels are tested; constituents with D6 and D2
charges are not covered.
\begin{table*}[t]
\centering
\small
\setlength{\tabcolsep}{4pt}
\renewcommand{\arraystretch}{1.15}
\begin{tabular}{lrrrrclcc}
\hline
Model & $\psi$ & $Y_*^2/Y_*^1$ & $c_{\rm eff}/6$ & $\tau$ & Minimum & Exponents $c$ & $r_5$ & $R_{\rm dec}$\\
\hline
Bicubic & $-18.43^\circ$ & $0.373$ & $0.366$ & $0.719$ & $s=+$ & $(0.643,\,0.924,\,1,\,1.225)$ & $0.862$ & $0.893$\\
 & $-40^\circ$ & $0.860$ & $0.569$ & $0.137$ & $s=+$ & $(0.080,\,0.943,\,1,\,1.429)$ & $0.818$ & $0.857$\\
 & $-60^\circ$ & $1.611$ & $0.507$ & $-0.412$ & $s=-$ & $(0.307,\,0.895,\,1,\,1.370)$ & $0.829$ & $0.866$\\
\hline
Degree 18 & $-85^\circ$ & $7.549$ & $0.031$ & $-1.112$ & $s=+$ & $(0.039,\,0.877,\,1,\,2.151)$ & $1.231$ & $1.164$\\
 & & & & & $s=-$ & $(0.877,\,1,\,1.039,\,1.151)$ & & \\
\hline
\end{tabular}
\caption{Axion-free attractors with D0--D4 charges and D4 class
$p=(\cos\psi,\sin\psi)$ outside the K\"ahler cone; $\psi=-18.43^\circ$ is
$p\propto(3,-1)$, and $c_{\rm eff}$ scales as $|p|^3$. The exponents follow from
Eq.~\eqref{fived:exponents} for the orientations $s$ of the momentum that give a strict
minimum; for degree 18 the branch of minima is shown.}
\label{tab:fived}
\end{table*}

\subsection{Vanishing central charge}\label{special:z0}
At $Z=0$ Eq.~\eqref{special:critical} reduces to $C_{abc}\bar f^b\bar f^c=0$, so the direction of $\bar f^a\neq0$ is a singular point of the complex projective cubic, and conversely every such direction occurs for suitable charges. For two moduli only the degenerate cubic $st^2$ has singular points; its four real modes then have $c=1$, as do all six modes of STU at $Z=0$~\cite{Bellucci:2007zi}. In the frame~\eqref{unal:frame} the radial component of Eq.~\eqref{special:critical} now gives $v\cdot v=2u^2$, so Eq.~\eqref{unal:radialbound} gives $\varrho\le2$, with equality only if $f$ is real up to a phase. In that case rotate to $f=e^{i\theta}U(1,\sqrt2,0,\ldots)$. Criticality fixes $T_{111}=\sqrt{2/3}$ and $T_{\alpha11}=0$ for $\alpha\ge2$. The first two complex directions then decouple, Eq.~\eqref{special:complexhessian} gives them four eigenvalues $h=2$, and every eigenvalue $\omega$ of $T_{1\alpha\beta}$ gives the pair $h=\sigma,\ \sigma(\sigma-1)$ with $\sigma=\frac23-\sqrt{8/3}\,\omega$. Now let the attractor be stable with positive exponents exactly $1$ and $2$, so that its spectrum lies in $\{0,2,6\}$ and contains $6$. If $\varrho<2$, the min--max principle gives $h_{\min}<2$, hence $h_{\min}=0$. If $\varrho=2$, the only pairs inside $\{0,2,6\}$ are $(0,0)$ and $(2,2)$, so the eigenvalue $6$ cannot occur. Such an attractor therefore has zero modes.

\paragraph{An integer example.} Such attractors exist, and with integer data. Take eight moduli and
\begin{equation}
\begin{aligned}
D={}&y^1\big[(y^2)^2-(y^3)^2-(y^4)^2\big]-y^2\big[(y^5)^2+6(y^6)^2+(y^7)^2+2(y^8)^2\big]+12\,y^3y^5y^6\\
&+4\,y^3y^7y^8+y^5\big[\eta_1(y^7)^2+\eta_2y^7y^8+\eta_3(y^8)^2\big]+y^6\big[\eta_4(y^7)^2+\eta_5y^7y^8+\eta_6(y^8)^2\big],
\end{aligned}
\label{z0:cubic}
\end{equation}
with the charges $q_3=2$, $p^4=-1$ and all others zero, so that $\mathcal W=2(t^3+t^1t^4)$. The intersection numbers $d_{abc}$ are integers for integer $\eta_i$, and even if $\eta_2$ and $\eta_5$ are even. Since $\mathcal W$ depends only on $t^1$, $t^3$ and $t^4$, the shifts of the other axions are symmetries at fixed charge. The locus $\mathcal M$ given by $t^1=i$, $t^3=t^4=0$, $y^5=\dots=y^8=0$ and $y^2=\ell>0$, with $x^2$ and $x^5,\dots,x^8$ free, consists of critical points with $D=\ell^2$, $Z=0$ and $V=Q^2=2$. There $2g=\mathrm{diag}(\frac12,\ell^{-2},\ell^{-2},\ell^{-2},\ell^{-1},6\ell^{-1},\ell^{-1},2\ell^{-1})$, and at fixed charge
\begin{equation}
\begin{aligned}
V={}&2+|\delta t^1|^2+\frac{2}{\ell^2}\big(|\delta t^3|^2+|\delta t^4|^2\big)+\frac{6}{\ell}\big[(\delta y^5)^2+6(\delta y^6)^2\big]+\frac{2}{\ell}\big[(\delta y^7)^2+2(\delta y^8)^2\big]+O(\delta^3).
\end{aligned}
\label{z0:jet}
\end{equation}
Hence $\ker\mathsf H=T\mathcal M$, the normal Hessian is positive, and the exponents are $c=(0^6,1^8,2^2)$ everywhere on $\mathcal M$: the flat directions are $y^2$ and the axions of $\mathcal M$, the fast directions $y^5$ and $y^6$, and the other eight are slow. The flat--slow--slow and flat--fast--fast components of $\nabla^3V$ vanish identically, as Appendix~\ref{App:flatthermal} implies for a constant spectrum. Only the components of $A$ along $y^7$ and $y^8$ enter $\Gamma(A,A)$: after the axions are shifted to zero, the symmetry $(t^7,t^8)\to-(t^7,t^8)$ of the prepotential and the charges removes the terms with one leg along these directions, and on its fixed set $t^7=t^8=0$, which is totally geodesic, the symmetry $(t^5,t^6)\to-(t^5,t^6)$ removes the fast couplings to two other slow legs. In orthonormal directions along the coordinate axes, with $(a_1,a_2)$ the components of $A$ along $y^7$ and $y^8$, the two components of $\Gamma(A,A)$ along $y^5$ and $y^6$ are
\begin{equation}
\begin{aligned}
\sqrt{\ell}\,\Gamma_5(A,A)&=\frac{5\eta_1+\eta_3+\eta_5}{5\sqrt2}\,a_1^2+\frac{7\eta_2+4\eta_4+2\eta_6}{10}\,a_1a_2+\frac{4\eta_1+5\eta_3+2\eta_5}{10\sqrt2}\,a_2^2,\\
\sqrt{\ell}\,\Gamma_6(A,A)&=\frac{6\eta_2+5\eta_4+\eta_6}{10\sqrt3}\,a_1^2+\frac{24\eta_1+12\eta_3+7\eta_5}{10\sqrt6}\,a_1a_2+\frac{12\eta_2+4\eta_4+5\eta_6}{20\sqrt3}\,a_2^2 .
\end{aligned}
\label{z0:gamma}
\end{equation}
The six coefficients are independent combinations of the $\eta_i$, so $\Gamma\neq0$ unless all $\eta_i$ vanish. For $\eta_1=1$ and the other $\eta_i=0$, Eq.~\eqref{eq:flatthermal} gives
\begin{equation}
F\supset-\frac{2Q}{9\ell}\Big[\big(a_1^2+\tfrac25a_2^2\big)^2+\tfrac{48}{25}\,a_1^2a_2^2\Big]\vartheta^5\log\vartheta .
\end{equation}
Over $\mathbb Q(\sqrt2,\sqrt3)$ the pairs $(y^5,y^6)$ and $(y^7,y^8)$ split into blocks on which $y^3$ multiplies the squares with the coefficients $\pm\sqrt6$ and $\pm\sqrt2$. Keeping one block of each gives a cubic in six moduli $(y^1,\dots,y^4,b,m)$, $y^1[(y^2)^2-(y^3)^2-(y^4)^2]-y^2(b^2+m^2)+y^3(\sqrt6\,b^2+\sqrt2\,m^2)+\eta\,bm^2$, with the same charges, the exponents $(0^4,1^7,2)$ and $\Gamma(A,A)=\frac{(7+4\sqrt3)\eta}{5\sqrt{2\ell}}\langle e_m,A\rangle^2e_b$ for unit vectors $e_b$ and $e_m$ along $y^b$ and $y^m$; its coefficients are irrational.

\paragraph{Checks.} We integrated the full extremal flow of Eq.~\eqref{fnd:contact} outward from the horizon, starting on the leaf of an endpoint, for $\eta=(1,0,0,0,0,0)$ and $(1,2,-1,1,0,2)$ at $\ell=1$ and $2$, along the $y^7$ direction, 24 random slow directions and a control along $x^1$, at three amplitudes each. Over seven $e$-folds of $z$ the fast components grow as $z^2\log z$ with the coefficients of Eq.~\eqref{z0:gamma} to $2.4\times10^{-14}$ after division by the squared amplitude, the control shows no logarithm, and the flat coordinates move at order $z^2$ without a logarithm. Degenerate unaligned minima with $Z\neq0$ behave differently in the one example we constructed, with three moduli: the spectrum $\{0,1^3,2^2\}$ and $\Gamma\neq0$ occur at a single point of a critical curve, and nearby points of the curve have non-integer exponents, one of them below one, so generic asymptotic data end elsewhere.

\section{Calabi--Yau geometries and exterior solutions}\label{App:CY}
\subsection{Intersection data and binary-cubic classification}
The classical type-IIA vector-multiplet prepotential is
$F=-\kappa_{abc}X^aX^bX^c/(6X^0)$, where $\kappa_{abc}$ are symmetric
triple intersections. Table~\ref{tab:cygeometry} specifies the divisor bases
and physical K\"ahler cones. Mixed monomials include their permutation
multiplicity in
\begin{equation*}
D=\kappa_{abc}y^ay^by^c/6.
\end{equation*}
\begin{table*}[t]
\centering
\small
\setlength{\tabcolsep}{3pt}
\begin{tabular}{llll}
\hline
Threefold and divisor basis & $(\kappa_{111},\kappa_{112},\kappa_{122},\kappa_{222})$
 & $D(y)$ & Cone \\
\hline
Bicubic in $\mathbb P^2\times\mathbb P^2$, $(J_1,J_2)$
 & $(0,3,3,0)$ & $\frac32y_1y_2(y_1+y_2)$ & $y_1,y_2>0$ \\
$\mathbb P^4_{1,1,1,6,9}[18]$, $(H,L)$
 & $(9,3,1,0)$ & $\frac12y_1(3y_1^2+3y_1y_2+y_2^2)$ & $y_1,y_2>0$ \\
$\mathbb P^4_{1,1,2,2,2}[8]$, $(H,L)$
 & $(8,4,0,0)$ & $\frac43y_1^3+2y_1^2y_2$ & $y_1,y_2>0$ \\
$\mathbb P^4_{1,1,2,2,6}[12]$, $(H,L)$
 & $(4,2,0,0)$ & $\frac23y_1^3+y_1^2y_2$ & $y_1,y_2>0$ \\
\hline
\end{tabular}
\caption{Classical geometric data. $J_1,J_2$ are the restricted ambient
hyperplanes; the weighted hypersurfaces are resolved and use the indicated
$(H,L)$ bases. The bicubic is a complete intersection
Calabi--Yau~\cite{Candelas:1987kf}; its intersections, cone,
$\chi=-162$ and $c_2\cdot J_a=36$ are in
Refs.~\cite{Hosono:1994ax,LarforsLukasRuehleSchneider2022,LeeLukas2025}.
The degree-18 data and $\chi=-540$ are from
Refs.~\cite{CandelasFontKatzMorrison1994,Hosono:1993qy}; the two controls are
from Ref.~\cite{CandelasDeLaOssaFontKatzMorrison1994}.}
\label{tab:cygeometry}
\end{table*}

For a real non-degenerate binary cubic, the number of real projective roots
is invariant under $GL(2,\mathbb R)$. Three distinct real roots can be sent
to $0,-1,\infty$, giving $uv(u+v)$; one real root and a conjugate pair give
$u(u^2+v^2)$. These are distinguished by positive and negative discriminant.
A double root gives $st^2$. A triple root makes the two-coordinate metric
degenerate and does not define a regular two-modulus theory. A real change
of frame acts on charges by a symplectic transformation, but generally is
not an integral charge-lattice duality.
The bicubic and degree-18 discriminants are $81/16$ and $-3/16$.
The bicubic K\"ahler cone is an entire $D>0,g>0$ component. For degree 18,
\begin{equation}
\begin{gathered}
U=y_1,\qquad W=\frac{2y_2+3y_1}{\sqrt3},\qquad D=\frac38U(U^2+W^2),\qquad U>0,\qquad W>\sqrt3U.
\end{gathered}
\label{cy:degree18frame}
\end{equation}
The other positive-metric component, $y_2<-3y_1$, is outside the physical
cone. For the controls, $t=y_1$, $s=y_2+2y_1/3$ gives $D=2st^2$ and
$D=st^2$, respectively. A nonzero discriminant also excludes a regular
classical non-BPS $Z=0$ critical point, whose charge direction would be a
singular point of the projective cubic (Appendix~\ref{special:z0}).

\subsection{Search statistics and representative attractors}
For each geometry, the search used the same 300 seeded integer charge
vectors in $[-8,8]^6$, with eight initial guesses per vector, and the
critical points were refined to high precision inside the physical cone.
Table~\ref{tab:cycounts} counts charge vectors, so its columns may overlap.
\begin{table*}[t]
\centering
\small
\setlength{\tabcolsep}{3pt}
\begin{tabular}{lrrrrrr}
\hline
Model & \shortstack{Spectrum $\{0,1,1,2\}$\\stable or flat} & \shortstack{Spectrum $\{0,1,1,2\}$\\quartic saddle}
 & \shortstack{Strict\\unaligned} & BPS & $Z=0$ & \shortstack{No root\\found} \\
\hline
Bicubic & $63\ (21.0\%)$ & $9\ (3.0\%)$ & $35\ (11.7\%)$ & 62 & 0 & 117 \\
Degree 18 & 0 & $42\ (14.0\%)$ & $38\ (12.7\%)$ & 17 & 0 & 102 \\
Degree 8 & $150\ (50.0\%)$ & 0 & 0 & 40 & 18 & 92 \\
Degree 12 & $161\ (53.7\%)$ & 0 & 0 & 31 & 16 & 92 \\
\hline
\end{tabular}
\caption{Charge-sample branch statistics; percentages use 300 charge
vectors per geometry. The numbers of refined stationary representatives,
including saddles, are respectively $185,309,208,208$. In the bicubic and
degree-18 models the non-BPS $Z\neq0$ representatives with spectrum
$\{0,1,1,2\}$ are exactly the aligned ones: all 114 satisfy
Eq.~\eqref{special:alignment} to better than $10^{-40}$, and none of the
301 unaligned representatives has this spectrum. In the $st^2$-type
controls these points lie on the non-BPS duality orbit, have exact flat
directions, and are generally not aligned in the cubic frame. Failure to
find a root is not a nonexistence result.}
\label{tab:cycounts}
\end{table*}

All 73 strict unaligned minima have $0<c_{\min}<1$, three non-integer
exponents, and one exponent equal to one. For the bicubic,
$c_{\min}$ ranges from $0.000234345$ to $0.985402$, with median $0.401310$;
for degree 18 the range is $0.000282937$ to $0.984636$, with median
$0.678285$. Thirty structured aligned examples and 74 charge-continuation
points supplement the random sample. In particular, stable aligned
phases in degree 18 exist despite their absence from this bounded sample.
These frequencies are not an invariant probability measure or an
exhaustive classification.

Write charges as $(q_0,q_1,q_2;p^0,p^1,p^2)$ and coordinates as
$(x_1,x_2;y_1,y_2)$. The following four representatives have positive metric,
$Z\ne0$, and lie strictly inside the K\"ahler cone:
\begin{itemize}
\item Bicubic unaligned (B-U). The charges and coordinates are, respectively,
\begin{equation*}
\begin{gathered}
(-5000,600,-400;6,30,-20),\qquad (5.007982,-5.608327;4.911028,3.425326).
\end{gathered}
\end{equation*}
Here $D=210.349614$, $Q^2=19012.696630$, and
\begin{equation}
\begin{aligned}
c=\bigl(&0.236765214295,0.532567719152,1,1.761248061965\bigr).
\end{aligned}
\label{cy:bicubicunaligned}
\end{equation}
The Hessian is strictly positive.
\item Degree-18 unaligned (18-U). The charges and coordinates are, respectively,
\begin{equation*}
\begin{gathered}
(0,-252,-288;-4,48,-42),\qquad (-1.779866,6.652279;3.594352,4.129435).
\end{gathered}
\end{equation*}
Here $D=180.325500$, $Q^2=30590.152479$, and
\begin{equation}
\begin{aligned}
c=\bigl(&0.187046454819,1,1.370175592003,3.076195687953\bigr).
\end{aligned}
\label{cy:degree18unaligned}
\end{equation}
The Hessian is strictly positive.
\item Bicubic aligned (B-A). The charges and coordinates are, respectively,
\begin{equation*}
(192,0,0;0,4,4),\qquad (0,0;4,4).
\end{equation*}
Here $D=192$, $Q^2=384$, $c=(0,1,1,2)$, and $\kappa/V_*=1/3>0$.
\item Degree-18 aligned (18-A). The charges and coordinates are, respectively,
\begin{equation*}
\begin{gathered}
(-4563,1350,378;-28,18,36),\qquad (0,0;3,6).
\end{gathered}
\end{equation*}
Here $D=175.5$, $Q^2=175500$, $c=(0,1,1,2)$, and
$\kappa/V_*=0.246402>0$.
\end{itemize}
Large volume follows from the exact transformation
\begin{equation}
\begin{gathered}
t\mapsto Lt,\qquad V\mapsto L^3V,\qquad (q_0,q_a,p^0,p^a)\mapsto(L^3q_0,L^2q_a,p^0,Lp^a).
\end{gathered}
\label{cy:dilation}
\end{equation}
An integer $L$ preserves integral charges in the cubic frame, and integer dilations bring all 910 stationary
representatives to $y_a\ge3$, $D\ge30$; the 73 strict unaligned minima, of volume up to $6.6$ before dilation, then
have $\min_ay_a\ge3.0$ and $98\le D\le1.2\times10^4$ (median $630$, $3\le L\le167$). A uniform rescaling of the
charges does not move the moduli.

\subsection{Aligned charges and the normalized quartic}
Let $f$ denote a $G_*$-unit zero-mode amplitude and define
\begin{equation}
 V_{\rm eff}=V_*+\kappa f^4+\cdots,\qquad
 \widehat\kappa=\frac{\kappa}{V_*}.
\label{cy:kappaconvention}
\end{equation}
Thus $\kappa$ has the units of $V$, whereas $\widehat\kappa$ is dimensionless.
For a D0--D4 point, $y\parallel p$ with $p_1,p_2>0$, the relaxed quartics are
\begin{align}
 k_{0,B}(p)
 &=\frac{9p_1^2p_2^2(p_1+p_2)^2}
 {4(p_1^2+p_1p_2+p_2^2)^3}>0,\nonumber\\
 k_{0,18}(p)
 &=-\frac{27p_1^2(3p_1^2+3p_1p_2+p_2^2)^2}
 {4p_2^3(3p_1+p_2)^3}<0.
\label{cy:k0}
\end{align}
Both depend only on the ray, so the same formula defines $k_0(y)$.
The $st^2$ controls have $k_0=0$. The massive-field exchange in
Eq.~\eqref{special:quarticexchange} is essential, as in the corrected
analysis of Ref.~\cite{NTT2007}.

At $x=0$, a convenient aligned charge chart is
\begin{align}
 p^a&=a y^a,\qquad q_a=avD_a,\nonumber\\
 p^0&=\frac{av(3+v^2)}{1-v^2},\qquad
 q_0=\frac{aD(1+3v^2)}{1-v^2}.
\label{cy:alignedcharges}
\end{align}
Here $a\ne0$ and $v\ne\pm1$. The potential and quartic are
\begin{align}
 V_*&=2a^2D(y)\frac{(1+v^2)^3}{(1-v^2)^2},\nonumber\\
 \frac{\kappa}{V_*}&=k_0(y)
 \left(\frac{1-v^2}{1+v^2}\right)^3.
\label{cy:phasequartic}
\end{align}
Axion translations generate the general point. Explicitly, put
$b=av$, $e=a(1+3v^2)/(1-v^2)$ and $c=a v(3+v^2)/(1-v^2)$. Then
\begin{align}
p^0&=c,\qquad p^a=a y^a+c x^a,\nonumber\\
q_a&=bD_a(y)-aD_{ab}(y)x^b-cD_a(x),\nonumber\\
q_0&=eD(y)-bD_a(y)x^a+\frac a2x^aD_{ab}(y)x^b+cD(x).
\label{cy:axioncharges}
\end{align}
For $x=\beta y$ the magnetic charges stay parallel to $y$ and
$q_a=(b-2a\beta-c\beta^2)D_a(y)$, which vanishes for real $\beta$ only if
$a^2+bc=a^2(1+v^2)^2/(1-v^2)>0$. Only the phase $|v|<1$ is therefore, up to
axion shifts, the D0--D4--D6 family of Ref.~\cite{NTT2007}, which has $q_a=0$
and moduli proportional to the magnetic charges.
In a binary normal form $D=y_1^3-2y_1y_2^2+\beta y_2^3$ with
$g_* =\operatorname{diag}(3/4,1)$, the unit null vector is
$(0,\cos\theta,0,-\sin\theta)$, where
\begin{equation}
\begin{gathered}
 \cos\theta=\frac{1-v^2}{1+v^2},\qquad
 \sin\theta=\frac{2v}{1+v^2},\qquad \widehat\kappa=\frac{32-27\beta^2}{96}\cos^3\theta.
\end{gathered}
\label{cy:binaryquartic}
\end{equation}
This is Eq.~\eqref{special:quartic} in a unit coordinate. The cubic null
coefficient vanishes. The charge-map determinant in this normal form at
$a=1$ is $144(1+v^2)^7/(1-v^2)^5\ne0$, so the stable phases form open
families in continuous charge space. The bicubic is quartic-stable for
$|v|<1$, whereas degree 18 is quartic-stable for $|v|>1$; the opposite
phases are saddles. The finite-charge limit $v\to\pm1$ with
$a\propto1-v^2$ gives the exact D0--D6 shape moduli of
Eq.~\eqref{special:d0d6flat}. A zero quartic elsewhere requires further
analysis.

\subsection{The quartically lifted direction: horizon approach and entropy}\label{App:marginal}
Let $f$ be the marginal coordinate of Eq.~\eqref{cy:kappaconvention}, canonically normalized, after the massive fields have been eliminated, and let $\widehat\kappa>0$, so that $v=V_{\rm eff}/Q^2=1+\widehat\kappa f^4+\cdots$. In the extremal radial gauge $\Delta=r^2$, set $s=-\log(r/Q)$ and $w=\log(R/Q)$. To the marginal orders displayed, the scalar equation, focusing equation and constraint are
\begin{align}
f_{ss}-f_s&=\tfrac12e^{-2w}v_f,\nonumber\\
w_{ss}+w_s+w_s^2&=-f_s^2,\nonumber\\
v&=e^{2w}\big[(1+w_s)^2+f_s^2\big],
\label{eq:IRmarginalflow}
\end{align}
so that $f_{ss}-f_s=2\widehat\kappa f^3+\cdots$. Excluding the growing homogeneous mode, as convergence at the horizon requires, gives
$f_s=-2\widehat\kappa f^3+\cdots$ and
$w=\widehat\kappa f^4/2+\cdots$.
The metric coefficient follows from the constraint. Since $\log(1/z)=s+w$,
\begin{equation}
f\sim\pm[4\widehat\kappa\log(z_*/z)]^{-1/2},\qquad
\frac{R^2}{Q^2}-1\sim\widehat\kappa f^4.
\label{eq:IRmarginalprofile}
\end{equation}
On the positive branch $f\sim(4\widehat\kappa s)^{-1/2}$, and consequently
\begin{equation}
\begin{gathered}
 \frac{R''}{R}\sim-\frac{1}{16\widehat\kappa r^2s^3},\qquad \frac RQ-1\sim\frac{1}{32\widehat\kappa s^2}.
\end{gathered}
\label{cy:logtidal}
\end{equation}
This logarithmic approach follows from the second-order system, without
assuming an analytic first-order flow; related marginal behavior was
identified in Ref.~\cite{TripathyTrivedi2005}. The exactly vanishing marginal source is a separate branch; nonzero sources enter this regime after a source-dependent crossover.

At finite temperature let $t_r=\rho/b_{\rm ne}$. The regular inner solution and its focusing integral give
\begin{align}
f(t_r)&=f_h+2\widehat\kappa f_h^3
\log\frac{t_r+1}{2}+\cdots,\nonumber\\
\frac{b_{\rm ne}R_h'}{R_h}
&=2\widehat\kappa^2f_h^6+o(f_h^6)+O(b_{\rm ne}/Q).
\label{eq:IRmarginalinner}
\end{align}
The horizon constraint $V_h=R_h^2-2b_{\rm ne}R_hR_h'$ then yields
$R_h^2/Q^2=1+\widehat\kappa f_h^4+o(f_h^4)$.
Matching fixes $f_h^2\sim[4\widehat\kappa L_T]^{-1}$, where
$L_T=\log(T_*/T)$. Hence
\begin{equation}
S_0=\pi Q^2,\qquad
S-S_0\simeq\frac{\pi Q^2}{16\widehat\kappa L_T^2},\qquad
C_Q\simeq\frac{\pi Q^2}{8\widehat\kappa L_T^3}.
\label{eq:IRmarginalentropy}
\end{equation}
The exterior sets $T_*$. Odd fifth-order terms can produce $L_T^{-5/2}$ corrections; for an even potential the first logarithmically enhanced correction can be $O(\log L_T/L_T^3)$. The analytic $O(T)$ entropy term is subleading. These are classical asymptotics and do not fix a quantum crossover scale.

\subsection{Full exterior equations and numerical results}
With $s$ and $w$ as above, the full scalar and Einstein equations,
including the constraint, are
\begin{align}
 D_s\phi_s-\phi_s&=\frac{e^{-2w}}{2Q^2}G^{-1}\partial V,\nonumber\\
 w_{ss}&=1+w_s-\frac{e^{-2w}V}{Q^2},\nonumber\\
 (1+w_s)^2+|\phi_s|_G^2&=\frac{e^{-2w}V}{Q^2}.
\label{cy:radialequations}
\end{align}
The boundary-value problem imposes asymptotic flatness and generic moduli
at the outer endpoint and decaying or center asymptotics at the horizon
endpoint. For the aligned solution, the outer scalar condition is
$\phi-\phi_s=\phi_\infty$ to leading order in $1/r$; the horizon condition
matches massive modes to the quadratic center graph and imposes
$f_s=-2\widehat\kappa f^3$ together with the radial constraint. All four
scalar equations and the Einstein equation are solved. Finite endpoints,
collocation tolerances, and fitting windows were varied.

\begin{table*}[t]
\centering
\small
\setlength{\tabcolsep}{3pt}
\begin{tabular}{lllll}
\hline
Exterior & \shortstack{Domain\\in $s$} & Measured decay
 & \shortstack{Constraint\\residual} & \shortstack{Equation\\residual} \\
\hline
B-U & $[-18,65]$ & \shortstack[l]{$c=0.2367655$\\$\pm0.0000008$}
 & \shortstack[l]{$<6\times10^{-13}$\\normalized} & $3.0\times10^{-10}$ \\
18-U & $[-18,65]$ & \shortstack[l]{$c=0.1870445$\\$\pm0.0000040$}
 & \shortstack[l]{$<6\times10^{-13}$\\normalized} & $3.0\times10^{-10}$ \\
\shortstack[l]{B-A,\\four fields} & $[-16,1000]$ & \shortstack[l]{$p=0.5014$\\on $[400,900]$}
 & \shortstack[l]{$2.29\times10^{-14}$\\absolute} & $<1.0\times10^{-11}$ \\
\hline
\end{tabular}
\caption{Full exterior solutions. The unaligned uncertainties are
fit-window systematics, not regression errors. For B-A,
$f\propto(s+s_0)^{-p}$; the finite-$s$ window drift exceeds the regression
error and is not included in the displayed value of $p$.}
\label{tab:cyexteriors}
\end{table*}

For the unaligned solutions, the measured tidal slopes are $-1.526469$
and $-1.625913$, approaching $2c-2=-1.526470$ and $-1.625907$.
The regression standard errors on $c$ are $5.7\times10^{-9}$ and
$3.3\times10^{-8}$, smaller than the systematic uncertainties in
Table~\ref{tab:cyexteriors}. Slow amplitudes are nonzero; faster raw
projections contain nonlinear forcing and are not independent homogeneous
amplitudes. The exact RN linear profile $(1+e^s)^{-c}$ gives an invertible
local amplitude map, and the nonlinear exteriors preserve slow excitation.
These two exteriors were computed in the undilated frame, with asymptotic
moduli $(x;y)_\infty=(0.512,-0.563;0.492,0.362)$ for B-U and
$(-0.291,1.124;0.563,0.717)$ for 18-U; the classical dilation maps them to
the representatives quoted above. Integrating outward from high-order
stable-manifold data, instead of solving a boundary-value problem, also gives
regular asymptotically flat B-U exteriors and, after removing the known nonlinear correction
$\propto(r/Q)^{c}$ to the local exponent,
$c=0.2367652143\pm5\times10^{-10}$ and a tidal exponent
$-1.5264696\pm1.2\times10^{-6}=2c-2$; with the slowest mode switched off,
the decay is governed by the next exponent, $0.532567719152$.

For B-A,
\begin{equation*}
\phi_\infty=(1.20,-0.96;4.40,4.32).
\end{equation*}
Four independent moduli perturbations yield a nonsingular map to interior
scalar values. At $r/Q=1$ its singular values are
\begin{equation*}
(0.91019,0.49979,0.48304,0.23665).
\end{equation*}
The fitted exponent drifts from $p=0.5038$ on $[100,400]$ to $0.5014$
on $[400,900]$. At $s=900$,
$r^2s^3(-R''/R)=0.17538$, approaching $3/16$.
The two finest solutions differ by less than $8\times10^{-11}$ at 20,002
comparison points through $s=400$. These results support local open
families of full exteriors. The horizon and infinity beyond the finite
numerical cutoffs use the derived asymptotic expansions; the entire
asymptotic basin has not been mapped.

\subsection{Perturbative \texorpdfstring{$\alpha'$}{alpha-prime} comparison}
For the imaginary-constant correction $F=-D+i\epsilon_{\alpha'}(X^0)^2$~\cite{CandelasDeLaOssaGreenParkes1991,AlexandrovKetovWakimoto2016},
\begin{equation}
\begin{gathered}
 \epsilon_{\alpha'}=\frac{\chi\zeta(3)}{16\pi^3},\qquad e^{-K}=8D-4\epsilon_{\alpha'},\qquad \mathcal W=q_0+q_at^a+p^aD_a(t)-p^0D(t)-2i\epsilon_{\alpha'} p^0,
\end{gathered}
\label{cy:alphaconvention}
\end{equation}
with $\epsilon_{\alpha'}=-0.392527818480$ for the bicubic and $\epsilon_{\alpha'}=-1.308426061598$ for degree 18.
Charges are held fixed after selecting the classical volume; exact-volume
choices in Table~\ref{tab:cyalpha} use continuous charge families. Spectra are
displayed to six decimal places.
\begin{table*}[t]
\centering
\small
\setlength{\tabcolsep}{3pt}
\begin{tabular}{lrl}
\hline
Branch & \shortstack{Classical\\$D$} & Corrected exponent spectrum \\
\hline
Bicubic D0--D4 & 30 & $(0.018261,1.012122,1.031007,2.018190)$ \\
 & 100 & $(0.005758,1.003834,1.009654,2.005751)$ \\
Bicubic unaligned B-U family & 30 & $(0.255314,0.529207,1.019268,1.759449)$ \\
 & 100 & $(0.242442,0.531283,1.005797,1.760781)$ \\
Degree-18 D0--D4 & 30 & $(0.052494,1.034528,1.092103,2.051843)$ \\
 & 100 & $(0.018261,1.012122,1.031007,2.018190)$ \\
Degree-18 unaligned 18-U family & 30 & $(0.081203,1.049730,1.309330,3.030075)$ \\
 & 100 & $(0.163748,1.010241,1.357161,3.067709)$ \\
\shortstack[l]{Degree-18 aligned $v=2$,\\continued saddle} & 30 & $(-0.048299,0.903993,1.089497,2.206443)$ \\
 & 100 & $(-0.012442,0.978216,1.024379,2.053753)$ \\
\hline
\end{tabular}
\caption{Fixed-charge continuations with the leading perturbative
prepotential correction. A negative displayed $c$ in the last two rows
marks the unstable continuation of the classical aligned point. The
bicubic at $D=30$ and the degree-18 model at $D=100$ have the same
$\epsilon_{\alpha'}/D$, since $\chi_{18}/\chi_B=10/3$; their D0--D4 spectra, computed
separately for the two cubic forms, agree to all twelve digits retained,
as Eq.~\eqref{cy:alphaD0D4} requires.}
\label{tab:cyalpha}
\end{table*}

On the aligned D0--D4 branch the corrected spectrum has a closed form. At fixed charges let
$d=\epsilon_{\alpha'}/D(y_*)$, with $D$ the classical cubic at the corrected attractor; then
$\epsilon_{\alpha'}/D_{\rm cl}=d\,(\beta_d/\nu_d)^{3/2}$, and the radial axion and saxion and the $n-1$
transverse ones have
\begin{equation}
\begin{gathered}
h_{\rm ax}=\frac{6(d^2+8)\beta_d}{(d+4)\lambda_d},\qquad h_{\rm ax}^\perp=\frac{6d(d-2)\beta_d}{(d+4)\lambda_d},\\
h_{\rm sax}=-\frac{2(5d^4-4d^3-384d^2+128d-64)}{(d+4)\lambda_d},\qquad h_{\rm sax}^\perp=-\frac{2(d-2)\beta_d}{\lambda_d},
\end{gathered}
\label{cy:alphaD0D4}
\end{equation}
with $\beta_d=5d^2+4d+8$, $\nu_d=2d^3+3d^2+36d+8$ and $\lambda_d=d^3+36d^2+24d+16$; at $d=0$ these are
$6,0,2,2$. The cubic coefficients of the transverse directions drop out through second order, so the spectrum
depends only on $\epsilon_{\alpha'}/D_{\rm cl}$ and $n$, and for $\chi<0$ the transverse axions are lifted,
$h_{\rm ax}^\perp\simeq-\frac32\epsilon_{\alpha'}/D_{\rm cl}$.

At a normalized aligned radial point with $D_*=1$, put
$\sqrt8\,\zeta=\mu+i\nu$, $\mathcal R=\mu^2+\nu^2>0$.
Keeping charges fixed and including the attractor displacement gives
\begin{equation}
\begin{gathered}
\left.\frac{dh_s}{d\epsilon_{\alpha'}}\right|_0
=\frac{15(\mu^2-\nu^2)}{2\mathcal R},\qquad \left.\frac{dh_f}{d\epsilon_{\alpha'}}\right|_0
=-\frac{15(3\mu^2+\nu^2)}{2\mathcal R}<0.
\end{gathered}
\label{special:alphaspectrum}
\end{equation}
For general classical volume the expansion parameter is $\epsilon_{\alpha'}/D_*$.
No aligned radial continuation retains $c_f=2$ to first order. Other
corrected unaligned resonances are not excluded.
Including the first-order displacement of the critical point, the zero
mode has the exact leading shift
\begin{equation}
 h_0=c_0(c_0+1)
 =-\frac{3\epsilon_{\alpha'}}{2D}\frac{1-v^2}{1+v^2}
 +O((\epsilon_{\alpha'}/D)^2).
\label{cy:alphazero}
\end{equation}
Since $\epsilon_{\alpha'}<0$, the quartic-stable bicubic phase acquires a positive
quadratic lift, whereas the quartic-stable classical degree-18 phase
acquires a negative one. Along the charge ray of the 18-A representative
($v=2$, $y\propto(1,2)$), dilated to classical volume $D=100$, the latter
continuation becomes a saddle, but a displaced strict minimum exists at
$(x_1,x_2;y_1,y_2)=(0.345162,-0.985718;2.807301,3.695468)$, with
corrected volume $96.04$ and spectrum
$(0.040598,0.978107,1.047994,2.054022)$. At $D=10$, the D0--D4 zero
exponents become $0.048183$ and $0.114216$, respectively. The 18-U ray
dilated only to classical volume $D=10$ instead reaches a minimum--saddle
fold at $0.409604188683$ of the full correction; at classical $D=30$ and $D=100$ the same ray survives as a
strict minimum (Table~\ref{tab:cyalpha}).
For the B-U and 18-U families at classical $D=100$, the largest spectral
shifts are $0.00580$ and $0.02330$, smaller than their classical
non-integer deviations, although corrections can be decisive near a
bifurcation.

\paragraph{Full sample.} Each of the 73 strict unaligned minima was dilated
by the smallest integer $L$ giving classical volume $D\ge100$, and
separately $D\ge1000$, and continued in the correction from $0$ to
$\epsilon_{\alpha'}$ at 70 digits. At $D\ge100$, 35 of 35 bicubic and 37 of 38
degree-18 minima remain strict minima, with largest shifts of $c_{\min}$
equal to $0.0053$ and $0.060$ (median $0.0021$ and $0.0084$). The
exception, with classical $c_{\min}=2.8\times10^{-4}$, loses stability
through a transcritical exchange at $53\%$ of the correction. At
$D\ge1000$ all 73 survive, with shifts of $c_{\min}$ below $0.00056$
(bicubic) and $0.0087$ (degree 18). No minimum--saddle fold occurs in
these continuations, and no other $c_{\min}$ leaves $(0,1)$.

\paragraph{Comparison with earlier work.} The lift~\eqref{cy:alphazero}
vanishes in the D0--D6 limit $v\to\pm1$ and is nonzero for D0--D4 ($v=0$)
and D2--D6 ($|v|\to\infty$) charges, in agreement with the behavior found
for $st^2$ in Ref.~\cite{Bellucci:2008tx} and, for D0--D4 and D0--D6
charges, for general cubics in Ref.~\cite{Dominic:2011st}. Its sign is that of $-\epsilon_{\alpha'}(1-v^2)/(1+v^2)$: for $\chi<0$, as for
both threefolds here, the D0--D4 flat directions are stabilized and the
D2--D6 ones destabilized, and the signs are reversed for $\chi>0$. In our convention the $\alpha'$ term
raises $e^{-K}$ for $\chi<0$, as in the K\"ahler potential of
Ref.~\cite{Becker:2002nn}. Reference~\cite{Dominic:2011st} states that
two-parameter D0--D4 attractors are stable only if $\det(D_{ab})>0$, with
$D_{ab}=D_{abc}p^c$. Inside the K\"ahler cone of a two-modulus model this
matrix has Lorentzian signature, since $D_{ab}y^b=2D_a$ and $D_{ab}$ is
negative definite on $\{D_a\delta y^a=0\}$, so the condition is not met there,
whereas we find a strict minimum for the bicubic D0--D4 point
(Table~\ref{tab:cyalpha}). The two results refer to opposite
signs of the correction. The large-charge expansion of Ref.~\cite{Dominic:2011st}
(their Eqs.~(4.7)--(4.8)) assumes $\gamma>0$, i.e.\ a correction that lowers
$|e^{-K}|$; this is our $\epsilon_{\alpha'}>0$, which with the sign of
Refs.~\cite{CandelasDeLaOssaGreenParkes1991,Becker:2002nn} is the case
$\chi>0$, where we also find the two-modulus D0--D4 flat directions
destabilized. For that sign the two computations agree. The stability conditions stated
in Ref.~\cite{Dominic:2011st} for two and three moduli are not met at
points with positive-definite metric, and the stable two-modulus D0--D4
points found here occur for the opposite sign, $\chi<0$. The conventions of Ref.~\cite{Dominic:2011st}
are not uniform in the overall sign of $e^{-K}$; the argument of the
logarithm in their Eq.~(3.2) is positive only for $D_{abc}y^ay^by^c<0$, as
in Ref.~\cite{NTT2007}, and in that convention their term $i\beta(X^0)^2$,
$\beta=-\zeta(3)\chi/(16\pi^3)$, is $-i\epsilon_{\alpha'}(X^0)^2$ here, in
line with the sign-independent statement above.
Reference~\cite{Dominic:2011iz} drops the $\chi$ term, citing a symplectic
transformation, so its statement that perturbative corrections do not
change the number of massless directions concerns the term
$\alpha_{0a}X^aX^0$; the $\chi$ term changes $e^{-K}$ and cannot be removed
by a real symplectic transformation.

\paragraph{Size of the neglected corrections.} The correction considered
here has relative size $|\epsilon_{\alpha'}|/D$, which the
dilation~\eqref{cy:dilation} suppresses as $L^{-3}$ and a uniform
rescaling of the charges at fixed moduli leaves unchanged. Four-derivative
couplings proportional to the second Chern class are not included; as in
the corrected entropy of supersymmetric black holes, their relative size
is $c_2\cdot y/(6Q^2)$, suppressed as $L^{-2}$ by the dilation and as
$N^{-2}$ by a uniform rescaling of the charges by $N$. For the bicubic,
$c(X)=(1+J_1)^3(1+J_2)^3/(1+3J_1+3J_2)$ gives
$c_2=3J_1^2+3J_2^2+9J_1J_2$ and $c_2\cdot(J_1,J_2)=(36,36)$. The
degree-18 hypersurface is an elliptic fibration over $\mathbb P^2$ with
section $\sigma$, $H=\sigma+3L$, so $c_2\cdot L=12\,c_1(\mathbb P^2)\cdot\ell=36$
and $c_2\cdot\sigma=c_2(\mathbb P^2)-c_1(\mathbb P^2)^2=-6$, hence
$c_2\cdot(H,L)=(102,36)$, with $\ell$ the line class of $\mathbb P^2$, as in Ref.~\cite{CandelasFontKatzMorrison1994}. At
the four representatives listed above,
$c_2\cdot y/(6Q^2)=2.6\times10^{-3}$ (B-U), $2.8\times10^{-3}$ (18-U),
$0.125$ (B-A) and $5.0\times10^{-4}$ (18-A), to be compared with
$|\epsilon_{\alpha'}|/D=1.9\times10^{-3}$, $7.3\times10^{-3}$, $2.0\times10^{-3}$ and
$7.5\times10^{-3}$. Both corrections are therefore small for the
unaligned representatives, whose classical deviations from integers are
of order one, while for B-A the four-derivative terms dominate and the
quadratic lift of its flat directions is not determined here; a uniform
rescaling of the B-A charges by $N\gtrsim8$, which leaves $|\epsilon_{\alpha'}|/D$
unchanged, reverses this ordering. Across the 73 unaligned minima dilated
to $D\ge100$, $c_2\cdot y/(6Q^2)$ ranges from $8\times10^{-6}$ to $0.16$ with
median $4.8\times10^{-3}$; it is below a tenth of the smallest Hessian
eigenvalue $c_{\min}(c_{\min}+1)$ for 68 of them, and a uniform rescaling by
at most $2.64$ achieves this for all. At $D\ge1000$ the median is
$1.2\times10^{-3}$, and a rescaling by at most $1.32$ suffices.
Worldsheet instantons contribute terms weighted by $n_\beta e^{-2\pi\beta\cdot y}$;
the genus-zero invariants of the curve classes generating the Mori cone
are $n_{1,0}=n_{0,1}=189$ for the bicubic~\cite{Hosono:1993qy,Hosono:1994ax}
and $n_{1,0}=540$, $n_{0,1}=3$ for the degree-18 hypersurface in the
$(H,L)$ basis~\cite{CandelasFontKatzMorrison1994}, so at the
representatives these weights are below $4\times10^{-6}$. The two exterior
solutions of Table~\ref{tab:cyexteriors}, mapped by the dilation to the
representative charges (dilation factors 10 and 6), keep $\min_ay_a\ge3.3$
along the whole flow, where the same bound on the instanton weights holds. Finally, the charges are quoted
in the cubic frame. In an integral symplectic basis the prepotential also
contains real quadratic terms, so lattice charges differ from these by a
rational symplectic transformation, which leaves the metric and the
spectra unchanged; the open sets of charges found here contain lattice
points after rescaling, and the statistics of Table~\ref{tab:cycounts}
refer to the cubic frame. These statements concern the
vector-multiplet theory and the specified families. Worldsheet
instantons, spacetime higher derivatives, and the complete quantum
horizon are not determined by this comparison.

\section{Related models}\label{App:Related}
This appendix collects two models outside the ungauged cubic theories of Appendix~\ref{App:Special}: the circle reduction of five-dimensional Einstein--Maxwell--Chern--Simons theory and a gauged supergravity.
\subsection{The circle-reduced Chern--Simons deformation}\label{special:emcs}
The two-scalar model obtained by reducing five-dimensional
Einstein--Maxwell--Chern--Simons theory has
$G=3(dx^2+dy^2)/(4y^2)$, $y>0$, and
\begin{align}
2V_\lambda={}&y^3(p^0)^2+3y(p^1-xp^0)^2+\frac{[q_0+xq_1+\lambda(3x^2p^1-x^3p^0)]^2}{y^3}+\frac{[q_1+\lambda(6xp^1-3x^2p^0)]^2}{3y}.
\label{special:emcspotential}
\end{align}
The convention $\lambda=1$ is minimal supergravity~\cite{HorowitzSantos2024}.
The shift $x\mapsto x+a$ acts on the charges by
\begin{align}
 p^1&\mapsto p^1+ap^0,\nonumber\\
 q_1&\mapsto q_1-6\lambda ap^1-3\lambda a^2p^0,\nonumber\\
 q_0&\mapsto q_0-aq_1+3\lambda a^2p^1+\lambda a^3p^0.
\end{align}
The dilation $(x,y)\mapsto\sigma(x,y)$, $\sigma>0$, acts with charge
weights
\begin{equation*}
\begin{gathered}
(q_0,q_1,p^0,p^1)\mapsto
(\sigma^{3/2}q_0,\sigma^{1/2}q_1,\sigma^{-3/2}p^0,\sigma^{-1/2}p^1).
\end{gathered}
\end{equation*}
Every regular attractor can therefore be placed at $(x,y)=(0,1)$, where
\begin{equation}
\begin{gathered}
 q_1(q_0+2\lambda p^1)=3p^0p^1,\qquad 9[(p^0)^2+(p^1)^2-q_0^2]-q_1^2=0.
\end{gathered}
\label{special:emcscritical}
\end{equation}
For $M_{ij}=\partial_i\partial_jV$, the entries are
\begin{align}
 M_{xx}&=3(p^0)^2+12\lambda^2(p^1)^2+6\lambda p^1q_0-2\lambda p^0q_1+q_1^2,\nonumber\\
 M_{xy}&=-3p^0p^1-2\lambda p^1q_1-3q_0q_1,\nonumber\\
 M_{yy}&=3(p^0)^2+6q_0^2+q_1^2/3.
\end{align}
Since $\mathsf H=2M/(3V_*)$, spectrum $(1,2)$ requires
$\operatorname{tr}M=12V_*$ and $\det M=27V_*^2$. For $p^1\ne0$, normalize
$p^1=1$. Exact elimination of the charge variables from these equations
and Eq.~\eqref{special:emcscritical} gives
\begin{equation}
 (\lambda^2-1)(4\lambda^2-9)=0.
\label{special:emcselimination}
\end{equation}
At $\lambda=3/2$ the only canonical solution is
$(q_0,q_1,p^0,p^1)=(-1,0,0,1)$; at $-3/2$ it is its electric-sign reversal.
Both are protected by axion parity. At $\lambda=1$, the spectral projectors
$P_1=(6I-\mathsf H)/4$, $P_2=(\mathsf H-2I)/4$ satisfy
\begin{equation}
 T_{ijk}T_{abc}(P_2)_{ia}(P_1)_{jb}(P_1)_{kc}=0,
 \qquad T=\nabla^3V,
\label{special:emcsprojector}
\end{equation}
on the critical and spectral locus. Since $G_*=(3/4)I$, this is a positive
multiple of the squared resonant norm. Electric-charge reversal gives
$\lambda=-1$. If $p^1=0$, normalization to $p^0=1$ leaves D0--D6 at every
$\lambda$ and D2--D6 only at $\lambda=\pm1$; D0--D6 has $T_{yxx}=0$
directly. If both magnetic charges vanish, criticality forces the excluded
zero-charge solution. These cases exhaust the real regular $(1,2)$ locus
and give $\Gamma=0$.
On the stable D0--D4 branches,
\begin{equation}
\begin{gathered}
 h_y=2,\qquad h_x=4\lambda^2+2s\lambda,\qquad c_x=\frac{|4\lambda+s|-1}{2},\qquad s=\operatorname{sgn}(q_0p^1).
\end{gathered}
\end{equation}
Thus $c_x=2\lambda$ for $s=+1$, $\lambda\ge0$. For the particular fixed
charges $(q_0,q_1,p^0,p^1)=(1,0.7,0.3,1)$, a numerical continuation gives
\begin{equation}
\begin{gathered}
(c_s,c_f)=(1,2) +(0.129450011136272,1.85485610053963)(\lambda-1) +O((\lambda-1)^2).
\end{gathered}
\label{special:emcscontinuation}
\end{equation}
These derivatives describe this charge vector, not a charge-independent law.

\subsection{Gauged supergravity: the \texorpdfstring{$SO(8)$}{SO(8)} STU model}\label{special:gauged}
With a scalar potential, $-U(\phi)$ added to the bracket of the action of Sec.~\ref{Sec:Resonance}, a static extremal horizon is $\mathrm{AdS}_2\times S^2$ with radii $L_2$ and $R_h$ that differ. At fixed charges the attractor conditions~\cite{Morales:2006gm,Cacciatori:2009iz,DallAgata:2010ejj} are
\begin{equation}
\begin{gathered}
\frac1{R_h^2}=\frac{V_*}{R_h^4}+\frac{U_*}2,\qquad
\frac1{L_2^2}=\frac{V_*}{R_h^4}-\frac{U_*}2,\qquad
d\Big(U+\frac{2V}{R_h^4}\Big)_{\!p}=0 ,
\end{gathered}
\label{gauged:attractor}
\end{equation}
with $R_h$ held fixed in the last one. Equivalently, $R_h^2$ is the critical value of $V_{\rm eff}=2V/(1+\sqrt{1-2UV})$~\cite{Bellucci:2008cb}, and linear scalar perturbations decay as $z^c$, $z=(r-r_h)/L_2$ in an affine radial coordinate, with
\begin{equation}
\begin{aligned}
c(c+1)\in\operatorname{spec}\mathsf H,\qquad
\mathsf H&=\frac{G^{-1}\nabla^2V_{\rm eff}}{2V_{\rm eff}}\Big|_p
=\frac{L_2^2}{4}\,G^{-1}\nabla^2\Big(U+\frac{2V}{R_h^4}\Big)\Big|_p ,
\end{aligned}
\label{gauged:exponents}
\end{equation}
so that the eigenvalues of $\mathsf H$ are the masses $m^2L_2^2$ in the throat. For $U=0$ this is the definition of Sec.~\ref{Sec:Resonance}. The radius mode no longer decouples: a deformation $R=R_h(1+bz+\cdots)$ has exponent exactly one and forces the scalars at the same weight, $(2-\mathsf H)v=L_2^2\,b\,G^{-1}dU(p)$ for the slope $v$ of $\phi$, so it enters the smoothness criterion together with the scalar modes; when $\mathsf H$ has the eigenvalue $2$ and $dU(p)$ has a component along its eigenspace, a logarithm appears already at first order.

We have applied these formulas to the STU truncation of $SO(8)$ gauged supergravity, the $U(1)^4$ sector of M-theory on $\mathrm{AdS}_4\times S^7$~\cite{DuffLiu1999,CveticEtAl1999}, with the six scalars and target metric of Eq.~\eqref{stu:metric}, coupling $g=1/L_{\mathrm{AdS}_4}$ and
\begin{equation}
U=-g^2\sum_{i=1}^3\big(2\cosh\varphi_i+e^{\varphi_i}\chi_i^2\big).
\end{equation}
For $g\neq0$ no static spherical attractor of finite area has a fast mode, so the $(1,2)$ resonance cannot occur:
\begin{equation}
m^2L_2^2<\frac{6(1+x)}{1+2x}<6,\qquad x=-\tfrac12U_*R_h^2>0,
\label{gauged:bound}
\end{equation}
for all six scalars, for arbitrary electric and magnetic charges and arbitrary horizon values of the axions. The proof uses the representation $4V=T^{\mathsf T}(M_1\otimes M_2\otimes M_3)T$, $U=-g^2\sum_i\operatorname{tr}M_i$, where $T$ is the vector of the eight charges and $M_i$ is the positive unimodular matrix of the $i$th dilaton--axion pair. Along a geodesic with unit tangent $v$ in the metric $4G$, $M_i\mapsto M_i^{1/2}e^{sA_i}M_i^{1/2}$ with commuting generators, $A_i^2=t_i^2$ and $\sum_it_i^2=1$, so that $\nabla^2V(v,v)\le(t_1+t_2+t_3)^2V\le3V$ and $\nabla^2U(v,v)<0$; Eq.~\eqref{gauged:attractor} then gives Eq.~\eqref{gauged:bound}. Among $4484$ numerically sampled dyonic attractors with nonzero axions, the largest eigenvalue is $5.979$.

On the electric branch with vanishing axions the spectrum is explicit. Every horizon value $\varphi_*$ and every $\varepsilon_g=g^2R_h^2>0$ occur, with charges $Q_I^2=R_h^2e^{-a_I\cdot\varphi_*}(1+\varepsilon_g\sum_ie^{a_{Ii}\varphi_{i*}})$, where $a_I$ are the four sign vectors $(+++)$, $(+--)$, $(-+-)$, $(--+)$. With $C_i=\cosh\varphi_{i*}$, $S_i=\sinh\varphi_{i*}$ and $R_h^2/L_2^2=1+2\varepsilon_g\sum_iC_i$, the dilaton block of $\mathsf H$ in an orthonormal frame is
\begin{equation}
\begin{gathered}
\mathsf H_{ii}=\frac{2L_2^2}{R_h^2}\big[1+\varepsilon_g(C_j+C_k)\big],\qquad
\mathsf H_{ij}=\frac{2L_2^2}{R_h^2}\,\varepsilon_gS_k ,
\end{gathered}
\end{equation}
for $i,j,k$ distinct. Both this matrix and $2-\mathsf H$ are strictly diagonally dominant with positive diagonal, so the three dilatons have $0<m^2L_2^2<2$, that is $0<c<1$. For example, $R_h=1$, $g=\frac14$, $e^{\varphi_*}=(2,1,1)$ and positive charges give $L_2^2=\frac{32}{45}$ and $m^2L_2^2=(\frac{14}9,\frac85,\frac{76}{45})$ for the dilatons and for the axions alike, that is $c=(0.8437,0.8601,0.8924)$, each twice. For other signs of the charges the axion block can have negative eigenvalues, which does not affect the bound. At $\varphi_*=0$ these spectra reduce to those of Sec.~5 of Ref.~\cite{CastroVerheijden2021}, where the gauging was found to lower the dimensions below their integer values at $g=0$. The four-charge black holes of Refs.~\cite{DuffLiu1999,CveticEtAl1999} are the analytic exteriors with these horizons. At the equal-charge solution the linear map from the three horizon amplitudes of the dilatons to their boundary coefficients has rank three, so generic boundary data excite modes with $0<c<1$ there. In this model the obstruction to smoothness is thus the non-integer exponent of Ref.~\cite{HKS2022AdS}, not a resonance. The statement concerns the six scalars of the truncation in the two-derivative theory; other fields of the compactification are not covered.

\section{Detuning and the infrared description}\label{App:Detuning}
\subsection{Detuned fast response}
Throughout this appendix, $D=u^2-k^2$ and $u\in[k,1]$.
A complementary check varies the fast scalar exponent before taking the low-temperature limit. In the canonical two-field calculation, replace the fast mass by $\nu(\nu+1)$ with $\nu=2+\delta$, while keeping the cubic potential coefficient $10\gamma x^2y$ fixed. The scalar operator is $L_\nu=\partial_u(D\partial_u)-\nu(\nu+1)$. Put
\begin{align}
\lambda_\nu&=\nu(\nu+1),&d_\nu&=6-\lambda_\nu,\nonumber\\
q_\nu(u)&=u^2-\frac{2k^2}{\lambda_\nu},&q_1&=q_\nu(1).
\end{align}
Since $L_\nu q_\nu=d_\nu u^2$, the exact regular, zero-source response is
\begin{equation}
Y_\nu=\frac{5\gamma}{d_\nu}\left[q_\nu(u)-q_1\frac{P_\nu(u/k)}{P_\nu(1/k)}\right].\label{eq:detunedY}
\end{equation}
The inverse exists for positive $\lambda_\nu$; integration by parts gives $\langle h,L_\nu h\rangle=-\int(Dh'^2+\lambda_\nu h^2)du$ for zero outer data. Apparent poles in \eqref{eq:detunedY} at $\nu=2$ are removable at fixed $k>0$.

Fixed-temperature Feynman--Hellmann variation of \eqref{eq:Routhian} gives
\begin{equation}
\partial_\gamma F\big|_{T,q,\phi_\infty}=\frac12\int_{b_{\rm ne}}^\infty\frac{\partial_\gamma V}{R^2}\,d\rho.
\end{equation}
At quartic source order, $Q^2d\rho/R_0^2=M_0du$ and $x=\epsilon u$. Integrating in $\gamma$ gives $F_4^{\rm ex}=(5M_0\gamma/2)\int u^2Y_\nu du$. A Green identity therefore yields the exact boundary kernel
\begin{align}
F_4^{\rm ex}(\nu,k)={}&\frac{25M_0\gamma^2}{2d_\nu}\,\mathcal B_\nu(k),\nonumber\\
\mathcal B_\nu(k)={}&\frac{1-k^5}{5}-\frac{2k^2(1-k^3)}{3\lambda_\nu}-\frac{(1-k^2)q_1}{d_\nu}(2-q_1D_\nu(k)),\label{eq:detunedkernel}\\
D_\nu(k)&=\left.\partial_u\log P_\nu(u/k)\right|_{u=1}.\nonumber
\end{align}
The independent quartic contact contributes an analytic term. The metric and area variations are accounted for by stationarity, not neglected.

The large-argument source and response coefficients of the regular Legendre function $P_c$ have the ratio
\begin{align}
\mathcal R(c)&=\frac{\Gamma_{\mathrm E}(-c-\tfrac12)\Gamma_{\mathrm E}(c+1)}{2^{2c+1}\Gamma_{\mathrm E}(-c)\Gamma_{\mathrm E}(c+\tfrac12)}=\tan(\pi c)\,\frac{\Gamma_{\mathrm E}(c+1)^2}{2^{2c+1}\Gamma_{\mathrm E}(c+\tfrac12)\Gamma_{\mathrm E}(c+\tfrac32)},\label{eq:knownkernel}
\end{align}
where $\Gamma_{\mathrm E}$ is Euler's gamma function, distinct from the resonant map, and the two forms are related by the reflection formula $\Gamma_{\mathrm E}(-c-\tfrac12)\Gamma_{\mathrm E}(c+\tfrac32)=\pi/\sin\pi(-c-\tfrac12)$ and its analogue for $\Gamma_{\mathrm E}(-c)\Gamma_{\mathrm E}(c+1)$. At integers $n\ge0$, reflection and duplication give
\begin{equation}
\mathcal R(n)=0,\qquad
\mathcal R'(n)=\frac{4^n}{(2n+1)\binom{2n}{n}^{\!2}},
\label{eq:IRintegerkernel}
\end{equation}
so that $\mathcal R(\nu)=\frac4{45}\delta+O(\delta^2)$ here. This is the known static kernel. In Ref.~\cite{BerensteinMancilla2023}, Eq.~(26), set boundary dimension $d=1$, $h=\nu+1$, and $r_h=(2z_0)^{-1}$; gamma duplication gives the same ratio, with the usual response normalization factor $2\nu+1$. Their Weyl-squared forcing differs from the slow-field-square forcing in \eqref{eq:detunedY}. Appendix D, Eqs.~(D.85)--(D.86), of Ref.~\cite{MSY2016} contains the equivalent free thermal factor $\Gamma_{\mathrm E}(1/2-\Delta)/\Gamma_{\mathrm E}(1-\Delta)$ with $\Delta=\nu+1$. Its direct convergence range is narrower than $\Delta\simeq3$; the complete exterior in \eqref{eq:detunedkernel} fixes the needed local subtraction here.

Separating the part analytic in $k^2$, denoted by $F_{\rm even}$, the two coalescing terms are
\begin{equation}
F_4^{\rm ex}-F_{\rm even}=\frac{M_0}{Q}\{C_5(\nu)k^5+C_\nu(\nu)k^{2\nu+1}+\cdots\},
\end{equation}
where
\begin{align}
C_5&=\frac{5Q\gamma^2(10-3\lambda_\nu)}{6\lambda_\nu(6-\lambda_\nu)},\nonumber\\
C_\nu&=-\frac{25Q\gamma^2(2\nu+1)\mathcal R(\nu)}{2(6-\lambda_\nu)^2}.\label{eq:detunedpair}
\end{align}
Their residues are $C_5=2Q\gamma^2/(9\delta)+O(1)$ and $C_\nu=-2Q\gamma^2/(9\delta)+O(1)$. Thus the static zero removes one of the two mixing denominators, and the remaining simple poles cancel between an analytic thermal term and a non-integer response. The limit is
\begin{equation}
-\frac{2Q\gamma^2}{9}k^5\frac{k^{2\delta}-1}{\delta}\longrightarrow-\frac{4Q\gamma^2}{9}k^5\log k.\label{eq:polepair}
\end{equation}
Deleting the analytic $k^5$ term before taking this limit would give an incorrect divergent result.

For $s=2\delta\log k$ in a fixed compact set, the corresponding crossover is
\begin{equation}
F_4^{\rm ex}-F_{\rm even}=-\frac{4Q\gamma^2}{9}k^5\log k\,\frac{e^s-1}{s}+O(Q\gamma^2 k^5),\label{eq:crossover}
\end{equation}
with the ratio defined as one at $s=0$. A uniform remainder follows from the exact source/response hypergeometric branches
\begin{align}
S_\nu(z)&={}_2F_1(-\nu/2,(1-\nu)/2;1/2-\nu;z),\nonumber\\
T_\nu(z)&={}_2F_1((\nu+1)/2,(\nu+2)/2;\nu+3/2;z).
\end{align}
Their coefficients are analytic near $(\nu,z)=(2,0)$ and $S_2(k^2)=q_1$. In the exact response term the factor
\begin{align}
&\frac{q_1^2}{S_\nu(k^2)\{S_\nu(k^2)+\mathcal R(\nu)k^{2\nu+1}T_\nu(k^2)\}}=1+\delta h(\delta,k)
\end{align}
has bounded $h$ for small $k$ and $|\delta|<\delta_0<1/2$. Derivatives of the additional powers of $k$ remain bounded because they are multiplied by positive powers of $k$. After multiplication by the simple pole, the deviation contributes only $O(k^5)$. This controls even detunings much smaller than $k^2$; a bound proportional to $k^2/\delta$ would not suffice. Replacing $k$ by $\vartheta$ changes only the stated remainder.

Equation~\eqref{eq:crossover} describes a crossover of classical finite-source coefficients. Changing the reference scale of $\log\vartheta$ shifts an analytic fifth-power term but leaves the logarithmic coefficient unchanged.

\paragraph{Shifted slow exponent.} If corrections also shift the slow exponent to $c_1=1+\delta_1$, the slow profile is the regular Legendre function of degree $c_1$, and its square forces the fast mode with radial weight $u^{2c_1}$. For the extremal operator $L^{(0)}_\nu=\partial_u(u^2\partial_u)-\nu(\nu+1)$,
\begin{equation}
L^{(0)}_\nu u^{2c_1}=(2c_1-\nu)(2c_1+\nu+1)\,u^{2c_1},
\end{equation}
so the mixing denominator $6-\lambda_\nu=(2-\nu)(3+\nu)$ of Eq.~\eqref{eq:detunedpair} becomes $(2c_1-\nu)(2c_1+\nu+1)$, and the coalescing pair is $k^{4c_1+1}$ and $k^{2\nu+1}$. To leading order in the detunings and for $|\delta_1\log k|,|\delta_2\log k|\ll1$, the pair combines as in Eq.~\eqref{eq:crossover}, with $k^5$ replaced by $k^{5+4\delta_1}$ and $s=2(\nu-2c_1)\log k=2(\delta_2-2\delta_1)\log k$; the uniform control of the remainder established above for $\delta_1=0$ is not extended here. An exact non-integer resonance survives on the surface $\delta_2=2\delta_1$. Whenever $\delta_1$ or $\delta_2$ is nonzero the horizon is non-smooth already at linear order, more strongly than through the resonance if $\delta_1\neq0$ or $\delta_2<0$. The shifted modes also contribute their own linear responses, Eq.~\eqref{eq:IRlinear}. With $\mathcal R(1+\delta_1)=\delta_1/3+O(\delta_1^2)$ and $\mathcal R(2+\delta_2)=4\delta_2/45+O(\delta_2^2)$ from Eq.~\eqref{eq:IRintegerkernel},
\begin{equation}
F\supset-\frac{Q}{2}\,\delta_1\|A\|^2\vartheta^{3+2\delta_1}-\frac{2Q}{9}\,\delta_2\|B\|^2\vartheta^{5+2\delta_2},
\end{equation}
where $A$ and $B$ are the slow and fast amplitudes. For $|\delta_1\log\vartheta|\ll1$ the first term contains $-Q\delta_1^2\|A\|^2\vartheta^3\log\vartheta$, which is smaller than the resonant logarithm only for $\vartheta\gg\frac32|\delta_1|\,\|A\|/\|\Gamma(A,A)\|$. For $|\delta_2\log\vartheta|\ll1$ the second contains $-\frac{4Q}{9}\delta_2^2\|B\|^2\vartheta^5\log\vartheta$, of the same form as the resonant term. The fast amplitude also enters at cubic source order, through the term $5M_0\langle B,\Gamma(A,A)\rangle\int_k^1x_1^2y_1\,du$ with the regular profiles $x_1$ and $y_1$; it is analytic at zero detuning, and to first order in the detunings it contributes
\begin{equation}
F\supset-\frac{8Q}{9}(2\delta_1+\delta_2)\langle B,\Gamma(A,A)\rangle\,\vartheta^5\log\vartheta .
\end{equation}
For $\delta_1=0$ the three terms combine into $-\frac{4Q}{9}\|\Gamma(A,A)+\delta_2B\|^2\vartheta^5\log\vartheta$. Here $B$ is the fast amplitude of Eq.~\eqref{eq:profiles}, without the induced part $\Gamma(A,A)/(\delta_2-2\delta_1)$. The resonant term therefore dominates only if $(2|\delta_1|+|\delta_2|)\,\|B\|\ll\|\Gamma(A,A)\|$. Together with the crossover window, the resonant logarithm thus keeps the coefficient of Eq.~\eqref{eq:main}, to leading order in the detunings and for $|\delta_i\log\vartheta|\ll1$, when
\begin{equation}
\begin{gathered}
\big|(\delta_2-2\delta_1)\log\vartheta\big|\ll1,\qquad (2|\delta_1|+|\delta_2|)\,\|B\|\ll\|\Gamma(A,A)\|,\qquad \vartheta\gg\tfrac32|\delta_1|\,\|A\|/\|\Gamma(A,A)\|.
\end{gathered}
\label{eq:window}
\end{equation}
For corrections suppressed by $\ell_{\rm hd}^2/Q^2$, with $\ell_{\rm hd}$ the length scale of the higher-derivative terms, these conditions are mild.

\subsection{Infrared dictionary}\label{App:IR}
The static operator $L_c$ identifies a throat scaling dimension $\Delta=c+1$. This identification does not make the full exterior dynamics an $\mathrm{AdS}_2$ theory. We use the classical throat generating functional, with standard quantization, without assuming a microscopic conformal quantum mechanics. Near extremality the throat is only nearly $\mathrm{AdS}_2$: thermal corrections from massive fields require backreaction beyond Jackiw--Teitelboim gravity~\cite{CastroHollanderMartinezVerheijden2026}, and on non-BPS branches with dimension-one operators the near-$\mathrm{AdS}_2$ correlators are not well defined~\cite{CastroVerheijden2021}. Put $N=Q^2$ and choose an orthonormal target frame. In the unit-radius Euclidean throat metric $\hat g$, spherical reduction at fixed charges gives
\begin{equation}
\begin{aligned}
I_E={}&\frac N2\int\sqrt{\hat g}\,
\big[(\partial x)^2+(\partial y)^2+2x^2+6y^2+10\Gamma_{iab}y^ix^ax^b\big]+\cdots,\\
\Gamma_{iab}={}&\frac{V_{;iab}}{20Q^2}.
\end{aligned}
\label{eq:IRaction}
\end{equation}
Fix the source convention
$Z[\alpha]=\langle\exp(+\int d\tau\,\alpha O)\rangle=e^{-I_{\rm ren}}$.
For dimensionless throat time $\tau$ and $\eta=u^{-1}$, the source, response and bulk-to-boundary kernel are
\begin{equation}
\begin{gathered}
\phi_\Delta=\alpha_\Delta u^{\Delta-1}
+\beta_\Delta u^{-\Delta}+\cdots,\qquad K_\Delta(\eta,\tau;t)=b_\Delta
\left[\frac{\eta}{\eta^2+(\tau-t)^2}\right]^\Delta,
\\
b_\Delta=\frac{\Gamma_{\mathrm E}(\Delta)}
{\sqrt\pi\,\Gamma_{\mathrm E}(\Delta-\tfrac12)}.
\end{gathered}
\label{eq:IRsource}
\end{equation}
The one-point function is
$\langle O_\Delta\rangle=N(2\Delta-1)\beta_\Delta+\text{local terms}$.
Thus
\begin{align}
\langle O_\Delta(t)O_\Delta(0)\rangle
&=\frac{\mathcal K_\Delta}{|t|^{2\Delta}},
&\mathcal K_\Delta&=N(2\Delta-1)b_\Delta,\nonumber\\
\mathcal K_2&=\frac{6N}{\pi},
&\mathcal K_3&=\frac{40N}{3\pi}.
\label{eq:IRtwopoint}
\end{align}
The normalized cubic contact integral is
$\mathcal D_{223}=4/(3\pi)$; before multiplying the three kernel normalizations, its value at $0,1,\infty$ is $\pi^2/8$.
The differentiated vertex in \eqref{eq:IRaction} is $10N\Gamma_{iab}$, with the conventions of Ref.~\cite{FreedmanMathurMatusisRastelli1998}. Therefore
\begin{equation}
\begin{aligned}
&\langle O_{2a}(t_1)O_{2b}(t_2)O_{3i}(t_3)\rangle=\frac{C^{(3)}_{abi}}
{|t_{12}|\,|t_{13}|^3|t_{23}|^3}, C^{(3)}_{abi}=-\frac{40N}{3\pi}\Gamma_{iab},\\
&\mathcal C_{ab}{}^i=\frac{C^{(3)}_{abi}}{\mathcal K_3}
=-\Gamma^i{}_{ab},\widehat C_{abi}=-\frac{\sqrt{10\pi/3}}{3\sqrt N}\Gamma_{iab}.
\end{aligned}
\label{eq:IRthreepoint}
\end{equation}
Here $O_{2a}(t)O_{2b}(0)\supset\mathcal C_{ab}{}^i|t|^{-1}O_{3i}(0)$, and the hat denotes unit two-point normalization. The sign in \eqref{eq:IRthreepoint} follows from the specified plus-source convention.

At $\Delta_3=2\Delta_2-1$, the factor $1/2!$ in conformal perturbation theory cancels the factor of two from the two sides of
$\int_{a<|t|<\ell}dt/|t|=2\log(\ell/a)$.
With energy scale $\mu=\ell^{-1}$ and the canonical scaling of sources factored out,
\begin{equation}
\begin{aligned}
\mu\partial_\mu g_3^i
&=-\mathcal C_{ab}{}^i g_2^ag_2^b
=\Gamma^i{}_{ab}g_2^ag_2^b,\\
g_2&=A(\xi),\\
g_3(z)&=B(\xi)+\Gamma(A(\xi),A(\xi))\log z,
\end{aligned}
\label{eq:IRRG}
\end{equation}
where $\mu/\mu_0=z$.
Dimensionless couplings $\lambda_\Delta=\mu^{\Delta-1}g_\Delta$ instead obey
\begin{equation}
\beta_{\lambda_2}=\lambda_2,\qquad
\beta_{\lambda_3}=2\lambda_3+\Gamma(\lambda_2,\lambda_2).
\label{eq:IRdimensionlessRG}
\end{equation}
Physical time is $t_{\rm phys}=Q\tau$, so
$g_\Delta^{(t_{\rm phys})}=Q^{\Delta-1}\alpha_\Delta$ and
$F=F_\tau/Q$. A local quadratic redefinition of the fast coordinate shifts $B$ but leaves the resonant coefficient unchanged.

\subsection{Linear response, the nonlinear law and exceptional exponents}\label{App:IRresponse}
For generic $c>0$, the horizon-regular solution $P_c(u/k)$ has the source-to-response ratio $\beta/\alpha=\mathcal R(c)k^{2c+1}$, with the kernel of Eq.~\eqref{eq:knownkernel}, whose first form is the one quoted in Sec.~\ref{Sec:CY}; the integrated thermal two-point function gives the same normalization through
\begin{equation}
b_{c+1}2^{-2c-1}\sqrt\pi\,
\frac{\Gamma_{\mathrm E}(-c-\tfrac12)}{\Gamma_{\mathrm E}(-c)}
=\mathcal R(c).
\label{eq:IRthermalidentity}
\end{equation}
For non-integer, non-half-integer $c$, the contributions at second order in the sources and fixed physical temperature are
\begin{align}
F_{\rm IR}&=-\frac Q2(2c+1)\mathcal R(c)|A_c|^2
\vartheta^{2c+1},\nonumber\\
S_{\rm IR}&=\pi Q^2(2c+1)^2\mathcal R(c)|A_c|^2
\vartheta^{2c}.
\label{eq:IRlinear}
\end{align}
They supplement the analytic thermal terms; at this source order $A_c=\xi_c$. They follow from scalar work: with
$h_c=P_c(u/k)/P_c(1/k)$ and
$\phi=\epsilon h_c+O(\epsilon^2)$,
\begin{equation}
\begin{aligned}
F_2(T_0(k))&=\frac{B_0}{2}h_c'(1),\\
B_0&=Q\sqrt{1-k^2},\\
h_c'(1)&=c+\text{analytic terms in }k^2-(2c+1)\mathcal R(c)k^{2c+1}+\cdots,
\end{aligned}
\label{eq:IRscalarwork}
\end{equation}
where the last line uses the large-argument form of the regular solution,
\begin{equation}
\begin{gathered}
P_c(t)=\frac{1}{N_c}\Big[t^cS_c(t^{-2})+\mathcal R(c)\,t^{-c-1}T_c(t^{-2})\Big],\qquad N_c=\frac{\sqrt\pi\;\Gamma_{\mathrm E}(c+1)}{2^c\,\Gamma_{\mathrm E}(c+\frac12)},
\end{gathered}
\label{eq:IRlegendre}
\end{equation}
with $P_c(1)=1$ and the hypergeometric branches $S_c$, $T_c$ defined above.

At the exceptional exponents these expressions change. For the positive-integer resonance $(m,2m)$, where the kernel vanishes by Eq.~\eqref{eq:IRintegerkernel}, the forced fast response and its analytic pole partner give the source-quartic identity~\eqref{eq:general_fast_coefficient}, in which the bracket equals $4^m/\binom{4m}{2m}$. Other channels may contribute for $m\ge2$; the replacement $a\mapsto A(\xi)$ for the complete $(1,2)$ result uses Appendix~\ref{App:Matching}. At $c_0=n+\tfrac12$, $n\ge0$, \eqref{eq:IRlinear} is replaced by the finite pole-canceled logarithm
\begin{equation}
\begin{gathered}
F_{\log}=Q|A_{c_0}|^2 a_{c_0}
\vartheta^{2c_0+1}\log\vartheta,\qquad a_{c_0}=-(2c_0+1)\mathop{\rm Res}_{c=c_0}\mathcal R(c),
\\
(a_{1/2},a_{3/2},a_{5/2})
=\left(\tfrac18,\tfrac9{128},\tfrac{225}{8192}\right).
\end{gathered}
\label{eq:IRhalfinteger}
\end{equation}
Only the accompanying analytic power depends on the logarithm's reference scale. For $c\to0^+$,
\begin{equation}
\begin{gathered}
\mathcal R(c)=c+(\log4-2)c^2+O(c^3),\qquad S_{\rm IR}=\pi Q^2|A_c|^2[c+O(c^2)]\vartheta^{2c}.
\end{gathered}
\label{eq:IRsmallc}
\end{equation}
The factor $\vartheta^{2c}$ must remain unexpanded when $c|\log\vartheta|$ is appreciable. This limit does not describe the nonlinear marginal flow.

\paragraph{All orders in the sources.} Equation~\eqref{eq:IRlinear} holds at every order in the sources when $A_c$ is the nonlinear amplitude of the extremal exterior. Let the attractor be isolated, with positive Hessian and no flat directions, and let $0<c<1$ be its smallest exponent, with eigenspace $E_c$ and orthogonal projection $\Pi_c$; write $c_2>c$ for the next exponent. No other condition on the spectrum is needed, and the higher exponents may resonate. Let $A_c(\xi)$ be defined by the extremal exterior with the same charges and asymptotic moduli, $\Pi_c(\phi-p)=A_c(\xi)z^c+o(z^c)$. Then, for $c\neq\frac12$ and in every finite source quotient,
\begin{equation}
\begin{aligned}
F\big|_{\vartheta^{2c+1}}&=-\frac Q2(2c+1)\mathcal R(c)\,\|A_c(\xi)\|^2,\\
S\big|_{\vartheta^{2c}}&=\pi Q^2(2c+1)^2\mathcal R(c)\,\|A_c(\xi)\|^2 ,
\end{aligned}
\label{eq:IRallorders}
\end{equation}
which are the complete coefficients of these powers. We give two derivations. Both rest on an extension of Lemmas L1 and L2 to real exponents, which we describe first.

\emph{Expansion.} In every finite source quotient the real-degree Legendre
inverse~\eqref{eq:matching-PQ-inverse} gives differentiable inner and outer
expansions with exponents $j+\sum_i m_i c_i$, $j,m_i\in\mathbb N_0$.
Only finitely many powers and logarithmic degrees occur below each cutoff.
For $a>-d-1$ the zero-Dirichlet inverse in \eqref{eq:matching-positive-Green}
maps $O(u^a)$ to $O(u^{\min(a,d)})$ uniformly down to $u=k$, with a factor
$1+|\log u|$ at $a=d$. The nonlinear forcing has $a=2c$. The resulting
scalar bounds are $\phi-p=O(u^c)$ and $\phi'=O(u^{c-1})$. Focusing gives
$g-1=O(u^{\min(2c,1)})$, with the same logarithmic factor at $2c=1$.
On compact outer intervals the coefficients converge to the controlled
extremal exterior with slow amplitude $A_c(\xi)$. Appendix~\ref{App:IRrealexpansion}
proves these statements, including the bounds at $c=\tfrac12$.

\emph{Hamilton--Jacobi derivation.} Equations~\eqref{eq:thermalHJ} and~\eqref{eq:thermalCS} hold for any spectrum, with $\Psi=Q\sum_n\psi_n\mathsf b^n$ now a sum over the exponents just described. Assign to a monomial in the normal-form coordinates the weight $\sum_im_ic_i$. The kernel of the operator in Eq.~\eqref{eq:thermalCS} consists of the polynomials of weight $n-1$, with the logarithmic combinations of Appendix~\ref{App:RGform} at resonances. A term $\mathsf b^n$ is thus either such a homogeneous solution or is forced by pairs $i+j=n$ of lower terms and, at $n=2$, by $1/(2w)$. There is no thermal correction with $0<n<1$, and $\psi_1$ is a constant, which vanishes; every pair containing $i=1$ therefore drops out. For $c\neq\frac12$ no pair contributes at $n=2c+1$. If $c<\frac12$, both members of a remaining pair exceed one, which would require $n>2$. If $c>\frac12$, there is no term with $1<n<2$, because a homogeneous solution of weight below one is linear in the sources and $\Psi$ has no linear term; both members are then at least two, which would require $n\ge4$. Hence $\psi_{2c+1}$ is a homogeneous solution of weight $2c$. Since $c$ is the smallest exponent, it is a quadratic form in the coordinate $X_c$, plus a term linear in a coordinate of exponent $2c$ if there is one, and the linear term vanishes. The quadratic form is fixed at second order by Eq.~\eqref{eq:IRscalarwork}, since $\Psi^{(2)}=\frac12B_0h_c'(1)|\xi_c|^2$. The coordinate $X_c$ is unique, because a nonlinear monomial cannot have the smallest weight, and $X_c(\xi)=A_c(\xi)$. This gives $\Psi|_{\mathsf b^{2c+1}}=-\frac Q2(2c+1)\mathcal R(c)\|A_c\|^2$. The change to fixed temperature is as in Appendix~\ref{App:RGform}: $\mathsf b=\vartheta[1+O(\vartheta)+O(\|A_c\|^2\vartheta^{2c})]$, so the coefficient of $\vartheta^{2c+1}$ in $F$ is the same, and $S=-\partial_TF$ gives Eq.~\eqref{eq:IRallorders}.

\emph{Matching derivation.} The second route follows Appendix~\ref{App:Matching}. No scalar inner degree lies below $c$, and the nonlinear forcing starts at $k^{2c}$, so the inner solution of degree $c$ is the regular homogeneous one, $k^cN_cA_cP_c(t)$, with its coefficient fixed by the extremal limit through Eq.~\eqref{eq:matching-connection}. By Eq.~\eqref{eq:IRlegendre} its decaying branch is the outer tail $k^{2c+1}\mathcal R(c)A_c\,u^{-c-1}$. A tail $k^ju^{-d-1}$ requires inner degree $j-d-1\ge c$, so no departure from even powers of $k$ before $j=2c+1$ carries unstable data; such a departure solves the homogeneous constrained Jacobi problem with zero boundary data and vanishes by the uniqueness argument of L4. At $j=2c+1$ the tail pairs with the extremal variation $u^c\,dA_c$ in the flux~\eqref{eq:matching-flux-normalization}, with Wronskian $(2c+1)u^{-2}$; the induced radius response is $O(u^{-1}\log^pu)$ and its pairing vanishes. Hence the non-analytic part of the scalar-work one-form is $-Q(2c+1)\mathcal R(c)\langle A_c,dA_c\rangle\,k^{2c+1}$, and integration from Reissner--Nordstr\"om gives Eq.~\eqref{eq:IRallorders} again. At cubic order the law can also be checked directly. For a single field with $W/Q=1+\frac c2x^2+w_3x^3$ one has $A_c=\xi-(3w_3/c)\xi^2+\cdots$ and $v_3=(2+6c)w_3x^3$, and the coefficient of $k^{2c+1}$ in $\int_k^1h_c^3\,du$ is $3(2c+1)\mathcal R(c)/[c(3c+1)]$, as Eq.~\eqref{eq:IRallorders} requires through $\Psi^{(3)}$ of Eq.~\eqref{eq:thermalHJinputs}.

\paragraph{Subleading terms and the temperature map.} The terms that follow in $\Psi$ are of three kinds: $\mathsf b^{1+\mathsf w}$, where $\mathsf w>2c$ is a sum of at least two exponents ($3c$, $4c$, $c+c_2$, $2c_2$, \ldots) and the coefficient is a polynomial of weight $\mathsf w$ in the amplitudes; the analytic term $\psi_2\mathsf b^2$; and forced terms of order $\mathsf b^{2+2c}$ and higher. Coincident weights can add logarithms; for instance, when $1/c$ is an integer the term $\mathsf b^2$ can acquire one, if its forcing of weight one is nonzero. With $s=(2c+1)^2\mathcal R(c)\|A_c\|^2$,
\begin{equation}
\begin{aligned}
\frac{S-S_0}{S_0}={}&s\,\vartheta^{2c}+o(\vartheta^{2c})-4\psi_2(\xi)\,\vartheta+\cdots,
\end{aligned}
\label{eq:IRtower}
\end{equation}
where the corrections to the first term come from the weights listed above, of relative order $\vartheta^{\min(c,\,c_2-c)}$ up to logarithms, and $-4\psi_2=2+O(\xi^2)$. For $c<\frac12$ the first term dominates, with $\mathcal R(c)>0$, once $\vartheta^{1-2c}\ll s/2$. For $\frac12<c<1$ it is a non-analytic correction to the linear term, with $\mathcal R(c)<0$. For $c<\frac12$ the relation between $k$ and the temperature is itself anomalous: $g_h^2-1=s\,k^{2c}+o(k^{2c})$ in Eq.~\eqref{fnd:physicalT}, so that
\begin{equation}
k=\vartheta\,\big[1+s\,\vartheta^{2c}+o(\vartheta^{2c})\big].
\label{eq:IRkmap}
\end{equation}
The leading coefficients in Eq.~\eqref{eq:IRallorders} are the same in both variables, but the next ones are not: the contribution of $s^2$ to the coefficient of the power $4c$ in $S/S_0-1$ is $\frac12s^2$ at fixed $\mathsf b$ and $(2c+\frac12)s^2$ at fixed $\vartheta$. Fits of numerical data must therefore use one variable consistently; those of Appendix~\ref{App:Numerics} use $\vartheta$. At $c=\frac12$ the exponent $2c+1$ coincides with that of the analytic term. The forcing $1/(2w)$ then has the resonant component $-\frac18\|X_{1/2}\|^2$ of weight one, and Eq.~\eqref{eq:thermalCS} gives
\begin{equation}
\begin{aligned}
F&\supset\frac Q8\,\|A_{1/2}(\xi)\|^2\,\vartheta^2\log\vartheta,\\
S-S_0&\simeq\frac{\pi Q^2}{2}\,\|A_{1/2}(\xi)\|^2\,\vartheta\log(1/\vartheta),
\end{aligned}
\label{eq:IRhalf}
\end{equation}
which is Eq.~\eqref{eq:IRhalfinteger} with $a_{1/2}=\frac18$, now with the nonlinear amplitude; in the matching derivation it comes from the term $\frac18t^{-3/2}\log t$ of $N_{1/2}P_{1/2}$.

\paragraph{Horizon values.} In the inner region $u=O(k)$ the mode is of order $k^c$ and, to leading order, $\Pi_c(\phi-p)=\Pi_c(\phi_h-p)\,P_c(u/k)$. Matching to $A_cz^c$ with Eq.~\eqref{eq:IRlegendre} gives the horizon value
\begin{equation}
\Pi_c(\phi_h-p)=N_c\,A_c(\xi)\,k^c\,[1+o(1)],
\label{eq:IRhorizonvalue}
\end{equation}
with relative corrections of order $k^c$, $s\,k^{2c}$ and $k$; $N_1=1$ and $N_2=\frac23$, in agreement with Eq.~\eqref{eq:linearprofiles}. In this region $\rho=b_{\rm ne}u/k\,[1+O(k)]$ is an affine parameter along the radial null geodesics, normalized at infinity, so that with $P_c'(1)=\frac12c(c+1)$ the tidal component at the horizon is
\begin{equation}
R_{kk}\big|_h=2\,|\partial_\rho\phi|_G^2=\frac{c^2(c+1)^2N_c^2}{2}\,\frac{\|A_c\|^2}{Q^2}\,k^{2c-2}\,[1+o(1)].
\label{eq:IRtidal}
\end{equation}
It grows without bound as $T\to0$ for $c<1$ and reaches $\ell^{-2}$ at $\vartheta\sim(\|A_c\|\ell/Q)^{1/(1-c)}$, the estimate of Sec.~\ref{Sec:CY}.

Equations~\eqref{eq:IRtower} and~\eqref{eq:IRtidal} share the amplitude. With
$\mathcal R(c)/N_c^2=\tan(\pi c)/[\pi(2c+1)]$, which follows from the reflection formula, their ratio is free of it:
\begin{equation}
\frac{S-S_0}{S_0\,Q^2\vartheta^2R_{kk}|_h}=H(c)+o(1),\qquad H(c)=\frac{2(2c+1)\tan\pi c}{\pi c^2(c+1)^2},
\label{eq:IRratio}
\end{equation}
for $0<c<\frac12$ and $A_c\neq0$, and by Eq.~\eqref{eq:IRhalf} it is $\frac{128}{9\pi^2}\log(1/\vartheta)+O(1)$ at
$c=\frac12$. With the specific heat $C_Q=T\partial_TS=2c(S-S_0)[1+o(1)]$ in place of $S-S_0$ the ratio is $2cH(c)$, which
tends to $4$ as $c\to0$; the quartically lifted direction of Appendix~\ref{App:marginal}, with a different radial
solution, gives the same value.

For the complete spectrum $\{1,2\}$ the inner fast field~\eqref{eq:matching-fast-core} at the horizon, $t=1$,
$H_2(1)=\frac23$, gives in normal coordinates
\begin{equation}
\Pi_2(\phi_h-p)=\mathsf D_h\,k^2\log k+O(k^2),\qquad \mathsf D_h=\tfrac23\Gamma(A,A),
\label{eq:IRfasthorizon}
\end{equation}
and an analytic change of coordinates changes only the $O(k^2)$ term. With Eqs.~\eqref{eq:main}
and~\eqref{eq:firstdivergence}, $C_\alpha=-Q[\xi^\alpha]\|\mathsf D_h\|^2=-\frac{Q^5}{36}[\xi^\alpha]K_{\rm ext}$, where
$K_{\rm ext}$ is the coefficient of $\log^2|\lambda|$ in $\nabla_k\nabla_kR_{kk}$ at the extremal horizon. Both limits
and the ratio~\eqref{eq:IRratio} hold at fixed small source (Appendix~\ref{App:IRrealexpansion}); in the solutions of
Appendix~\ref{App:Numerics}, fits of $\Pi_2(\phi_h-p)/k^2$ against $\log k$ give
$|{-Q\|\mathsf D_h\|^2/C_F}-1|<1.3\times10^{-8}$ for all 19 nonzero signals.

\subsection{Real-exponent extension of L1 and L2}\label{App:IRrealexpansion}
Fix a finite source Taylor quotient, analytic $G,V$, fixed charges and
normalized outer data at an isolated attractor. Let the positive exponents
be $c_i$, with smallest exponent $0<c<1$. The following construction
also gives the coefficient bounds at $c=1/2$. The coincident thermodynamic
power at that value is given in \eqref{eq:IRhalf}.

\paragraph{Regular scalar inverse.}
As in L1, differentiation at fixed $k>0$ gives $L_d\phi_\alpha=f_\alpha$
for $d=c_i$, and $g_\alpha''=h_\alpha$, with lower source orders on the
right. The fixed-charge constraint still gives \eqref{eq:matching-metric-slope}.
In the variable $t=u/k$, put $\Delta(t)=t^2-1$.
Writing $\varsigma=t-1$, $y=\sum y_n\varsigma^n$ and $f=\sum f_n\varsigma^n$,
coefficient comparison gives
\begin{equation*}
 2(n+1)^2 y_{n+1}= f_n+\{d(d+1)-n(n+1)\}y_n,\qquad n\ge0 .
 \tag{\theequation a}\label{pf:horizon-rec}
\end{equation*}
For analytic $f$ this determines an analytic germ for each $y_0$, since
the multiplier of $y_n$ tends to $-1/2$ and the $f_n$ have a geometric bound.
Take $P_d(1)=1$ in the homogeneous equation. For $t=\cosh x>1$,
put $\zeta=\cosh x+\sinh x\cos\theta$. Then
$P_d(t)=\pi^{-1}\int_0^\pi \zeta^d\,d\theta>0$.
Indeed the radial operator in $x$ applied to $\zeta^d$ is
$d\,\partial_\theta(\sin\theta\,\zeta^{d-1})/\sinh x$.
The integral has value one at $x=0$ and is analytic and even there.
Dominated convergence after division by $t^d$ gives
\begin{equation*}
 \begin{aligned}
 \lim_{t\to\infty}t^{-d}P_d(t)
 &=\frac{2^{d+1}}{\pi}\int_0^{\pi/2}\cos^{2d}\theta\,d\theta=\frac{2^d\Gamma_{\mathrm E}(d+1/2)}
        {\sqrt\pi\,\Gamma_{\mathrm E}(d+1)}=N_d^{-1}.
 \end{aligned}
 \tag{\theequation b}\label{pf:normalization}
\end{equation*}
Choose the real decaying Legendre solution
$Q_d=P_d\int_t^\infty [\Delta(s) P_d(s)^2]^{-1}ds$.
Differentiation gives $\Delta(t)(P_dQ_d'-P_d'Q_d)=-1$ and
$L_d Q_d=0$. Its leading term is $N_dt^{-d-1}/(2d+1)$, and its
horizon singularity is $-\tfrac12\log(t-1)+O(1)$.
The Wronskian gives \eqref{eq:matching-PQ-inverse} by differentiation.
The combined expression is bounded and tends to zero at $t=1$. Since $Q_d$
is unbounded there, \eqref{pf:horizon-rec} identifies it with the analytic
particular solution with $y_0=0$. Subtracting $P_d(t)Y_f(L)/P_d(L)$,
where $L=1/k$, imposes zero outer data, exactly as in L1.

\paragraph{The two expansions at infinity.}
On $t^\rho$, $L_d$ acts by
$(\rho-d)(\rho+d+1)t^\rho-\rho(\rho-1)t^{\rho-2}$.
In $t^\rho\sum_{n\ge0}a_nt^{-2n}$ choose $a_0=1$,
$\rho=d$ or $-d-1$. For $n\ge1$ the recursion is
\begin{equation*}
 (\rho-2n-d)(\rho-2n+d+1)a_n
       =(\rho-2n+2)(\rho-2n+1)a_{n-1}.
 \tag{\theequation c}\label{pf:infinity-rec}
\end{equation*}
The decaying branch always exists. The coefficient ratio tends to one
unless the series terminates, so the series converge for $t>1$.
The growing ladder hits the other root precisely when $2d+1=2N$ with $N$
a positive integer, that is, at half-integral $d$. At integral $d$ there
is no collision and the series terminates. At $d=N-1/2$ every numerator
factor through $n=N$ is nonzero.
Put $\sigma=-d-1$ and let $q_-$ be the monic decaying
solution. Adding $\kappa q_-\log t$ to the growing series cancels the
obstruction when
$(2\sigma+1)\kappa=(\sigma+2)(\sigma+1)a_{N-1}$.
This follows by applying $L_d$ to $q_-\log t$. The remaining forcing
is an ordinary series and all subsequent recursion denominators are nonzero.
Reduction of order on $q_-$ near infinity gives convergence because
the integrand $1/(\Delta(t)q_-^2)$ is $t^{2d}$ times a convergent series in
$t^{-2}$, with at most one $t^{-1}$ term. This locates the single logarithm
on the decaying ladder of the growing solution.

\paragraph{Radial germs and endpoint division.}
Use radial germs which are finite sums of $t^\rho(\log t)^p f(1/t)$
with $f$ holomorphic at zero and $p$ a nonnegative integer.
Products and $t\partial_t$ preserve this class. Ordinary differentiation
adds a factor $t^{-1}$. Write $\tau_t=\log t$. For $m\ne-1$ a primitive is
\begin{equation*}
 \begin{aligned}
 \int t^m\tau_t^p\,dt
 &=t^{m+1}\sum_{i=0}^p
 \frac{(-1)^i p!\,\tau_t^{p-i}}{(p-i)!(m+1)^{i+1}} .
 \end{aligned}
 \tag{\theequation d}\label{pf:primitives}
\end{equation*}
For $m=-1$ the primitive is $\tau_t^{p+1}/(p+1)$. In each real-power ladder
the nonzero denominators are
bounded away from zero and eventually grow linearly. Taylor bounds therefore
permit integration and differentiation on smaller disks in $1/t$. The
Green formula, metric integrations and finite target jets give the exact finite
separation $\sum A_r(L)B_r(t)$ by the induction of L1. Regular inputs may be
kept separate since the combined Green inverse is regular on each such input.
Endpoint division requires one further operation. By \eqref{pf:normalization}
and \eqref{pf:infinity-rec}, $P_d(L)=N_d^{-1}L^d(1+e_d)$, where the
positive orders of $e_d$ belong to $2\mathbb N_{>0}$ or
$2d+1+2\mathbb N_0$, possibly with a logarithm. The geometric series for
$(1+e_d)^{-1}$ converges for large $L$. Only finitely many products contribute
through any prescribed order, even at a half-integer $d$.

\paragraph{Allowed thermal powers.}
To see why subtraction at $L$ introduces no differences of unrelated
exponents, put
\begin{equation*}
 \mathcal S=\left\{j+\sum_i m_i c_i:\ j,m_i\in\mathbb N_0\right\}.
 \tag{\theequation e}\label{pf:powers}
\end{equation*}
In each separated product track an endpoint monomial $L^{-b}$ and radial
power $t^\rho$, suppressing logarithms. The two requirements are
$b\in\mathcal S$ and $b-\rho\in\mathcal S$. These are precisely the
inner and fixed-$u$ orders. The normalized homogeneous $P_d(t)/P_d(L)$
satisfies both by the two ladders and the geometric series.
Products add both orders, and $t\partial_t$ preserves them. The factor
$1-t^{-2}$ in the connection term with two Euler derivatives adds either
zero or two to the outer order.

For one radial forcing, a particular scalar inverse at infinity uses the
forced powers $\rho-2n$. Indeed at a power $t^\nu$ with coefficient
polynomial in $\tau_t$, the diagonal operator is
$(\partial_{\tau_t}+\nu-d)(\partial_{\tau_t}+\nu+d+1)$. It is invertible on polynomials
unless one constant vanishes, in which case one integration in $\tau_t$ suffices.
The off-diagonal term lowers the power by two, by \eqref{pf:infinity-rec}.
In the convergent germs furnished by the Green formula, any ladder outside
the forcing must therefore start at an indicial root.
An arbitrary free growing solution can be subtracted as a multiple of $P_d$
before imposing the outer condition. The residual free ladder starts at
$-d-1$. Thus forced powers add $2n$ to $b-\rho$, and a free decaying power
gives $b+d+1+2n\in\mathcal S$. At a forced root the logarithms stay at
that same power. This choice of particular solution preserves horizon
regularity because the subtracted $P_d$ is regular.
If its radial part takes a power $\nu$ at $L$, the Dirichlet subtraction
multiplies $L^{-(b-\nu)}$ by $P_d(t)/P_d(L)$, which again satisfies both
requirements. This cancellation is essential when $d$ exceeds $b$.

For the metric use $t^2g_{tt}=-g|t\phi_t|_G^2$. A particular $Z_0(t,L)$
can be chosen with just the forced powers by setting the two free constants
at infinity to zero. The Euler operator on a coefficient at power $\nu$
is $(\partial_{\tau_t}+\nu)(\partial_{\tau_t}+\nu-1)$. The normalized solution at
positive source order is
\begin{equation*}
 Z_0(t,L)-Z_0(L,L)+(t/L-1)
          \{g_\alpha'(1)-[t\partial_t Z_0(t,L)]_{t=L}\}.
 \tag{\theequation f}\label{pf:metric-endpoint}
\end{equation*}
Here $g_\alpha'(1)$ is the $u$ slope in \eqref{eq:matching-metric-slope}.
It uses only outer orders of lower coefficients and the analytic factor $1-k^2$.
Evaluating at $L$ uses $b-\nu$, and $t/L$ has inner and outer orders $1,0$.
This closes both inductions. Since $\min(1,c_i)>0$ and the target is
finite-dimensional, $\mathcal S$ meets bounded intervals in finite sets.
Taylor remainders of the radial germs have the same bounds under any fixed
number of Euler derivatives, up to the displayed logarithmic powers.
Truncating the reciprocal after degree $M$ in $e_d$ leaves the exact
remainder $(-e_d)^{M+1}/(1+e_d)$.
Euler derivatives preserve its power bound since they preserve the positive
order of $e_d$. In particular, if $\gamma$ is the first order in $\mathcal S$
above a cutoff, the $i$-th $k$ derivative of the remainder is
$O(k^{\gamma-i}(1+|\log k|)^p)$ for some finite $p$.
These estimates apply at fixed inner $t$ and uniformly on each fixed outer
$\delta\le u\le1$ in the respective expansions. Horizon evaluation uses
regularity at $t=1$. For the temperature and scalar-work operations of L1,
Taylor expansion in the relative change $k/k_0(\vartheta)-1$ uses Euler
derivatives and preserves $\mathcal S$, where $k_0$ is the analytic RN inverse.
The implicit coefficient of this relative change is a unit since
$k\vartheta_0'(k)/\vartheta_0(k)\to1$.

\paragraph{Uniform bounds and the extremal limit.}
Let $h_d(u)=N_d k^dP_d(u/k)$. Positivity, \eqref{pf:horizon-rec} and the
differentiated infinity expansion give $h_d\asymp u^d$ and
$|h_d'|\le C u^{d-1}$ on $k\le u\le1$, uniformly for small $k$.
Use \eqref{eq:matching-positive-Green} with this $h_d$. If $|f(u)|\le C u^a$
and $a>-d-1$, then on $[k,2k]$ one has
$|J(s)|\le C(s-k)s^{a+d}$, and on $[2k,1]$ integration of $v^{a+d}$
gives $|J(s)|\le C s^{a+d+1}$. In both intervals
\begin{equation*}
 \begin{aligned}
 \frac{|J(s)|}{(s^2-k^2)h_d(s)^2}&\le C s^{a-d-1},\\
 |y(u)|&\le C u^d\int_u^1 s^{a-d-1}\,ds .
 \end{aligned}
 \tag{\theequation g}\label{pf:green-bound}
\end{equation*}
Hence $y=O(u^{\min(a,d)})$ for $a\ne d$ and $y=O(u^d\Lambda_u)$ for
$a=d$, $\Lambda_u=1+|\log u|$. Differentiating the Green formula gives
$y'=(h_d'/h_d)y+J/[(u^2-k^2)h_d]$, so the corresponding derivative bound
loses one power. Nonzero fixed outer data add a multiple of $h_d(u)/h_d(1)$.

Induct in total source order, starting with these homogeneous scalars and
the zero linear metric perturbation. Suppose the lower scalar coefficients
and their first derivatives are $O(u^c)$ and $O(u^{c-1})$, and the lower
metric coefficients satisfy the bounds derived next. The nonlinear potential
force is $O(u^{2c})$ by stationarity and Taylor expansion.
The connection term is $O(u^{2c})$ since $|D|\le u^2$. The factor
$g^{-2}-1$ multiplies a force $O(u^c)$ and is $O(u^c)$ because
$\min(2c,1)>c$, also allowing the logarithm at $2c=1$.
Thus the scalar inverse applies with $a=2c$. Even at $d=2c$, the bounds
$u^{2c}\Lambda_u=O(u^c)$ and $u^{2c-1}\Lambda_u=O(u^{c-1})$
give the asserted scalar estimates.

Focusing gives $g_\alpha''=O(u^{2c-2})$ and the outer constraint gives
$g_\alpha'(1)=O(1)$. Integrating back gives $g_\alpha'=O(u^{2c-1})$
for $2c<1$, $O(\Lambda_u)$ for $2c=1$, and $O(1)$ for $2c>1$.
At the horizon \eqref{fnd:horizondata} reads, at positive source order,
\begin{equation*}
 \begin{aligned}
 &2g_\alpha(k)-2k g_\alpha'(k)=\big[(v-1)-(g-1)^2+2k(g-1)g'\big]_{\alpha,u=k}=O(k^{2c}).
 \end{aligned}
 \tag{\theequation h}\label{pf:real-horizon}
\end{equation*}
The two metric products use only lower orders. For $2c<1$ they are
$O(k^{4c})$, for $2c=1$ they are $O(k^2\Lambda_k^2)$, and for
$2c>1$ they are $O(k^2)$. The potential starts at quadratic order.
Adding $2kg_\alpha'(k)$ and integrating $g_\alpha'$ from $k$ to $u$
now gives $g_\alpha=O(u^{\min(2c,1)})$, with $O(u\Lambda_u)$ at equality.
This closes the simultaneous induction, uniformly down to $u=k$ at each
fixed source order.

For convergence repeat the L2 induction on $[\delta,1]$. The homogeneous
$h_d,h_d'$ converge there to $u^d,du^{d-1}$. Extending the integrand of $J$
by zero below $k$ gives the integrable majorant $C v^{d+2c}$, and the lower
source orders converge on compact subintervals. Dominated convergence in
the numerator is uniform in its upper limit by the $L^1$ bound on the
difference. The outer Green integral and its derivative then converge
uniformly on $[\delta,1]$. The metric slope converges by the outer constraint,
and two integrations of focusing with $g_\alpha(1)=0$ give the same compact
convergence for the metric. The bounds at zero give a controlled endpoint,
including $u\phi'\to0$ and $ug'\to0$. The limit has the prescribed
outer data and satisfies the constraint. The stable graph and normalized
reconstruction of Appendix~\ref{App:Branches} therefore identify it,
coefficientwise as in L2, with the unique controlled extremal exterior.
In $E_c$ the limiting Green integral divided by $u^c$ has a finite limit
and an $O(u^c)$ remainder by \eqref{pf:green-bound} with $a=2c,d=c$.
Finally $z=u/g=u(1+o(1))$. Thus the coefficient of $u^c$ is the
$A_c(\xi)$ defined by $\Pi_c(\phi-p)=A_c(\xi)z^c+o(z^c)$.
All estimates concern a fixed finite source quotient.

\paragraph{Fixed small source.} The estimates are also uniform in the source. For an isolated attractor with
$0<c\le\frac12$, in an analytic chart linearized in a Hessian eigenbasis, use the norms $\sup u^{-c}(|q|+u|q'|)$
for $q=\phi-p$ and $\sup w^{-1}(|m|+u|m'|)$ for $m=g-1$, with $w=u^{2c}$, or $u(1+|\log u|)$ at $2c=1$. On a ball of
radius $R$ the nonlinear scalar forcing is $O(R^2u^{2c})$, which the bound~\eqref{pf:green-bound} returns to the
ball, the focusing inverse bounds the metric by $CR^2w$, and differences gain a factor $CR$. Choosing $R$ and then
a common complex source ball makes the fixed-point map a contraction for all $0<k\le k_0$: the regular solutions
are holomorphic in the source, with $(S/S_0-1)/k^{2c}$, or $(S/S_0-1)/[k\log(1/k)]$ at $c=\frac12$, bounded
independently of $k$. Since every Taylor coefficient converges as $k\to0$ (Appendix~\ref{App:IRresponse}), Cauchy
estimates give the limit of the function itself at each fixed sufficiently small real source:
$(S/S_0-1)/\vartheta^{2c}\to s$ for $0<c<\frac12$, $(S/S_0-1)/[\vartheta\log(1/\vartheta)]\to\frac12\|A_{1/2}\|^2$ at
$c=\frac12$, and likewise for $\Pi_c(\phi_h-p)/k^c\to N_cA_c$, $R_{kk}|_h$ and the ratio~\eqref{eq:IRratio}. With
$M\to M_{\rm ext}(\xi)$, integrating $\partial_{b_{\rm ne}}\Psi=-\frac12\log(S/S_0)$ at fixed source gives
$F-M_{\rm ext}+S_0T=-\frac{Qs}{2(2c+1)}\vartheta^{2c+1}+o(\vartheta^{2c+1})$, or
$\frac Q8\|A_{1/2}\|^2\vartheta^2\log\vartheta+o(\vartheta^2\log\vartheta)$ at $c=\frac12$, without differentiating a
remainder. The norms $\sup(|q|/u+|q'|)$ and $\sup(|m|/u+|m'|)$ give in the same way the fixed-source limits of
$\mathsf D_h$ and $K_{\rm ext}$ for the complete spectrum $\{1,2\}$; its free energy would need a uniform bound
$O(\vartheta^5|\log\vartheta|)$ on the non-analytic remainder, and for $\frac12<c<1$ the linear entropy must first be
subtracted uniformly, which we have not shown.

\section{Finite-source numerical method}\label{App:Numerics}
\subsection{Models and boundary-value problem}
The finite-source calculations set $Q=1$ and solve the complete equations on $u\in[k,1]$. Three potentials are used. The exponential model is Eq.~\eqref{eq:exponentialmodel}, and the delayed model is Eq.~\eqref{eq:countermodel} with $\gamma=\frac35$ and $\lambda=\frac15$; both have flat target metric. For the protected $t^3$ model,
\begin{align}
v_{\rm P}(x,y)&=\frac12\left[y^3+3x^2y
+\frac{(1-x^3)^2}{y^3}+\frac{3x^4}{y}\right],\nonumber\\
G_{ij}d\phi^id\phi^j&=\frac{3}{4y^2}(dx^2+dy^2).
\label{eq:numProtected}
\end{align}
The first two attractors are at the origin; the protected attractor is $(x,y)=(0,1)$. Its exponents are $(1,2)$ and its ordinary projected cubic is $2/(5\sqrt3)$, while its covariant resonance map vanishes.

Chebyshev collocation uses
\begin{align}
s&=\frac{\log[(u+k)/(2k)]}{L},
&L&=\log\frac{1+k}{2k},\nonumber\\
a(u)&=\frac{u-k}{u+k},
&\partial_u(D\partial_u)&=\frac1{L^2}
\left(a\partial_s^2+L\partial_s\right).
\label{eq:numGrid}
\end{align}
Horizon regularity is imposed at $u=k$ itself, with no displaced endpoint. The outer conditions are $\phi(1)=\xi$, $g(1)=1$, and the horizon constraint is imposed along with the scalar regularity conditions. The non-extremality parameter $k$ is an additional Newton unknown enforcing the prescribed physical temperature
\begin{equation}
\vartheta=\frac{k\sqrt{1-k^2}}{(1+k)^2g_h^2}.
\label{eq:numTemperature}
\end{equation}
The target-connection terms are retained in the protected model. The calculations use 80 decimal digits and 80--208 collocation intervals; analytic Jacobians give fixed-source derivatives with respect to $k$ for the first-law check.

The extremal calculation solves the stable Hamilton--Jacobi branch and its amplitude equation,
\begin{equation}
\begin{gathered}
v=w^2+G^{ij}w_iw_j,\qquad w(p)=1,\qquad G^{ij}w_i\partial_jH^a=wH^a,\qquad A^a(\xi)=H^a(\xi),
\end{gathered}
\label{eq:numAmplitude}
\end{equation}
with the canonical slow normalization. Here $w=W/Q$, $z=u/g$, and $z(1)=1$. This determines $A(\xi)$ independently of the thermal fits. Truncations of the series at orders 100 and 140 differ by at most $1.38\times10^{-64}$ in $A$, and direct integration of the exact equations agrees to 29--41 digits. For example,
\begin{equation}
\begin{aligned}
&A_{\rm E_a}(0.3)=0.31005940695317581180678244915206412311,
\end{aligned}
\label{eq:numAmplitudeExample}
\end{equation}
and the series $A_{\rm E_a}(\epsilon)=\epsilon+a_3\epsilon^3+O(\epsilon^5)$ has $a_3=\frac{277}{750}$, as in Eq.~\eqref{eq:a3}.

There are 48 temperatures for each of 22 sources, hence 1,056 accepted points, covering $10^{-3}\le\vartheta\le10^{-1}$. Across those points the maximum constraint residual is $9.77\times10^{-51}$ and
\begin{equation}
\left|\frac{dF}{dk}+\frac{S}{2\pi}\frac{d\vartheta}{dk}\right|
\le1.23\times10^{-47}.
\label{eq:numFirstLaw}
\end{equation}
Grid and precision comparisons and a shooting calculation support errors below $10^{-28}$ at the points compared.

\subsection{Coefficient extraction and uncertainty}
At fixed source write $F_{\log}=C_F\vartheta^5\log\vartheta$. The entropy relation is
\begin{equation}
S_{\log}=-2\pi C_F\vartheta^4(5\log\vartheta+1),
\qquad C_S=-10\pi C_F.
\label{eq:numEntropyCoefficient}
\end{equation}
Free energy and entropy are fitted separately, with no predicted coefficient used as a prior or constraint; since both come from the same solutions and obey the first law to $10^{-47}$, their agreement is a consistency check of the fits rather than an independent measurement. The primary fit subtracts the exact analytic Reissner--Nordstr\"om background and uses the 42 temperatures in $[0.001,0.06]$. Its 37 free-energy basis functions consist of analytic powers $0$--$27$, $\vartheta^5\log\vartheta$, and $\vartheta^j(\log\vartheta)^p$ for $6\le j\le9$, $p=1,2$. The entropy fit uses analytic powers $0$--$26$, the leading $\vartheta^4\log\vartheta$, and logarithmic orders $5$--$8$. Column-normalized normal equations are solved at 160-digit precision.

The retained variations use analytic degrees $21$--$29$, upper endpoints $0.03$--$0.08$, a lower endpoint of $0.0012$, no background subtraction, and a free leading squared logarithm. Additional fits allowing cubic logarithms at orders $6$--$9$ extend to $0.09$. A full-interval ninth-degree fit has appreciable truncation bias despite small residuals and is excluded as an estimate. For a diagnostic leading term
$\vartheta^5(C_1\log\vartheta+C_2\log^2\vartheta)$, the entropy coefficient must be converted with
\begin{equation}
C_1=-\frac{S_{\log1}}{10\pi}-\frac25C_2.
\label{eq:numLogSquaredConversion}
\end{equation}
The input-error amplification, measured by the $\ell^1$ norm of the regression weights for the desired coefficient, is about $10^{31}$ in the primary free-energy fit and reaches $2.21\times10^{37}$ in the extended fits. The error envelope $\Delta C$ is twice the largest retained free-energy, entropy, basis or interval deviation from the primary free-energy estimate, with the conditional numerical floor $10^{-48}$ times this amplification included. It is an empirical systematic envelope, not a statistical confidence interval or a rigorous asymptotic remainder. A common bias outside the tested bases is not excluded.

\subsection{Results}
Table~\ref{tab:finiteSources} gives every source and the prediction
$C_{\rm pred}=-\tfrac49\|\Gamma(A(\xi),A(\xi))\|^2$.
For the 19 sources with a nonzero prediction, the largest central relative discrepancy of the free-energy coefficients is $1.26\times10^{-8}$, and that of the entropy coefficients is $1.87\times10^{-8}$; the largest relative envelope is $4.11\%$. The maximum relative difference between the independently extracted free-energy and entropy coefficients is $6.05\times10^{-9}$.

\begin{table*}[!t]
\centering
\footnotesize
\begin{tabular}{crrrrr}
\hline
Source & $\epsilon$ & $C_F$ & $C_{\rm pred}$ & $C_F/C_{\rm pred}$ & $\Delta C$ \\
\hline
$\mathrm E_a$ & 0.02 & $-3.4153509\times10^{-8}$ & $-3.4153509\times10^{-8}$ & $0.999999999$ & $4.41\times10^{-11}$ \\
$\mathrm E_a$ & 0.04 & $-5.4742559\times10^{-7}$ & $-5.4742559\times10^{-7}$ & $0.999999999$ & $4.41\times10^{-11}$ \\
$\mathrm E_a$ & 0.08 & $-8.8210969\times10^{-6}$ & $-8.8210969\times10^{-6}$ & $1.000000000$ & $1.77\times10^{-10}$ \\
$\mathrm E_a$ & 0.12 & $-4.5186748\times10^{-5}$ & $-4.5186748\times10^{-5}$ & $1.000000000$ & $4.20\times10^{-10}$ \\
$\mathrm E_a$ & 0.16 & $-1.4518672\times10^{-4}$ & $-1.4518672\times10^{-4}$ & $1.000000001$ & $8.04\times10^{-10}$ \\
$\mathrm E_a$ & 0.2 & $-3.6203726\times10^{-4}$ & $-3.6203726\times10^{-4}$ & $1.000000003$ & $1.39\times10^{-9}$ \\
$\mathrm E_a$ & 0.3 & $-1.9716888\times10^{-3}$ & $-1.9716888\times10^{-3}$ & $1.000000008$ & $4.48\times10^{-9}$ \\
$\mathrm E_g$ & 0.02 & $-1.3528262\times10^{-8}$ & $-1.3528262\times10^{-8}$ & $0.999999999$ & $4.41\times10^{-11}$ \\
$\mathrm E_g$ & 0.04 & $-2.0958421\times10^{-7}$ & $-2.0958421\times10^{-7}$ & $0.999999999$ & $4.41\times10^{-11}$ \\
$\mathrm E_g$ & 0.08 & $-3.1503495\times10^{-6}$ & $-3.1503495\times10^{-6}$ & $0.999999999$ & $1.70\times10^{-10}$ \\
$\mathrm E_g$ & 0.12 & $-1.5024456\times10^{-5}$ & $-1.5024456\times10^{-5}$ & $1.000000000$ & $3.97\times10^{-10}$ \\
$\mathrm E_g$ & 0.16 & $-4.4857735\times10^{-5}$ & $-4.4857735\times10^{-5}$ & $1.000000000$ & $7.46\times10^{-10}$ \\
$\mathrm E_g$ & 0.2 & $-1.0374875\times10^{-4}$ & $-1.0374875\times10^{-4}$ & $1.000000001$ & $1.26\times10^{-9}$ \\
$\mathrm E_g$ & 0.3 & $-4.6455254\times10^{-4}$ & $-4.6455254\times10^{-4}$ & $1.000000003$ & $3.72\times10^{-9}$ \\
$\mathrm D$ & 0.05 & $-3.3156225\times10^{-9}$ & $-3.3156225\times10^{-9}$ & $1.000000013$ & $1.36\times10^{-10}$ \\
$\mathrm D$ & 0.1 & $-1.9842755\times10^{-7}$ & $-1.9842755\times10^{-7}$ & $1.000000001$ & $5.97\times10^{-10}$ \\
$\mathrm D$ & 0.15 & $-2.1451533\times10^{-6}$ & $-2.1451533\times10^{-6}$ & $1.000000002$ & $1.58\times10^{-9}$ \\
$\mathrm D$ & 0.2 & $-1.1607133\times10^{-5}$ & $-1.1607133\times10^{-5}$ & $1.000000004$ & $3.55\times10^{-9}$ \\
$\mathrm D$ & 0.3 & $-1.2796178\times10^{-4}$ & $-1.2796178\times10^{-4}$ & $1.000000011$ & $1.69\times10^{-8}$ \\
$\mathrm P$ & 0.05 & $-1.6334040\times10^{-17}$ & $0$ & --- & $5.06\times10^{-11}$ \\
$\mathrm P$ & 0.15 & $-1.6213879\times10^{-16}$ & $0$ & --- & $5.04\times10^{-10}$ \\
$\mathrm P$ & 0.3 & $-9.0017922\times10^{-16}$ & $0$ & --- & $2.84\times10^{-9}$ \\
\hline
\end{tabular}
\caption{Finite-source coefficients at $Q=1$. Boundary values are
$\mathrm E_a:(x,y)=(\epsilon,0)$,
$\mathrm E_g:(x,y)=(4\epsilon/5,3\epsilon/5)$,
$\mathrm D:(x_1,x_2,y)=(\epsilon,\epsilon,0)$, and
$\mathrm P:(x,y)=(\epsilon,1)$.
The ratio is undefined when $C_{\rm pred}=0$. Every entry uses the primary fit to the 42 temperatures with $10^{-3}\le\vartheta\le0.06$.
The displayed digits are central estimates; $\Delta C$ is the systematic envelope defined in the text, which includes variations with higher logarithmic powers.}
\label{tab:finiteSources}
\end{table*}

\paragraph{Coefficient-level convergence.} Because the regression weights amplify input errors by $\|w\|_1=2.52\times10^{31}$ for the primary free-energy fit ($2.30\times10^{26}$ for the entropy fit), the numerical contribution to the coefficients was checked directly. For the sources D with $\epsilon=0.05$ (the largest central discrepancy) and $\epsilon=0.3$, E$_a$ with $\epsilon=0.3$, E$_g$ with $\epsilon=0.02$, and P with $\epsilon=0.3$, all 48 thermal points were recomputed at the same targets with $n+32$ collocation intervals, 110 digits and Newton tolerance $10^{-75}$. Over the fitted temperatures the free-energy and entropy values change by at most $2.5\times10^{-52}$ and $3.8\times10^{-52}$. Propagated through the weights, these differences bound the change of $C_F$ by at most $6.3\times10^{-21}$ in absolute terms and, source by source, by at most $1.4\times10^{-15}$ of the prediction; the actual relative changes of $C_F$ for the four nonzero predictions are below $7.6\times10^{-17}$, and the protected coefficient changes by $3\times10^{-25}$. The refined fits reproduce the central discrepancies quoted above. The numerical solutions therefore do not limit the comparison.

\paragraph{Earlier logarithms.} The primary free-energy basis contains no $\vartheta^j\log\vartheta$ with $j<5$, and the entropy basis none with $j<4$. Adding $\vartheta^3\log\vartheta$ and $\vartheta^4\log\vartheta$ to the free-energy fits, and in a second variant also $\vartheta^2\log\vartheta$, with the powers lowered by one in the entropy fits (conversion $S=-\partial_TF$), for all 22 sources gives early coefficients at most $7.1\times10^{-15}$ of the low-order analytic coefficients and $2\times10^{-9}$ of the fifth-order logarithmic coefficient, and changes $C_F$ by at most $8.3\times10^{-6}$ in relative terms for the nonzero predictions. The data thus show no logarithm below order five, within the resolution allowed by the conditioning of these enlarged fits. The protected model is consistent with zero, with the largest reported bound $|C_F|\le2.84\times10^{-9}$.

For the exponential axis, the leading quartic source approximation first differs from the resummed prediction by more than $1\%$ at $\epsilon=0.12$, and by $12.36\%$ at $\epsilon=0.3$; including the known sixth-order term reduces the latter discrepancy to $0.707\%$. In the generic exponential direction, the leading quartic approximation differs by $3.35\%$ at $\epsilon=0.02$ and $52.36\%$ at $\epsilon=0.3$. In the delayed model the quartic source coefficient vanishes, while the leading sixth-order expression is $-0.2304\epsilon^6$. It differs from the resummed prediction by $8.58\%$ at $\epsilon=0.05$ and $31.26\%$ at $\epsilon=0.3$.

A restricted fit at the largest exponential-axis source gives an apparent leading $\vartheta^5\log^2\vartheta$ coefficient of approximately $-1.36\times10^{-11}$. Allowing subleading cubic logarithms changes it substantially, and the extended envelope contains zero. No robust additional leading logarithm is resolved.

\subsection{Finite-temperature Calabi--Yau solutions}\label{App:NumericsCY}
The law~\eqref{eq:IRallorders} was tested on the two unaligned exteriors of Appendix~\ref{App:CY}, in the dilated frames of the representatives 18-U and B-U (dilation factors 6 and 10), where $Q=174.900407316$ and $137.886535346$. The four real scalars $(x_1,x_2;y_1,y_2)$ have the metric $g\oplus g$ and the potential $V=|Z|^2+g^{a\bar b}D_aZ\,\overline{D_bZ}$ of the cubic prepotential, and the exponents are those of Eqs.~\eqref{cy:degree18unaligned} and~\eqref{cy:bicubicunaligned}; the slowest one is $c=0.1870464548$ (18-U) and $0.2367652143$ (B-U), and the next distinct ones are $c_2=1$ and $0.5325677192$. Two kinds of boundary data are used. The generic sources are the dilated asymptotic moduli of the exteriors of Table~\ref{tab:cyexteriors},
\begin{equation*}
\begin{gathered}
(-1.744958698,6.742542636;3.376556186,4.299715832),\\
(5.121384172,-5.632383758;4.922035261,3.623918470),
\end{gathered}
\end{equation*}
for 18-U and B-U. The source families are $\phi_\infty=\operatorname{Exp}_p(\epsilon e)$ with a fixed unit vector $e$ whose component along the slowest eigenvector is $0.05$, so that the linear amplitude is $\xi_c=0.05\,\epsilon$, for eight values $2.5\times10^{-5}\le\epsilon\le0.1$; the direction $e$ was chosen before the thermal fits.

\paragraph{Method.} With $q=(\log g,\phi)$, the difference $\delta=q_k-q_0$ between the thermal profile and the extremal profile with the same boundary data is solved from the exact equation that follows from Eq.~\eqref{eq:compactfull} and its extremal limit, by piecewise Chebyshev collocation in $\log u$ on a mesh shared by the two profiles; horizon regularity~\eqref{fnd:horizondata} is imposed at $u=k$. The mass difference $M-M_{\rm ext}$ is read from the derivative of $\delta$ at infinity, the entropy from the horizon area, and the temperature from Eq.~\eqref{fnd:physicalT}. The extremal amplitude $A_c$ is obtained separately, from the extremal profile alone: it is the intercept of $z^{-c}\,\Pi_c(\phi-p)$ fitted against $1,z^c,z^{2c},z^{3c}$ deep in the throat. There are 51 values of $k$ between $10^{-3}$ and $10^{-18}$ for each generic source and for six members of each family, shorter series for the two smallest sources, and additional points down to $k=10^{-24}$, 856 solutions in all. Two checks bound the numerical error. Refining the mesh and the collocation order changes $A_c$, the entropy and the free energy by at most $1.6\times10^{-10}$ in relative terms. Separate solutions of the extremal and thermal problems at 48 and 64 decimal digits, with direct subtraction of the two masses, agree with the entropy, the free energy and the mass difference to $3.5\times10^{-11}$ at the points checked, down to $k=10^{-24}$.

\paragraph{Fits.} The exact Reissner--Nordstr\"om functions at the same $\vartheta$ are subtracted, and $S/S_0$ and $F$ are fitted separately as functions of the physical temperature. The coefficients $C_S$ of $\vartheta^{2c}$ and $C_F$ of $\vartheta^{2c+1}$ are free in the fits; the prediction enters only in the comparison afterwards. The central fit uses $10^{-16}\lesssim\vartheta\lesssim10^{-6}$ and, besides the leading power, the powers $3c,4c,5c,1,2,2c_2$ for the entropy and the same powers raised by one for the free energy, as allowed by Eq.~\eqref{eq:IRtower}. The variations change the temperature window and the number of powers of the slowest mode, and add the mixed powers $c+c_2$ and $2c+c_2$. In a further set of fits the leading exponent is also free.

\begin{table*}[t]
\centering
\footnotesize
\setlength{\tabcolsep}{3pt}
\begin{tabular}{llllllll}
\hline
Model & Source & $A_c$ & $\xi_c$ & $C_S/C_S^{\rm pred}$ & $C_F/C_F^{\rm pred}$ & $p_S$ & $p_F$ \\
\hline
18-U & generic & $0.02632561963$ & $0.02565347944$ & $1.000000007$ & $1.000000004$ & $0.374092911$ & $1.374092909$ \\
 & $\epsilon=0.05$ & $0.003002672490$ & $0.0025$ & $1.000000007$ & $1.000000004$ & $0.374092911$ & $1.374092909$ \\
 & $\epsilon=0.10$ & $0.006636180304$ & $0.005$ & $1.000000007$ & $1.000000004$ & $0.374092911$ & $1.374092909$ \\
\hline
B-U & generic & $0.02330843089$ & $0.02449501280$ & $0.999999999$ & $1.000000000$ & $0.473530430$ & $1.473530430$ \\
 & $\epsilon=0.05$ & $0.004060061223$ & $0.0025$ & $0.999999985$ & $0.999999991$ & $0.473530423$ & $1.473530426$ \\
 & $\epsilon=0.10$ & $0.01119923262$ & $0.005$ & $0.999999978$ & $0.999999986$ & $0.473530419$ & $1.473530423$ \\
\hline
\end{tabular}
\caption{Finite-temperature solutions on the unaligned Calabi--Yau branches. $A_c$ is the nonlinear amplitude of the slowest mode in the extremal exterior and $\xi_c$ the linear projection of the boundary data. The ratios compare the fitted coefficients of $\vartheta^{2c}$ in $S$ and of $\vartheta^{2c+1}$ in $F$ (central fits) with Eq.~\eqref{eq:IRallorders} evaluated with $A_c$. The last two columns are the exponents found when the leading power is fitted, to be compared with $2c=0.3740929096$ (18-U) and $0.4735304286$ (B-U) and with $2c+1$.}
\label{tab:cythermo}
\end{table*}

\paragraph{Results.} Table~\ref{tab:cythermo} lists the two generic sources and two members of each family. For the 18 source series the central fitted coefficients agree with Eq.~\eqref{eq:IRallorders} to $2.3\times10^{-8}$, and to $4.9\times10^{-7}$ across the fit windows and bases with the exponent fixed; the central fits with a free exponent reproduce $2c$ and $2c+1$ to $10^{-8}$. Stress fits that omit needed terms or use only the warmest data deviate by up to $2.4\times10^{-4}$, and by up to $4.2\times10^{-3}$ with the exponent free. For B-U the term $\vartheta^{2c_2}$, with $2c_2=1.0651354383$ close to the analytic power, is needed at this accuracy. The linear amplitude does not reproduce the data. With $|\xi_c|^2$ in place of $\|A_c\|^2$, the prediction underestimates the magnitudes of both coefficients by the factors $1.44257$ (18-U) and $2.63746$ (B-U) at $\epsilon=0.05$ and by $1.76156$ and $5.01691$ at $\epsilon=0.1$, equal to $A_c^2/\xi_c^2$ in each case, and they approach the linear prediction as the source decreases: at $\epsilon=2.5\times10^{-5}$ the differences are $2.5\times10^{-4}$ and $6.2\times10^{-4}$. The horizon values test Eq.~\eqref{eq:IRhorizonvalue} without any fit in temperature: for the six sources of Table~\ref{tab:cythermo} the ratio $\Pi_c(\phi_h-p)/(N_cA_ck^c)$, with $N_c=1.087934$ (18-U) and $1.099900$ (B-U), differs from one by at most $3.6\times10^{-4}$ at $k=10^{-12}$, $2.7\times10^{-5}$ at $k=10^{-18}$ and $2.1\times10^{-6}$ at $k=10^{-24}$. The first law, evaluated from the raw sequences by finite differences in $\log\vartheta$ before any fit, holds to $2.6\times10^{-8}$ for $dF=-S\,dT$ and to $9.4\times10^{-10}$ for $dM=T\,dS$, and the separately fitted $C_S$ and $C_F$ of the six sources satisfy $C_S=-2\pi Q(2c+1)C_F$ to $8.0\times10^{-9}$. As above, the fit envelopes measure the sensitivity to the windows and bases tried; subleading mixed coefficients are not determined. Without any fit, the ratio~\eqref{eq:IRratio} formed from the
horizon gradient and area of the generic solutions departs from $H(c)$ by $-3.2\times10^{-8}$ (B-U) and
$1.7\times10^{-6}$ (18-U) at $k=10^{-24}$.

\paragraph{An integral D0--D4 charge.} For the charges $(1440,0,0;0,8,-6)$ on the bicubic
(Appendix~\ref{special:fived}), $Q=42.3095841$, we imposed $\phi_\infty=p+0.07\,e_c$, with $e_c$ the unit slow
eigenvector, and solved at $k=10^{-4},10^{-8},10^{-12}$ on two meshes and at 48 digits, with agreement to
$2\times10^{-11}$. The extremal amplitude $\|A_c\|=0.0701782222$ gives $s=0.00111463784$. With
$E_S=(S-S_0)/(S_0s\vartheta^{2c})$ and $E_F=(F-M_{\rm ext}+S_0T)/[-Qs\vartheta^{2c+1}/(2(2c+1))]$ the three
temperatures give $E_S=3.473$, $1.00329$, $0.9999955$ and $E_F=2.588$, $1.00209$, $0.9999955$, and the
ratio~\eqref{eq:IRratio} over $H(0.142464)=14.823637$ is $3.486$, $1.00351$, $1.0000120$; all three are within $1\%$
of one below $k\simeq4.4\times10^{-8}$. At these charges the range where $\vartheta^{2c}$ dominates the linear
entropy, $\vartheta\lesssim2.8\times10^{-5}$, and the range where a probe of Planck size has $R_{kk}|_h\ell^2<1$,
$\vartheta\gtrsim5.0\times10^{-5}$, do not overlap; after a uniform scaling of the charges by $10^4$, which leaves $c$
and $A_c$ unchanged, the anomalous entropy at $\vartheta=10^{-7}$ is 56 times the linear term while
$R_{kk}|_h\ell^2\simeq4\times10^{-4}$. Here $|\epsilon_{\alpha'}|/D_*=1.7\times10^{-4}$ and $c_2\cdot y_*/(6Q^2)=0.062$,
which falls as $N^{-2}$ under a uniform scaling by $N$.

\section{Reproducibility and availability}\label{App:Reproducibility}
The scripts, input parameters, final data and run logs that regenerate every computed number, table and figure of this paper, together with independent reimplementations of the main computations, are openly available at \url{https://github.com/Ailun-Zhang/horizon-resonances}~\cite{HRcode}. The repository documents the software, the precision settings and the run instructions. The Calabi--Yau exterior solutions of Appendix~\ref{App:CY} are computed in double precision from attractor data refined to high precision.

\providecommand{\href}[2]{#2}\begingroup\raggedright\endgroup

\end{document}